\documentclass[aps,prc,showpacs,amsmath,amssymb,onecolumn,superscriptaddress,letterpaper,nofootinbib, showkeys]{revtex4-2}

\usepackage{hyperref}
\usepackage{graphicx}
\usepackage{dcolumn}
\usepackage{bm} 
\usepackage{xcolor}
\usepackage[mathlines]{lineno}
\usepackage[T1]{fontenc}
\usepackage{orcidlink}
\usepackage[parfill]{parskip}
\usepackage{xspace}
\usepackage{booktabs}
\usepackage{array}
\usepackage{nicefrac}
\usepackage{wrapfig}
\usepackage{rotating}
\usepackage{soul}
\usepackage{textcomp}
\usepackage{multirow}

\usepackage{comments}
\usepackage{definitions}

\begin{document}

\preprint{}

\title{Study of deeply virtual Compton scattering at the future Electron-Ion Collider}
\author{E.~C.~Aschenauer\,\orcidlink{0000-0002-3856-5475}}
\affiliation{Department of Physics, Brookhaven National Laboratory, Upton, New York 11973, USA}
\author{V.~Batozskaya\,\orcidlink{0000-0003-1089-9200}}
\affiliation{National Centre for Nuclear Research (NCBJ), Pasteura 7, 02-093 Warsaw, Poland}
\author{S.~Fazio\,\orcidlink{0000-0002-4321-1946}}
\affiliation{University of Calabria and INFN-Cosenza, 87036 Rende (CS), Italy}
\author{A.~Jentsch\,\orcidlink{0000-0001-7326-6991}}
\affiliation{Department of Physics, Brookhaven National Laboratory, Upton, New York 11973, USA}
\author{J.~Kim\,\orcidlink{0000-0003-1018-810X}}
\affiliation{Department of Physics, Brookhaven National Laboratory, Upton, New York 11973, USA}
\author{K.~Kumerički\,\orcidlink{0000-0001-9758-5647}}
\affiliation{Department of Physics, Faculty of Science, University of Zagreb, HR-10000 Zagreb, Croatia}
\author{H.~Moutarde\,\orcidlink{0000-0003-3143-6213}}
\affiliation{IRFU, CEA, Université Paris-Saclay, 91191 Gif-sur-Yvette, France}
\author{K.~Passek-K.\,\orcidlink{0000-0001-6520-3640}}
\affiliation{Division of Theoretical Physics, Ruđer Bošković Institute, HR-10002 Zagreb, Croatia}
\author{D.~Sokhan\,\orcidlink{0000-0002-7204-5860}}
\affiliation{University of Glasgow, Glasgow G12 8QQ, UK}
\author{H.~Spiesberger\,\orcidlink{0000-0002-1674-3736}}
\affiliation{PRISMA+ Cluster of Excellence, Institut für Physik, Johannes Gutenberg-Universität, 55099 Mainz, Germany}
\author{P.~Sznajder\,\orcidlink{0000-0002-2684-803X}}
\affiliation{National Centre for Nuclear Research (NCBJ), Pasteura 7, 02-093 Warsaw, Poland}
\author{K.~Tezgin\,\orcidlink{0000-0001-9492-9512}}
\affiliation{Department of Physics, Virginia Tech, Blacksburg, VA 24061, U.S.A.}
\affiliation{Department of Physics, Brookhaven National Laboratory, Upton, New York 11973, USA}

\date{\today}

\begin{abstract}
This study presents the impact of future measurements of deeply virtual Compton scattering (DVCS) with the ePIC detector at the electron-ion collider (EIC), currently under construction at Brookhaven National Laboratory. The considered process is sensitive to generalized parton distributions (GPDs), the understanding of which is a cornerstone of the EIC physics programme. Our study marks a milestone in the preparation of DVCS measurements at EIC and provides a reference point for future analyses. In addition to presenting distributions of basic kinematic variables obtained with the latest ePIC design and simulation software, we examine the impact of future measurements on the understanding of nucleon tomography and DVCS Compton form factors, which are directly linked to GPDs. We also assess the impact of radiative corrections and background contribution arising from exclusive $\pi^0$ production.
\end{abstract}

\keywords{quantum chromodynamics, QCD, generalized parton distributions, GPD, deeply virtual Compton scattering, DVCS, Electron-Ion Collider, EIC, ePIC detector}

\maketitle

\section{Introduction}
\label{sec:intro}


Factorization theorems formulated within quantum chromodynamics (QCD) allow for the definition of reaction-independent objects describing hadrons. Among them, generalized parton distributions (GPDs)~\cite{Muller:1994ses,Radyushkin:1997ki,Ji:1996nm,Ji:1998pc} offer a rigorous framework that can be used to investigate the three-dimensional structure of nucleons. In particular, GPDs allow for nucleon tomography~\cite{Burkardt:2000za}, i.e. for the determination of the spatial distribution of partons carrying a specific fraction of a hadron's momentum. Tomography enables the depiction of nucleons as spatially extended objects composed of quarks and gluons, and it combines the partial information obtained by the one-dimensional parton distribution functions (PDFs) and elastic form factors. Another important feature of GPDs is their connection to the energy-momentum tensor form factors~\cite{Ji:1996ek,Ji:1996nm}. This connection enables the determination of the total angular momentum carried by quarks of a given flavour and gluons, potentially contributing to a better understanding of the spin structure of the nucleon. The relation between GPDs and the energy-momentum tensor can also be used to determine the ``mechanical'' properties of systems composed of partons, such as the distribution of pressure inside nucleons~\cite{Polyakov:2002yz, Goeke:2007fp, Polyakov:2018zvc, Lorce:2018egm}. 

The extraction of GPDs is based on information obtained from exclusive reactions, where the states of all incoming and outgoing particles are reconstructed in the experiment. This also includes the target hadron, which is required to remain intact during the interaction, thereby allowing the investigation of coherent changes in its state. The effectiveness of the GPD formalism in describing hadronic structure has been confirmed thanks to the development of theory and phenomenology methods, as well as the global experimental program conducted at laboratories such as DESY, CERN, and JLab (for a review of available results see for instance Refs.~\cite{Favart:2015umi, Kumericki:2016ehc, dHose:2016mda}). Nevertheless, the current state of reconstructing GPDs and other related quantities, such as the nucleon tomography and total angular momentum of partons, remains inadequate to describe the three-dimensional structure of the nucleon. This inadequacy is mainly due to insufficient experimental data, which motivates new measurements in colliders and with fixed target experiments. Among a variety of new proposals, such as future JLab experiments~\cite{Accardi:2023chb} and the electron-ion collider in China~\cite{Anderle:2021wcy}, this work focuses on the future electron-ion collider (EIC) and its experiments to be constructed at Brookhaven National Laboratory (BNL)~\cite{AbdulKhalek:2021gbh}. In our analysis, we aim to assess the possible impact of future EIC experimental data on extracting Compton form factors, which are convolutions of GPDs, as well as nucleon tomography. The EIC offers multiple advantages for our purposes as it will be the first collider of its class offering polarized beams (including light nuclei), a variable collision center of mass energy, and high luminosity. Currently, one experimental collaboration has been formed at the EIC, called ePIC~\cite{ref:ePIC_detector}.

The physics case for the EIC has been extensively documented in multiple reports. To date, the yellow report~\cite{AbdulKhalek:2021gbh} provides the most comprehensive source of information, aiming to define the requirements for the new machine to accomplish specific goals in exploring the internal structure of nucleons and nuclei. In this analysis, we expand on what was previously presented in the yellow report for deeply virtual Compton scattering (DVCS) off the proton. We stress another study of this type, which was published in Ref.~\cite{Aschenauer:2013hhw}. To the best of our knowledge, this has been the only other study of this type during the last decade. The improvements with respect to that former analysis include the use of state-of-the-art simulation software based on the up-to-date ePIC detector design, a modern Monte Carlo generator called EpIC\footnote{The names of the event generator EpIC and of the EIC detector ePIC had been independently chosen and should not be confused.}~\cite{Aschenauer:2022aeb}, a study of radiative corrections, an estimation of the  $\pi^0$ background, a new phenomenology analyses of nucleon tomography, and the extraction of DVCS sub-amplitudes (Compton form factors), involving machine learning techniques to minimize the model dependency. The goal of this analysis is to provide a new milestone in the preparation of DVCS measurements at the EIC and a new reference point for future phenomenological applications related to this process. 

This article is organized as follows. In Sect.~\ref{sec:theory} we review the description of DVCS and elements of the GPD framework important for our phenomenology analysis. In Sect.~\ref{sec:simulations} we describe the simulation tools used in this analysis and their set-up. Obtained results are presented in Sect.~\ref{sec:analysis}: we start with distributions of relevant kinematic variables, then we discuss radiative corrections and the $\pi^0$ background, finally we present the extraction of nucleon tomography information and Compton form factors from selected asymmetries. The conclusions are provided in Sect.~\ref{sec:conclusions}.  
\section{Process and theory} 
\label{sec:theory}

\begin{wrapfigure}[19]{r}{0.25\textwidth}
\begin{center}
\includegraphics[width=0.25\textwidth]{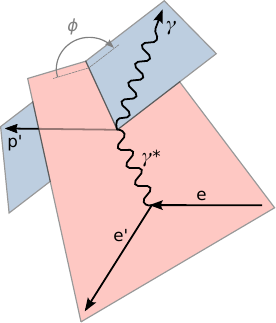}
\end{center}
\caption{The proton-at-rest frame: definition of the azimuthal angle~$\phi$.}
\label{fig:kinematics}
\end{wrapfigure}

The process under consideration is exclusive electroproduction of a single photon off a nucleon,
\begin{equation}
e(k) + p(p) \to e'(k') + p'(p') + \gamma(v)\,,
\label{eq:process}
\end{equation}
where the symbols in parentheses denote the four-momenta of the respective particles. If the target is not polarized transversely (which is the case considered in this study), the process can be described by three invariants and one scattering angle depicted (and defined) in Fig.~\ref{fig:kinematics}:~~~~~~~~~~~~~~~~~~~~~~~~~~~~~~~~~~~~~~~~~~~
\begin{equation}
    \frac{\dd^4 \sigma}{\dd \xB\, \dd Q^2\, \dd t\, \dd \phi} = \frac{\alpha_\text{em}^3\,\xB\,y^2}{8\,\pi\,Q^4\,\sqrt{1+\varepsilon^2}}\,\left| \mathcal{T} \right|^2 \,,
    \label{eq:process_cs}
\end{equation}
where $\alpha_\text{em}$ denotes the electromagnetic fine structure constant, $Q^2=-(k-k')^2=-q^2$ the negative four-momentum squared of the virtual photon, $\xB=Q^{2}/(2p\cdot q)$ the fraction of the proton momentum carried by the quark struck by the virtual photon in the infinite-momentum frame (the Bjorken variable), $y = (p\cdot q)/(p\cdot k)$ the inelasticity variable, $t=(p'-p)^2$ the squared four-momentum transfer at the proton vertex, and $\varepsilon = 2\,\xB\,M\,/Q$ the kinematic factor with $M$ standing for the proton mass. The reaction~\eqref{eq:process} can be described by two interfering sub-processes: DVCS and Bethe-Heitler (BH), where the latter is a pure electromagnetic process not probing the partonic content. The total amplitude squared is therefore:
\begin{equation}
    \left| \mathcal{T} \right|^2 = \left| \mathcal{T}_\text{DVCS} \right|^2 + \left| \mathcal{T}_\text{BH} \right|^2 + \mathcal{I} \, ,
    \label{eq:single_photon_production_amplitude}
\end{equation}
where $\mathcal{I}$ is the interference term: 
\begin{equation}
    \mathcal{I} = \mathcal{T}_\text{DVCS}\,\mathcal{T}_\text{BH}^{*} + \mathcal{T}_\text{DVCS}^{*}\,\mathcal{T}_\text{BH} \, .
\end{equation}
The DVCS and BH sub-process amplitudes, $\mathcal{T}_\text{DVCS}$ and $\mathcal{T}_\text{BH}$, can be parametrized in terms of experimentally accessible Compton (CFFs) and elastic form factors, respectively~\cite{Belitsky:2012ch}. This parametrization also includes the dependence on the charge and polarization of the beam and target particles.

The CFFs can be represented as convolution integrals of GPDs parametrizing the off-forward matrix elements of quark and gluon bilinear operators~\cite{Muller:1994ses,Radyushkin:1997ki,Ji:1996nm,Ji:1998pc} with hard perturbative coefficient functions, the latter being calculated within perturbative QCD. At leading order (LO) and including leading twist (LT) only, this relationship reads: 
\begin{equation}
    \Bigg\{ \substack{ \displaystyle \mathcal{F}(\xi,t) \\ \displaystyle \mathcal{\widetilde{F}}(\xi,t) } \Bigg\}
     = \sum_q e_q^2 \int_{-1}^{1} \dd x \Bigg[\frac{1}{\xi-x-i\epsilon}\mp \frac{1}{\xi+x-i\epsilon} \Bigg] \Bigg\{ \substack{ \displaystyle F^q(x,\xi,t) \\ \displaystyle \widetilde{F}^q(x,\xi,t) } \Bigg\} \, ,
     \label{eq:GPDtoCFF}
\end{equation}
where $q\in\{u,d,s\}$ denotes a light quark flavor (gluons do not contribute at LO), with $e_{q}$ representing the corresponding electric charge in units of the positron charge. The $F^q$ ($\widetilde{F}^q$) objects stand for the unpolarized (polarized) chiral-even GPDs $H^q$ or $E^q$ ($\widetilde{H}^q$ or $\widetilde{E}^q$) which give rise to the associated CFFs, generically denoted by $\mathcal{F}$ ($\mathcal{\widetilde{F}}$) and representing $\mathcal{H}$ or $\mathcal{E}$ ($\widetilde{\mathcal{H}}$ or $\widetilde{\mathcal{E}}$). In this regard, CFFs serve as a bridge between experimental results and the partonic interpretation of the nucleon. CFFs can also be used to directly quantify how hadrons respond to non-local QCD probes using techniques based on Froissart-Gribov projections~\cite{Semenov-Tian-Shansky:2023ysr}. The extraction of CFFs from data will be one of the topics discussed below. 
\begin{wrapfigure}[15]{r}{0.25\textwidth}
\begin{center}
\includegraphics[width=0.25\textwidth]{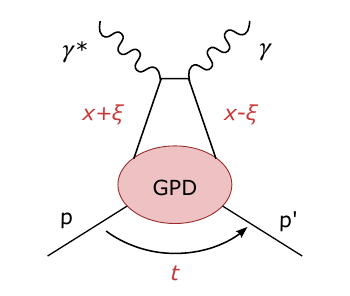}
\end{center}
\caption{DVCS at the partonic level described with GPDs (leading twist/leading order contribution).}
\label{fig:gpd}
\end{wrapfigure}

The GPDs depend on three kinematic variables illustrated in Fig.~\ref{fig:gpd}. These are: $x$, representing the average longitudinal momentum fraction carried by the active parton; $\xi$, describing the longitudinal momentum transfer; and the previously introduced $t$, describing the four-momentum transfer to the nucleon. For DVCS described with LT accuracy $\xi = \xB / (2 - \xB)$, which for EIC kinematics ($\xB \ll 1$) gives $\xi \approx \xB/2$. Similarly to other objects describing the partonic content of hadrons, like one-dimensional parton distribution functions (PDFs), GPDs also depend on the factorization scale, $\muF^2$. The DVCS hard-scattering sub-process, 
whose contribution is convoluted with GPDs as shown in Eq.~\eqref{eq:GPDtoCFF}, when calculated in higher orders, also depends on $\muF^2$, as well as on the coupling constant renormalization scale, $\muR^2$. In phenomenological applications, these two scales are often chosen to be equal to each other and denoted by $\mu^2$, corresponding to the typical scale of the process, which in DVCS is $Q^2$. The scale dependence in Eq.~\eqref{eq:GPDtoCFF} is suppressed for brevity.

GPDs encode a wealth of new information about the distribution of quarks and gluons. Specifically, when there is no collinear momentum transfer at the partonic level, $\xi\to 0$, but a non-zero total four-momentum transfer occurs in the interaction, $t\neq 0$, GPDs encode the spatial distribution of quarks and gluons within the plane transverse to the nucleon's motion in the infinite momentum frame~\cite{Burkardt:2002ks, Burkardt:2002hr}. In particular, for an unpolarized target, the distribution of unpolarized quarks inside the nucleon is derived from the Fourier transform of GPD $H^q$ as follows:   
\begin{equation}
    q \big(x,\vec{b}_\perp \big) = \int \frac{\dd^2 \vec{\Delta}_\perp}{(2\pi)^2} e^{-i\vec{b}_\perp\cdot\vec{\Delta}_\perp} H^q \big(x,0,t=-\vec{\Delta}_\perp^2 \big)\,.
    \label{eq:nucleon_tomography}
\end{equation}
Relying on a few reasonable assumptions, nucleon tomography can be directly extracted in the low-$\xB$ domain from the DVCS cross-section. We will explore this subject and the possibility of such direct extraction of tomography information at the EIC in Sect.~\ref{subsec:tomography}. 

In addition to nucleon tomography, GPDs allow for unique access to elements of the QCD energy-momentum tensor form factors~\cite{Ji:1996ek,Ji:1996nm}. This feature enables access to the nucleon's total angular momentum in terms of quarks and gluons, and therefore can contribute to a better understanding of the spin structure of the nucleon. Moreover, the connection between GPDs and the QCD energy-momentum tensor form factors enables the exploration of the associated energy-momentum tensor densities, including the so-called ``mechanical'' forces, induced in partonic systems~\cite{Polyakov:2002yz, Goeke:2007fp, Polyakov:2018zvc, Lorce:2018egm}. This, however, requires a very good knowledge of CFFs $\mathcal{H}$ (both their real and imaginary parts) over a wide range of $\xB$, as the extraction of mechanical forces relies on the dispersion relations. Moreover, CFFs must be measured over a wide range of $Q^2$, allowing one to decouple the relevant information through the analysis of evolution effects. Some of the analyses addressing these topics are found in the literature~\cite{Kumericki:2019ddg, Dutrieux:2021nlz,Burkert:2023atx}. The EIC will crucially contribute to our knowledge of CFFs in the small and intermediate $\xB$ domain, complementing information obtained from low-energy experiments. In addition, the long range in $Q^2$ offered by the EIC will allow us to have a much better grasp of the evolution effects. In Sect.~\ref{subsec:asymmetries}, we demonstrate the expected impact of the EIC on the extraction of CFFs from one of the observables.
\section{Simulation tools}
\label{sec:simulations}

This study is based on samples of Monte Carlo (MC) events generated with the EpIC~\cite{Aschenauer:2022aeb,www_epic} generator. In Sect.~\ref{subsec:epic}, we summarize the conditions and physics assumptions used in the generation, as well as the additional cuts applied at the analysis level. To assess the detector effects, particularly the geometrical acceptances and energy smearing, we use the detector response simulation suite described in Sect.~\ref{subsec:detectors}.

\subsection{EpIC Monte Carlo generator and generation conditions} 
\label{subsec:epic}

EpIC~\cite{Aschenauer:2022aeb,www_epic} is a versatile MC generator dedicated to exclusive reactions. The generator is built upon the PARTONS framework~\cite{Berthou:2015oaw,www_partons} and represents an advanced version of the primitive toyMC generator used in the analysis included in the EIC yellow report~\cite{AbdulKhalek:2021gbh}. EpIC can generate events for multiple exclusive processes and incorporates QED radiative corrections, currently within the framework of collinear approximation~\cite{Kripfganz:1990vm}. EpIC can be utilized for simulations in both collider and fixed-target kinematic setups, with the capability of saving the generated events in multiple formats (including HepMC3~\cite{Buckley:2019xhk,www_hepmc3}).

Similarly to the PARTONS framework, EpIC uses a modular structure. This feature enables for the separation of parts of the code responsible for performing distinct tasks, such as the random generation of kinematic configurations according to a given cross section, thereby aiding in navigation within the already complex project. Furthermore, this modular structure enables the creation of multiple modules of the same type, which can differ, for instance, in the computational algorithms or details of the process description. This setup provides an easy way to select between different modules.

The modules used in the generation of MC events for this study are executed in the following order:
\begin{itemize}
    \item Cross section for the process~\eqref{eq:process} evaluated in terms of Compton and elastic FFs using a set of unpublished analytic expressions developed by Guichon and Vanderhaeghen (also used, for instance, in~Ref.~\cite{Kroll:2012sm}). The parameterizations of Sachs elastic FFs used are based on the dipole ansatz. 
    \item Factorisation, $\muF^2$, and renormalisation, $\muR^2$, scales evaluated as $\muF^2 = \muR^2 = Q^2$.
    \item The skewness variable, $\xi$, expressed in terms of $\xB$ as $\xi = \xB/(2-\xB)$.
    \item DVCS CFFs evaluated from a set of lookup tables containing pre-computed values obtained by convoluting the Goloskokov-Kroll (GK) GPD model~\cite{Goloskokov:2005sd,Goloskokov:2008ib} with leading order / leading twist coefficient functions describing the hard scattering part of DVCS. The size of a single table, storing either real or imaginary part of a given CFF, is $70 \times 61 \times 60$, where the dimensions correspond
    to $10^{-6} < \xB < 0.95$, $0.7~\GeV^2 < Q^2 < 5\cdot 10^3~\GeV^2$ and $0 < |t| < 1.6~\GeV^2$ ranges, respectively. The algorithm called linterp~\cite{www_linterp} has been used for the interpolation between the nodes of the lookup tables.
    \item The FOAM algorithm \cite{Jadach:2005ex} is used for the random generation of kinematic configurations (i.e., kinematics of events) according to the cross section for the process~\eqref{eq:process} and a given radiative correction model. The algorithm has been set up with the following values: nCells = $8000$, nSamples = $1600$, and nBins = $1600$ (for the explanation of these parameters see FOAM's documentation).
    \item Radiative corrections generated for the lepton using a collinear approximation, including initial and final state radiation, and combinations of them (for more information see Sect.~\ref{subsec:rc}). 
    \item Generated events saved in both HepMC3~\cite{www_hepmc3} and a variation of PYTHIA6 ASCII formats (the latter for easy use in the detector simulations).
    
\end{itemize}
MC events have been generated in the following kinematic ranges:
\begin{itemize}
    \item $10^{-5} < \xB < 0.95$,
    \item $0.0001 < y < 0.95$,
    \item $0.7~\GeV^2 < Q^2 < 1000~\GeV^2$,
    \item $0.01~\GeV^2 < |t| < 1.6~\GeV^2$,
    \item $0.03~\rad < \phi < 2\pi-0.03~\rad$ (in the target rest frame).
\end{itemize} 
The cut on the azimuthal angle $\phi$ prevents direct probing of the BH cross section in the region where it becomes singular.  

The technical cut-off parameter
$\epsilon$ that separates soft- from hard-photon radiation is:
\begin{itemize}
    \item $\epsilon = 10^{-4}$
\end{itemize}
Here, $\epsilon$ describes an infrared cut-off: real collinear photons emitted by an electron with energy $E$ are allowed in the event simulation if their energy is above $\epsilon \times E$. The contribution of soft photons, combined with virtual corrections, is taken into account inclusively, i.e.\ events for this contribution are generated with zero photon momentum. The cut-off parameter should be chosen small enough such that it is below the energy resolution of the experimental set-up. For more information see Sec.~\ref{subsec:rc} and~Refs.~\cite{Kripfganz:1990vm,Aschenauer:2022aeb}.  

Additional kinematic cuts applied at the level of the analysis are:
\begin{itemize}
    \item $1~\GeV^2 < Q^2 < 100~\GeV^2$, 
    \item $0.01 < y < 0.6$ (in the analysis of nucleon tomography, see Sect.~\ref{subsec:tomography}),
    \item $0.01 < y < 0.85$ (in the extraction of Compton form factors, see Sect.~\ref{subsec:asymmetries}),
    \item $0.00001 < \xB < 0.7$, 
    \item $0.03~\GeV^2 < |t| < 1.5~\GeV^2$.
\end{itemize}
These cuts restrict the kinematic domain to that which is suitable for the physics analysis. For instance, the range $y < 0.01$ is excluded to mitigate the severe impact of energy smearing, while $y > 0.6$ is omitted due to the dominance of the BH sub-process, which washes out the useful physics signal. The latter restriction can be relaxed in the case of the analysis of Compton form factors from polarized data, as in this case, the pure BH contribution (related to $\left| \mathcal{T}_\text{BH}\right|$ in Eq.~\eqref{eq:single_photon_production_amplitude}) cancels out in the asymmetry and the signal arises from the interference.

MC samples have been generated for three beam energy configurations previously considered in the yellow report. These are:
\begin{itemize}
    \item $E_e = 5~\GeV \times E_p = 41~\GeV$,
    \item $E_e = 10~\GeV \times E_p = 100~\GeV$,
    \item $E_e = 18~\GeV \times E_p = 275~\GeV$,
\end{itemize}
where $E_e$ and $E_p$ denote energies of electron and proton beams (in the laboratory frame), respectively. For brevity and because of the redundancy with respect to other beam energy setups, we have resigned from showing results for $E_e = 5~\GeV \times E_p = 100~\GeV$, which has also been considered in the yellow report. 

Given the focus of this study on demonstrating the impact of future EIC data on the measurement of $t$-profiles of the DVCS cross section and $A_{\mathrm{LU}}$ beam spin asymmetries, MC samples have been generated solely for two electron beam polarization states and an unpolarized proton beam. We use two types of MC samples: one generated for only the DVCS sub-process, and one generated for both DVCS and BH sub-processes, including their interference.

Unless stated otherwise, results shown in this manuscript correspond to the integrated luminosity $\mathcal{L} = 10~\fb^{-1}$. The total cross sections together with the corresponding number of events for each beam energy configuration are given in Table~\ref{tab:cross_sections}. 
\begin{table}[!ht]
\centering
\begin{tabular}{P{0.1\textwidth}P{0.1\textwidth}P{0.18\textwidth}P{0.18\textwidth}P{0.18\textwidth}P{0.18\textwidth}}
\toprule
\multirow{2}{*}{$E_e$ [$\GeV$]} & \multirow{2}{*}{$E_p$ [$\GeV$]} & \multicolumn{2}{c}{$0.01<y<0.6$} &  \multicolumn{2}{c}{$0.01<y<0.85$} \\ 
& & $\sigma~/~\sigma_{\mathrm{DVCS}}$ [$\nb$] & $N~/~N_{\mathrm{DVCS}}$ (in M)& $\sigma~/~\sigma_{\mathrm{DVCS}}$ [$\nb$] & $N~/~N_{\mathrm{DVCS}}$  (in M)\\
\midrule
$5$& $41$& $0.83~/~0.36$& $8.3~/~3.6$& $1.72~/~0.39$& $17.2~/~3.9$\\
$10$& $100$& $0.85~/~0.38$& $8.5~/~3.8$& $1.76~/~0.41$& $17.6~/~4.1$\\
$18$& $275$& $0.90~/~0.43$& $9.0~/~4.3$& $1.79~/~0.45$& $17.9~/~4.5$\\ \bottomrule
\end{tabular}
\caption{Integrated cross sections for the $e + p \to e' + p' + \gamma$ process, $\sigma$, for given electron ($E_e$) and proton ($E_p$) beam energies, and conditions specified in Sect.~\ref{subsec:epic}, including additional kinematic cuts applied at the analysis level (separately for two cuts on $y$). The corresponding cross sections for the pure DVCS sub-process are denoted by $\sigma_{\mathrm{DVCS}}$, while the resulting number of events (in millions) for $\mathcal{L} = 10~\fb^{-1}$ are represented by $N$ and $N_{\mathrm{DVCS}}$.}
\label{tab:cross_sections}
\end{table}
 
\subsection{Simulation of detector response}
\label{subsec:detectors}

In order to assess the experimental prospects for DVCS at the future Electron-Ion Collider, generated MC samples were all processed through a full detector simulation constructed in the EICROOT~\cite{ref:EICROOT} simulation framework, which combines ROOT TGeo~\cite{ROOT} geometry definitions with GEANT4~\cite{GEANT4} simulations. The detector geometry is based on the EIC project detector, ePIC~\cite{ref:ePIC_dd4hep}, but the full simulation framework for the ePIC detector was not in a stable state for processing the full simulations at the time of the present study.

For reconstruction of the EpIC events, the main ePIC barrel detector was simulated using a parametrized detector acceptance and response (for the scattered electron and photons), while the outgoing hadron ``far-forward'' beam-line needed for reconstruction of the scattered proton was fully-simulated with the updated geometry including the beam-line magnets, fields, detector components, etc. The description of the full detector simulation is outlined below.

\subsubsection{Parametrized response of the ePIC barrel detector}

In order to simulate the impact of reconstructing the scattered electron and DVCS photon, a parametrization of the ePIC barrel detector was used which includes the various energy and angular resolutions of the electromagnetic calorimeters for photons (and electrons), and the momentum resolution of the tracking system used to extract the electron momentum. Additionally, acceptance gaps were introduced for the electron tracking to simulate the effect of service routing for the detectors, and an overall conservative efficiency factor of $95\%$ for photons and $90\%$ for electrons was applied to account for possible detector inefficiencies (discussion below). Backgrounds from various sources (other than Bethe-Heitler) were not included in these studies.

The angular resolution for all of the calorimeters is assumed to be $1\,\mrad$ (except in the analysis of $\pi^{0}$ contamination, see Sec.~\ref{subsec:pi0}). The angular resolution of the calorimeters is dependent upon the granularity and the energy of the particle being measured. However, here a choice is made to use an average of what is achievable for the calorimeters with modern reconstruction techniques. The energy resolutions depend on the range of pseudorapidity, $\eta$, defining the three different electromagnetic calorimeter subsystems. These ranges and energy resolutions are summarized in Table~\ref{tab:EMCAL_resolutions}, where the electron endcap is a $\rm{PbWO}_{4}$ crystal calorimeter, the barrel calorimeter is a combination of scintillating fiber and silicon-imaging calorimetery, and the hadron endcap is comprised tungsten power and scintillating fibers. Momentum resolution on the electron reconstruction was assumed to be $2.5\%$, and electron momentum and energy were reconstructed using both the calorimeter and tracking information detailed here. The use of both tracking and calorimeter information is why we assume a slightly worse efficiency for electron detection since in a full simulation or real detector the matching of information between detectors leads to some loss in overall efficiency.

\begin{table}[!ht]
\begin{center}
\begin{tabular}{>{\raggedright}P{0.18\textwidth}P{0.18\textwidth}P{0.18\textwidth}}
\toprule
Detector & $\eta$ acceptance & $\sigma_{E}/{E}$ \\
\midrule
 Electron Endcap    &  $[-3.5, -1.0]$  & $\displaystyle\frac{1\%}{\sqrt{E}} \oplus 1\%$ \\[10pt]
 Barrel Imaging  &  $[-1.0, 1.0]$   & $\displaystyle\frac{7\%}{\sqrt{E}} \oplus 1\%$  \\[10pt]
 Hadron Endcap    &   $[1.0, 3.5]$   & $\displaystyle\frac{12\%}{\sqrt{E}} \oplus 2\%$ \\
\bottomrule
\end{tabular}
\end{center}
\caption{\label{tab:EMCAL_resolutions} Summary of the geometric acceptance and energy resolution for reconstruction of DVCS photons and scattered electron energy in the parametrized ePIC central detector
 \cite{AbdulKhalek:2021gbh}. } 
\end{table}

\subsubsection{Full detector simulation of the far-forward region}

\begin{figure}[!ht]
\centering
\includegraphics[width=0.8\textwidth]{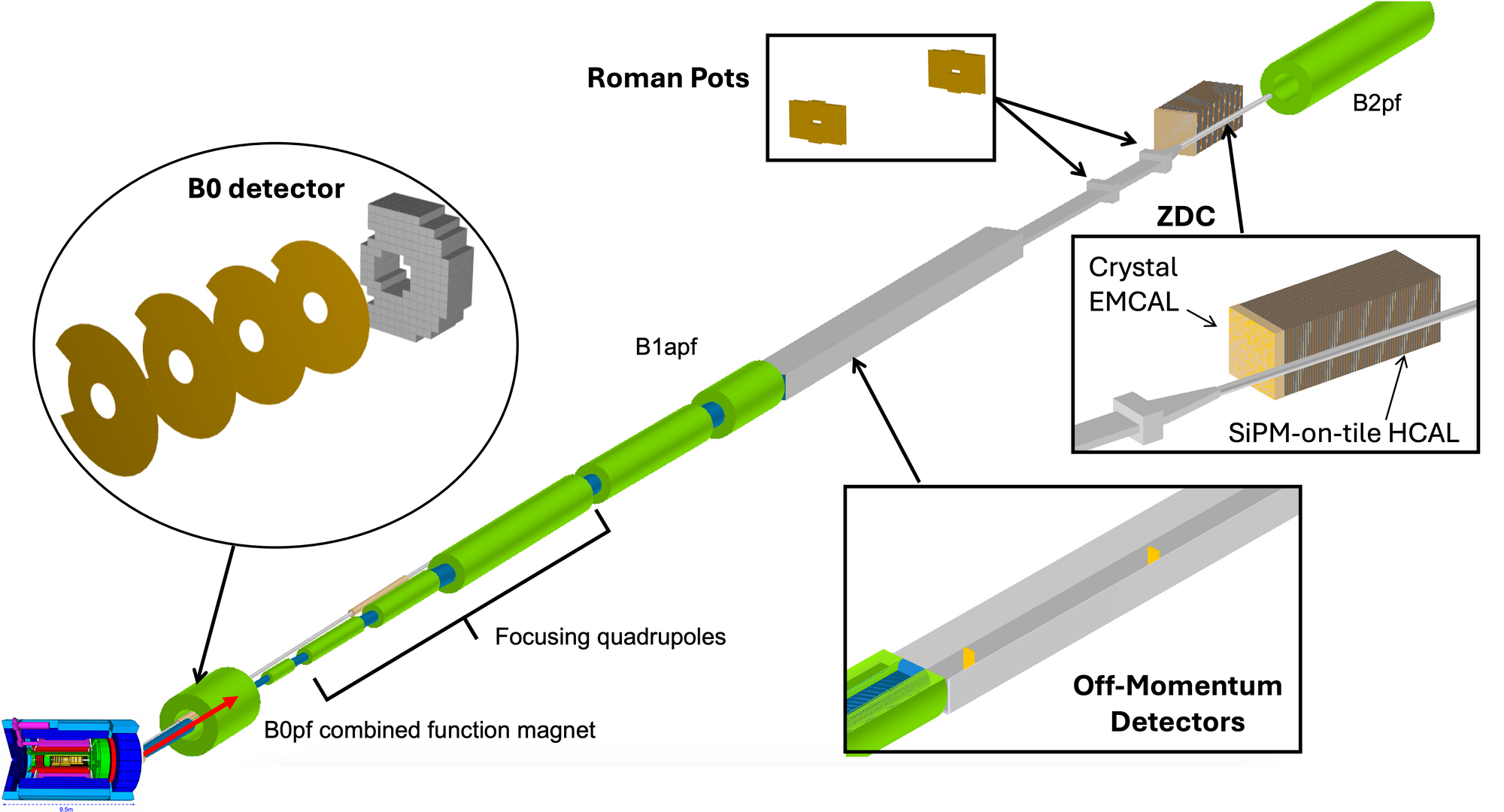}
\caption{Lattice of the hadron beam section including the far-forward detector suite for the ePIC experiment. The detector layout shown above reflects the geometry setup in the ePIC framework, with the image generated therein~\cite{ref:ePIC_dd4hep}. Given the state of the reconstruction software for the ePIC framework at the time of the simulations performed for the present article, a standalone GEANT4 simulation using EICROOT~\cite{ref:EICROOT} with the same ePIC far-forward detector geometry was used. The four subsystems are the B0 spectrometer, named for the magnet called B0pf where the detectors reside (zeroth bending magnet in the proton forward direction), the Off-Momentum Detectors (OMD), the Roman Pots (RP), and the Zero-Degree Calorimeter (ZDC). The magnet B1apf separates neutral particles going into the Zero-Degree-Calorimeter (ZDC) from charged hadrons and the beam particles and the B2apf magnet is another bending dipole magnet, listed just for reference. The main ePIC detector is depicted at the bottom left-hand corner of the figure, for reference (not to scale).}
\label{fig:farForwardDetectors}
\end{figure}

Table~\ref{tab:FFDetectors_acceptance} summarizes the geometric acceptance for far-forward scattered protons and neutrons achieved with the present design~\cite{AbdulKhalek:2021gbh}. In the following we summarize the main features of the subsystems as relevant to the present study,
in the order in which they appear when moving away from the interaction point, see Fig.~\ref{fig:farForwardDetectors}.
Details can be found in Refs.~\cite{ref:EICCDR,AbdulKhalek:2021gbh, ATHENAProp}. 

\begin{table}[!ht]
\begin{center}
\begin{tabular}{>{\raggedright}P{0.2\textwidth}P{0.18\textwidth}P{0.18\textwidth}P{0.18\textwidth}}
\toprule
Detector & Used for & $\theta$ acceptance [$\mrad$] & $x_{L}$ acceptance \\
\midrule
 B0 tracker    & $p$ & $5.5 - 20.0$  & N/A \\
 Off-Momentum  & $p$ & $0.0 - 5.0$   & $0.45 - 0.65$  \\
 Roman Pots    & $p$ & $0.0 - 5.0$   & $0.6 - 0.95 ^*$ \\
 Zero-Degree Calorimeter & $n$ & $0.0 - 4.0$   & N/A \\
\bottomrule
\end{tabular}
\end{center}
\caption{\label{tab:FFDetectors_acceptance} Summary of the geometric acceptance for far-forward scattered protons and neutrons
in polar angle $\theta$ and longitudinal momentum fraction $x_{L} = {p_{z, p'}}/{p_{z, beam}}$,
provided by the baseline EIC far-forward detector design \cite{AbdulKhalek:2021gbh}, where $p_{z, p'}$ is the longitudinal momentum of the scattered proton, and $p_{z, beam}$ is the default momentum for the central proton beam (e.g. 275 GeV).\\
\rule{0ex}{3ex}\footnotesize{$^{*})$ The Roman Pots acceptance at high values of $x_{L}$ depends on the beam optics choice for the accelerator.}} 
\end{table}

\subparagraph{B0 spectrometer and Calorimeter.} The B0 spectrometer is comprised of four layers of AC-LGAD silicon with 
$\sim 15-20\,\mu\mathrm{m}$ 
spatial resolution, and provides fast timing ($\sim 35\,\mathrm{ps}$) 
needed to correct for the influence of the rotation of the proton beam bunches to account for the 25 mrad crossing angle. This rotation of the bunch induces an effective smearing of the vertex position as a function of the longitudinal position of the particular proton within the bunch. The fast timing information allows the analyzer to establish where within the bunch the scattered proton originated and correct for this effect. The whole tracking system is embedded in the first dipole magnet after the interaction point for the ePIC detector (B0pf magnet), creating an overall tracking spectrometer. This subsystem
is optimized for reconstructing charged particles with polar scattering angles $5.5\,\mrad < \theta < 20.0\,\mrad$, which covers 
 $0.6\,\GeV < p_{T} < 2\,\GeV$ for protons with $E_{p} \sim 100\,\GeV$. It should be noted that this acceptance is not azimuthally symmetric; symmetric coverage is only provided up to $\sim 13\,\mrad$ due to physical constraints within the magnet.

In addition to the tracking detector in the B0 magnet, there is also a crystal electromagnetic calorimeter, which is not used in standard DVCS studies, but has relevance in backward $u$-channel DVCS.

\subparagraph{The Off-Momentum Detectors.} The Off-Momentum Detectors are described in detail in the EIC yellow report and in a previous study of free neutron structure~\cite{Jentsch:2021qdp}. Their primary role is to reconstruct protons with a large longitudinal momentum loss compared to the nominal beam momentum ($p_{\mathrm{proton}}/p_{\mathrm{beam}} \lesssim 60\%$). This subsystem is not of prime importance for the present study of DVCS, where most protons have a longitudinal momentum close to that of the beam momentum.

\subparagraph{Zero-Degree Calorimeter.} The Zero-Degree Calorimeter (ZDC), described in Refs.~\cite{AbdulKhalek:2021gbh} and~\cite{Jentsch:2021qdp}, consists of both an electromagnetic and hadronic calorimeter, important for tagging photons from various high energy exclusive processes, and spectator neutrons from charge exchange reactions or nuclear breakup.  The performance requirements are documented in~\cite{AbdulKhalek:2021gbh}. This detector is not used in studies of DVCS.

\subparagraph{Roman Pots.} The Roman pot detectors are designed to maximize geometric acceptance, especially at very low $p_{T} \sim 0.2\,\GeV$, which places them within a distance of $1\,\mathrm{cm}$ or less from the hadron beam core. These detectors are comprised of the same AC-LGAD silicon technology as is used in the B0 spectrometer. However, the two detectors need a different reconstruction approach.  For the Roman pots, a detailed description of the beamline magnets must be used to extract a transport matrix which describes the proton trajectories from the IP to the Roman pots. The detector setup has been studied extensively in~\cite{AbdulKhalek:2021gbh} and little has changed in the basic usage of these detectors in the full simulations except for refinements in the matrix transport calculations, which are reflected in the present study.

\subparagraph{Smearing effects on reconstruction.} The detector and beam effects used in this analysis have been extensively studied in previous efforts~\cite{AbdulKhalek:2021gbh, ATHENAProp}. The beam effects (e.g., transverse momentum smearing due to angular divergence generated by focusing the beam, longitudinal momentum spread of the beam) tend to play the dominant role in smearing of the momentum transfer, $t$, except in the case of the B0 tracking system where rigid $\sim 100\,\GeV$ proton tracks are more challenging to reconstruct in the magnetic field of the spectrometer. The various effects from the operation of the beam, as well as effects driven by detector technology choices are all included in the present study.

\section{Analysis and results}
\label{sec:analysis}

In this section, we discuss the analysis methods used and present the obtained results. We begin with Sect.~\ref{subsec:kinematics}, which shows basic kinematic distributions. Additionally, we illustrate the coverage of the phase space by the ePIC detector in comparison to other existing and previous experiments. In Sect.~\ref{subsec:rc}, we address the effect of radiative corrections, while in Sect.~\ref{subsec:pi0}, we examine the contamination caused by misidentified events originating from exclusive $\pi^0$ production. Finally, in Sects.~\ref{subsec:tomography} and \ref{subsec:asymmetries}, we discuss the extraction of tomography information from distributions as a function of the variable $t$, and the extraction of CFFs from the beam spin asymmetries, respectively.

\subsection{Distributions of kinematic variables}
\label{subsec:kinematics}

We start by showing basic kinematic distributions. They are: \emph{i}) separately obtained for each beam energy configuration considered in this study; \emph{ii}) made for MC events generated as described in Sect.~\ref{subsec:epic} and processed through the ePIC simulation described in Sect.~\ref{subsec:detectors}, \emph{iii}) obtained after applying the cuts specified in Sect.~\ref{subsec:epic} (except for the cut on the plotted variable, if applicable), including $0.01 < y < 0.85$; \emph{iv}) obtained for an integrated luminosity of $\mathcal{L} = 10~\fb^{-1}$, each.

\begin{figure}[!ht]
\centering
\includegraphics[width=\plotWidthThree]{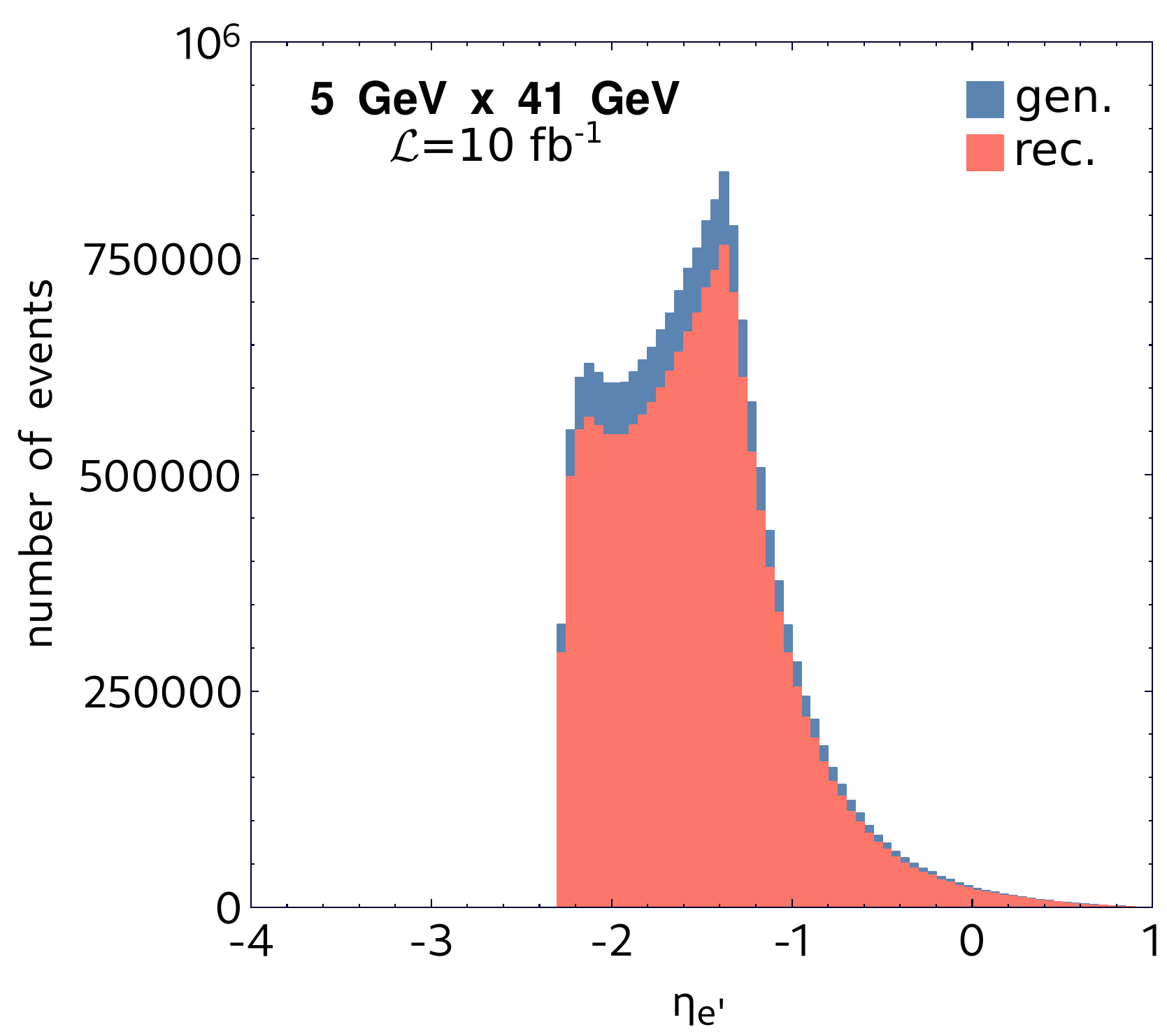}
\includegraphics[width=\plotWidthThree]{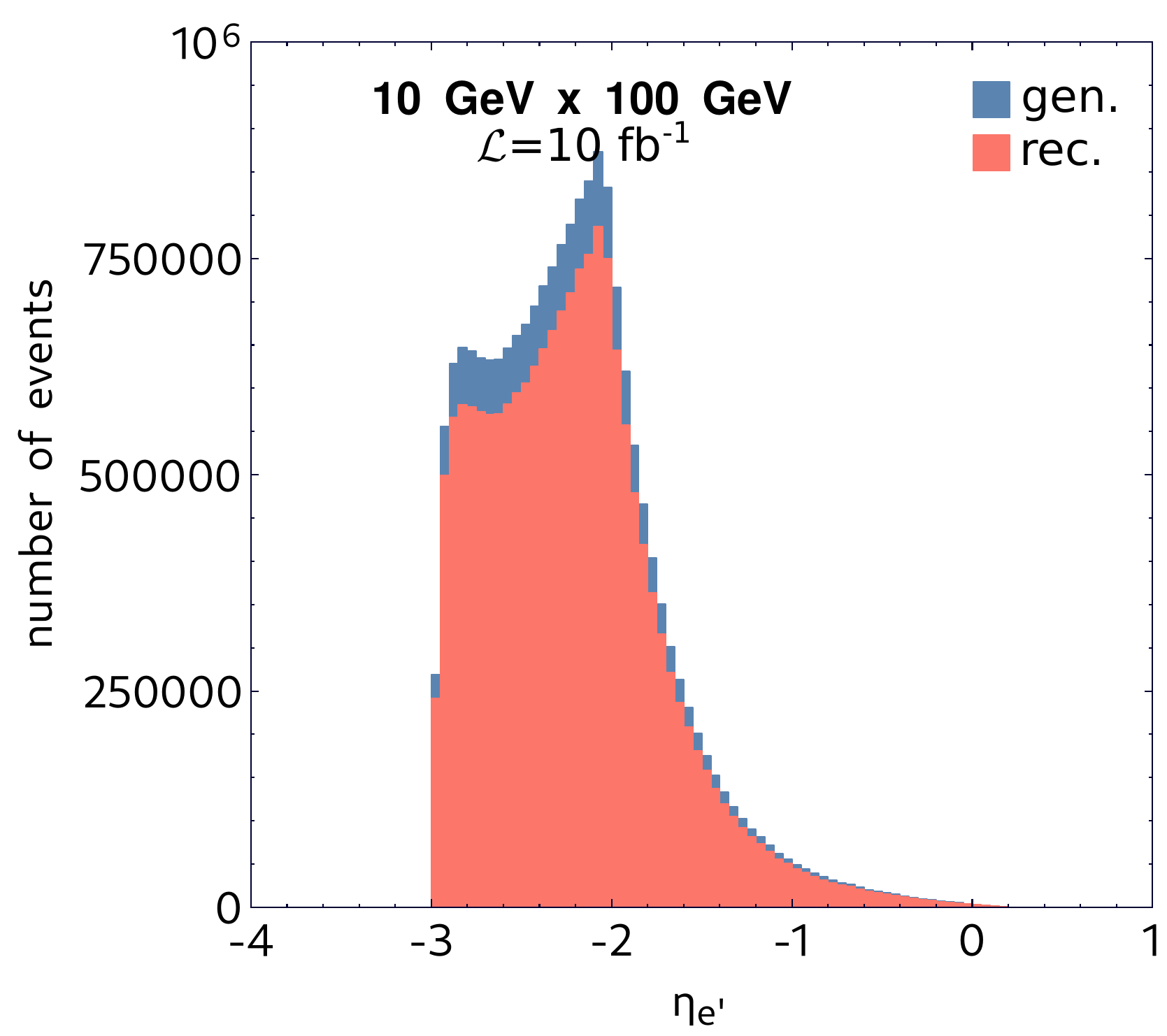}
\includegraphics[width=\plotWidthThree]{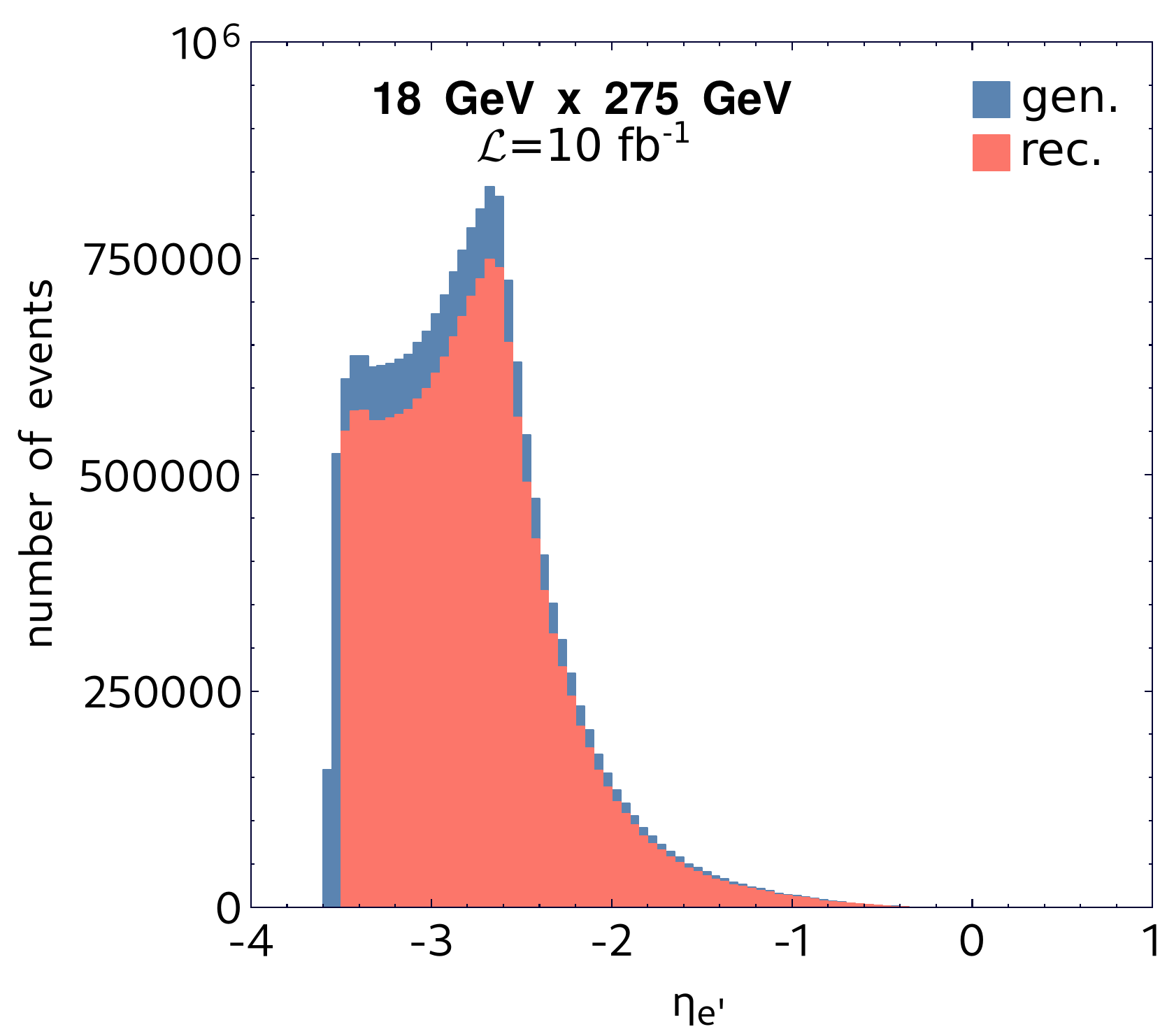} \\ 
\includegraphics[width=\plotWidthThree]{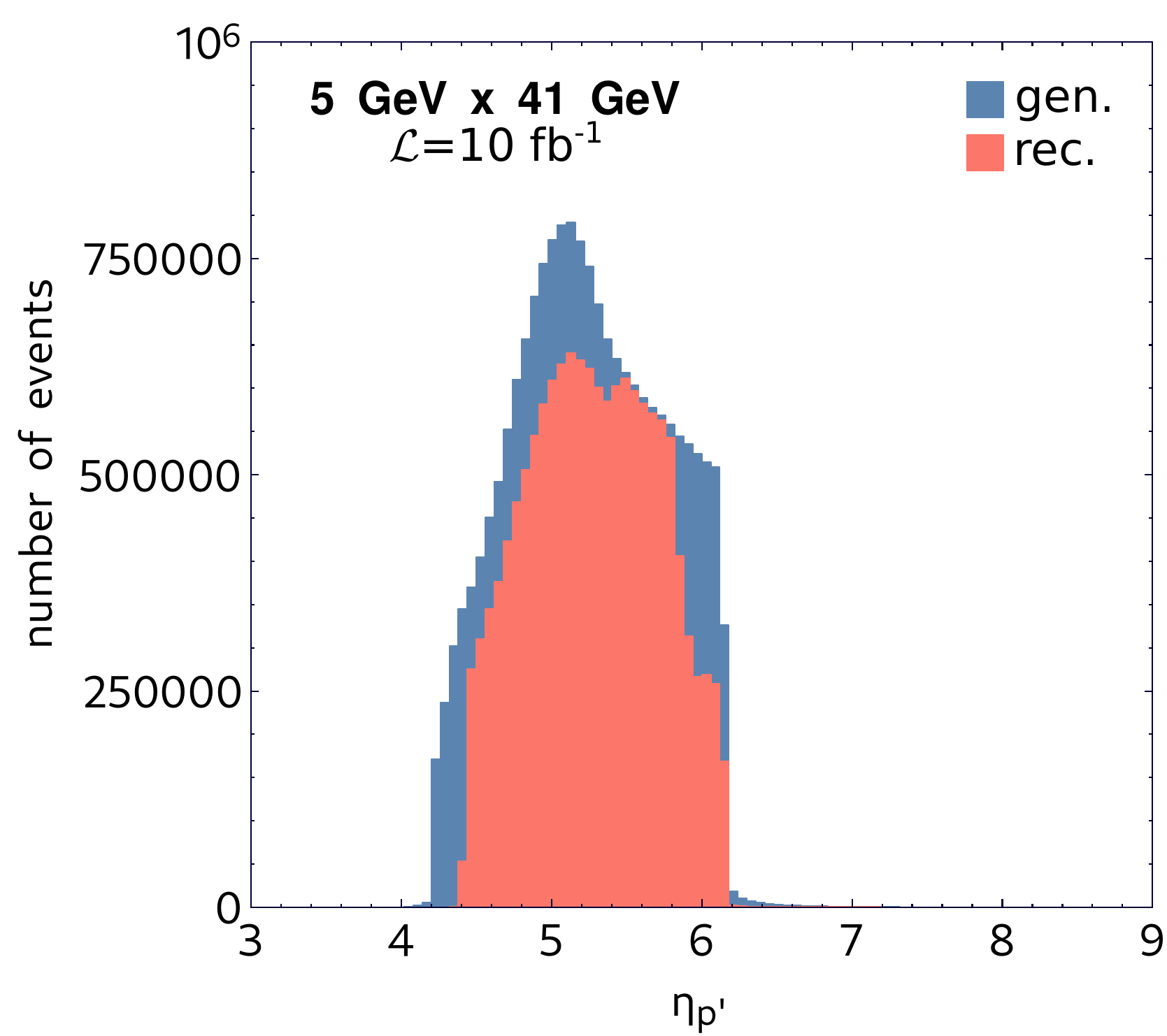}
\includegraphics[width=\plotWidthThree]{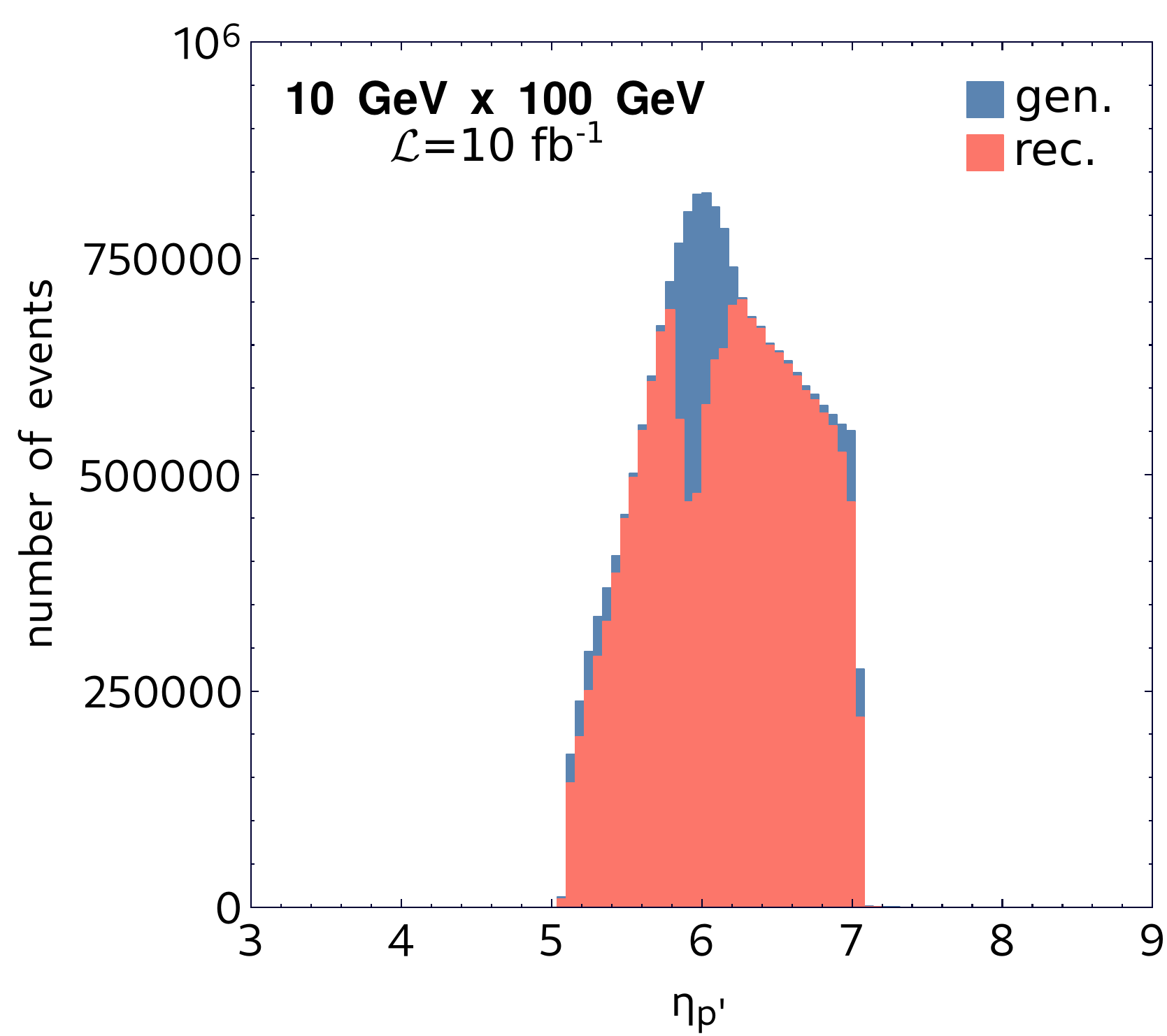}
\includegraphics[width=\plotWidthThree]{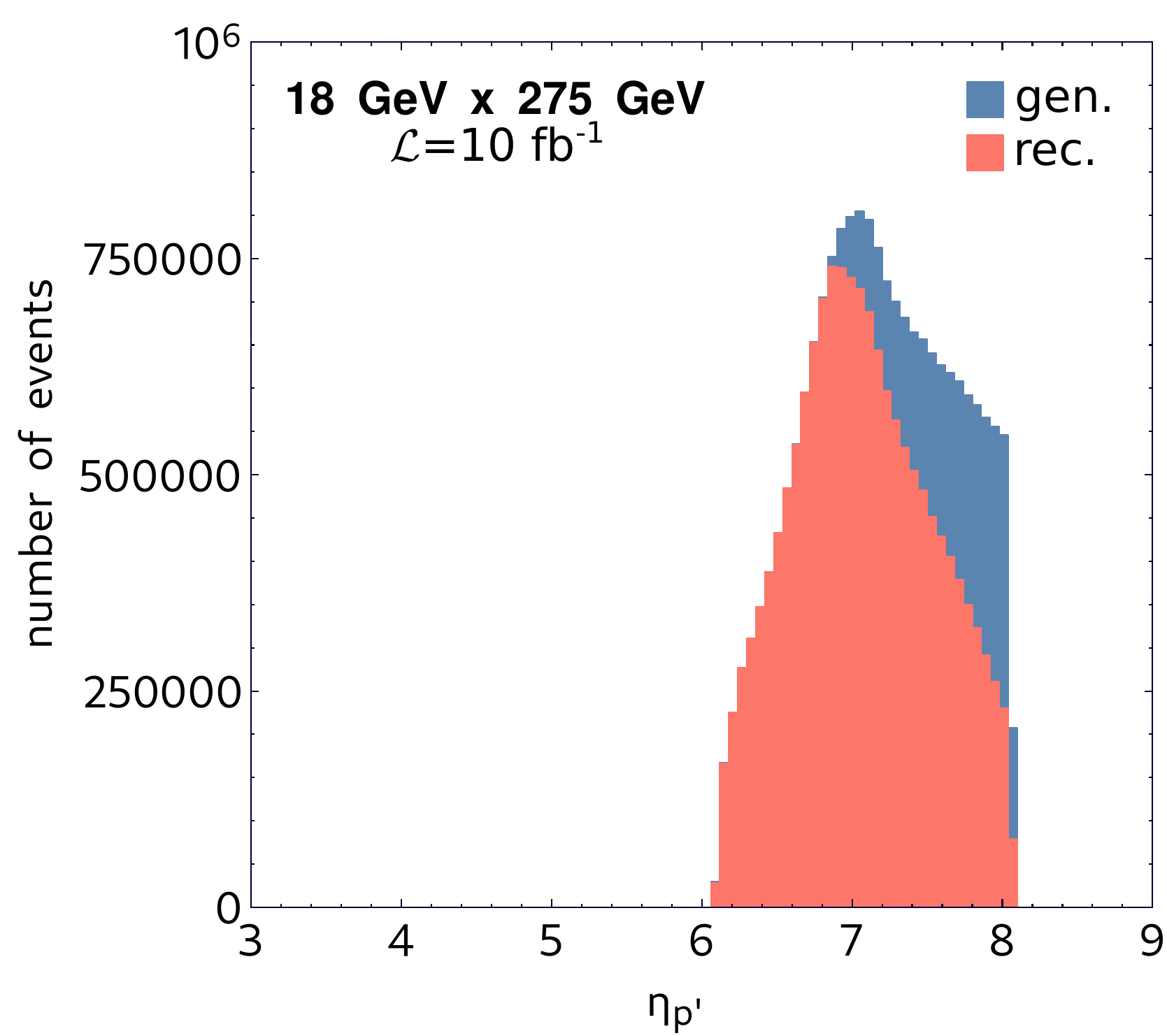} \\ 
\includegraphics[width=\plotWidthThree]{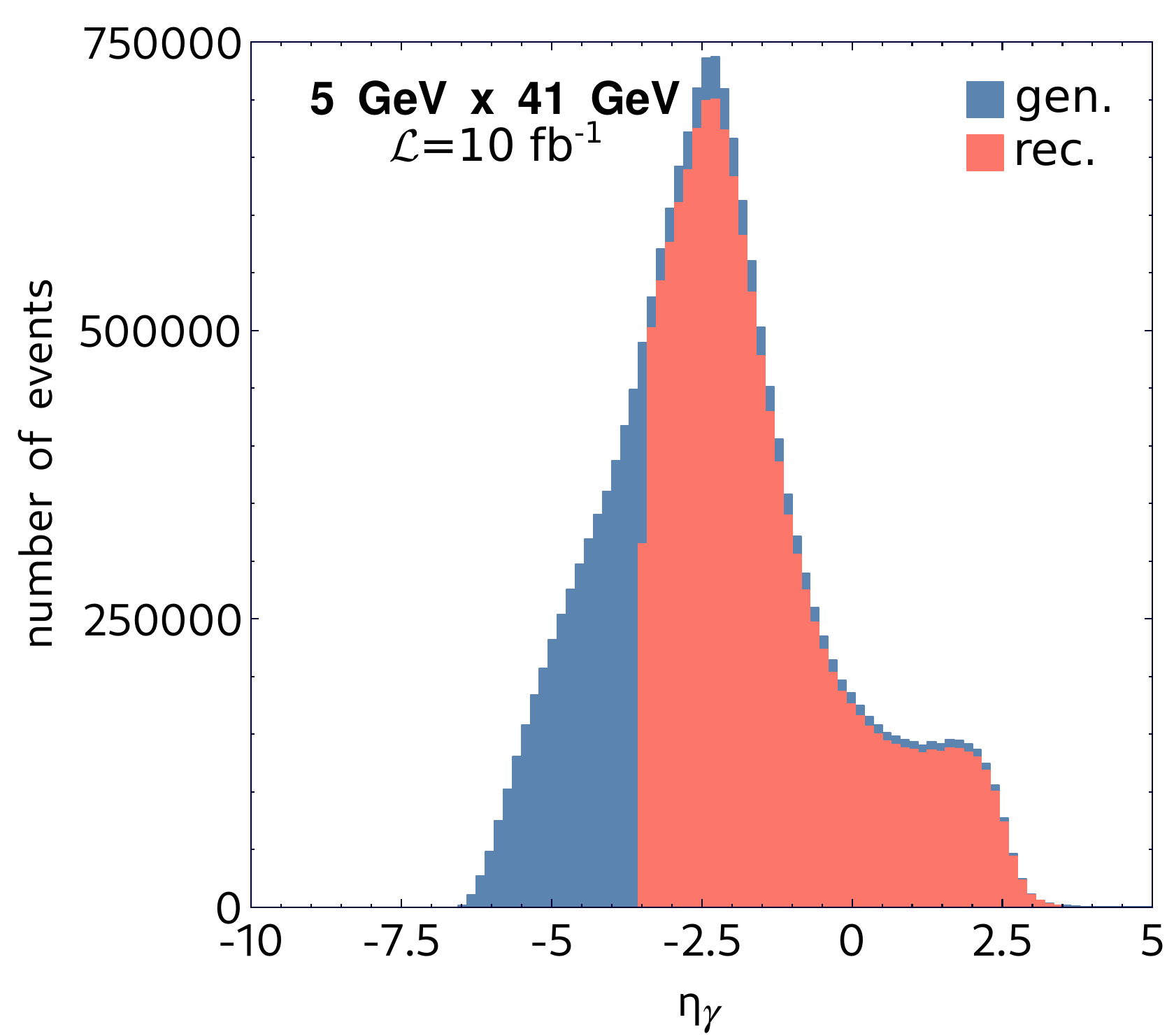}
\includegraphics[width=\plotWidthThree]{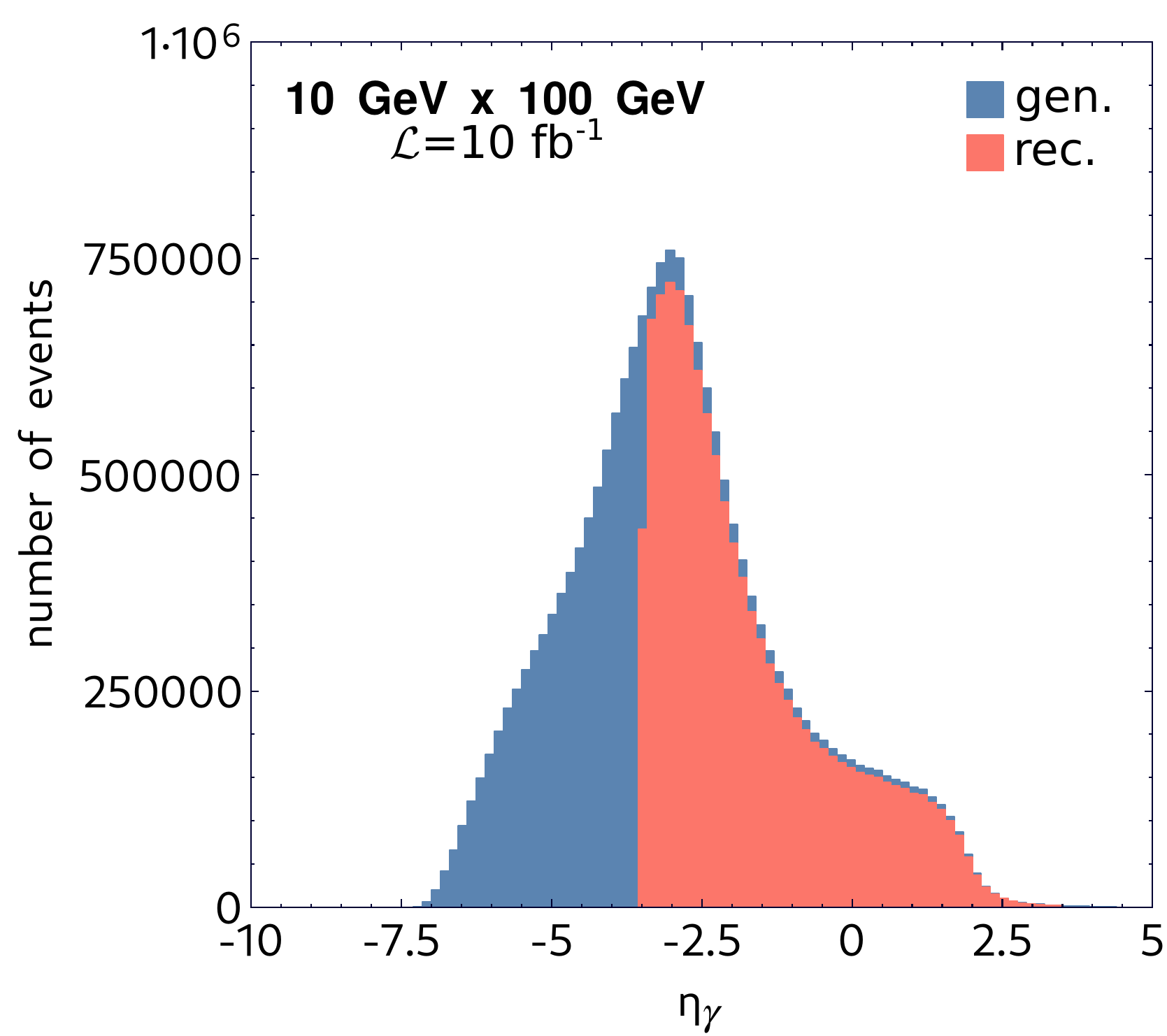}
\includegraphics[width=\plotWidthThree]{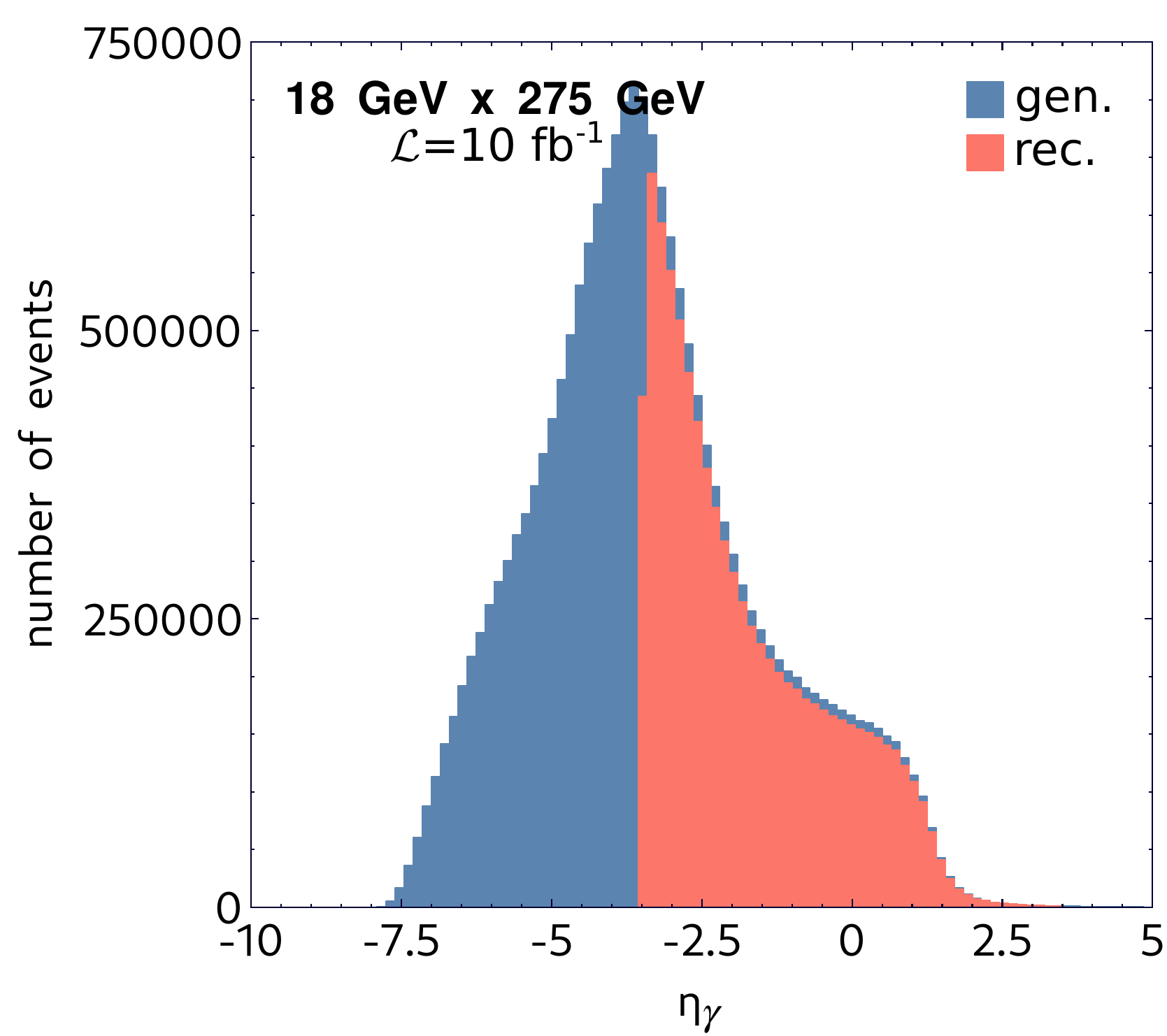} 
\caption{Distributions of pseudo-rapidity of scattered electron ($\eta_{e'}$), scattered proton ($\eta_{p'}$) and produced photon ($\eta_{\gamma}$) for generated (blue) and reconstructed (red) MC events containing the DVCS and Bethe-Heitler contributions to the cross section. The various beam energies are indicated in the plots.}
\label{fig:eta}
\end{figure}

\begin{figure}[!ht]
\centering
\includegraphics[width=\plotWidthThree]{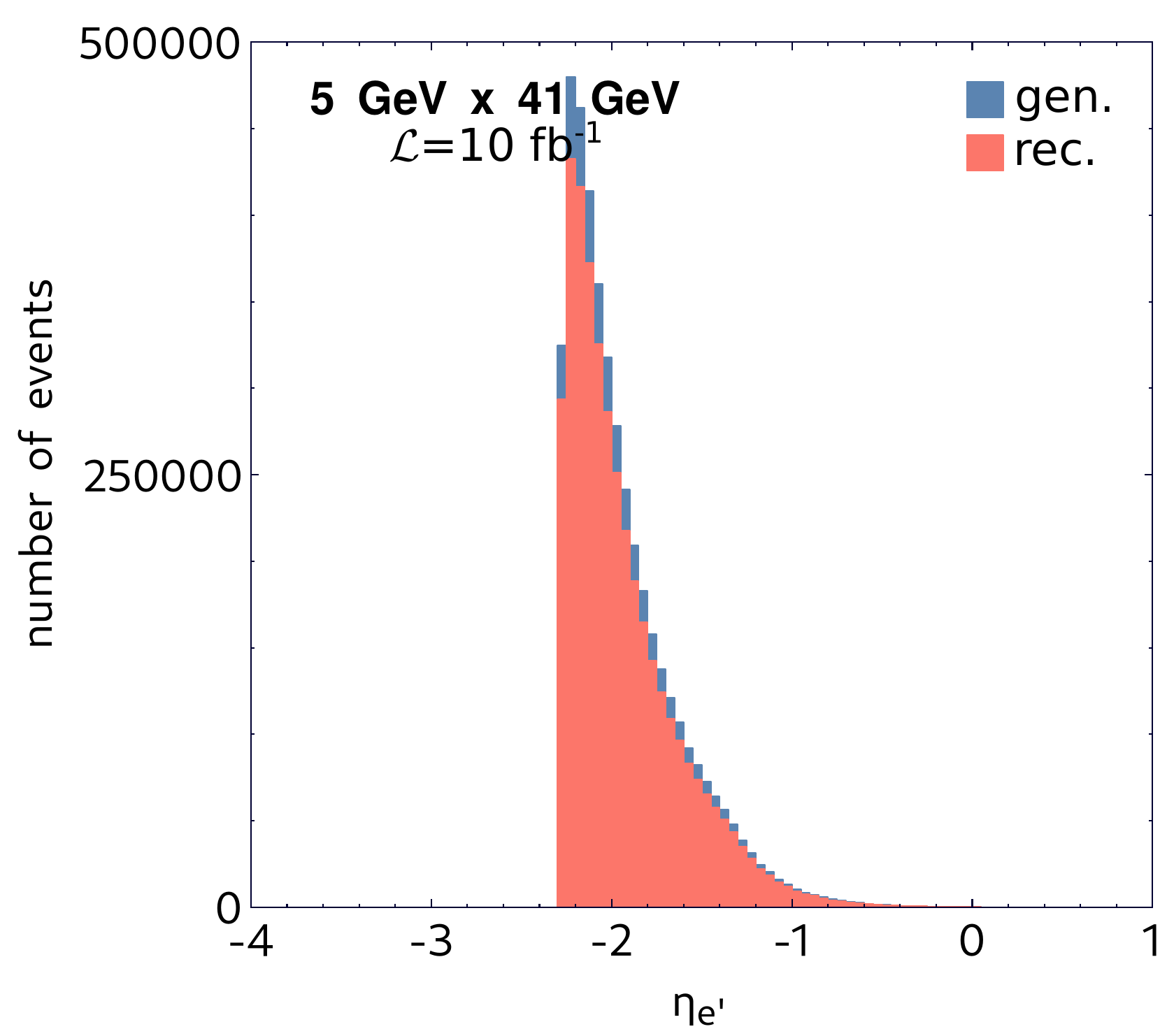}
\includegraphics[width=\plotWidthThree]{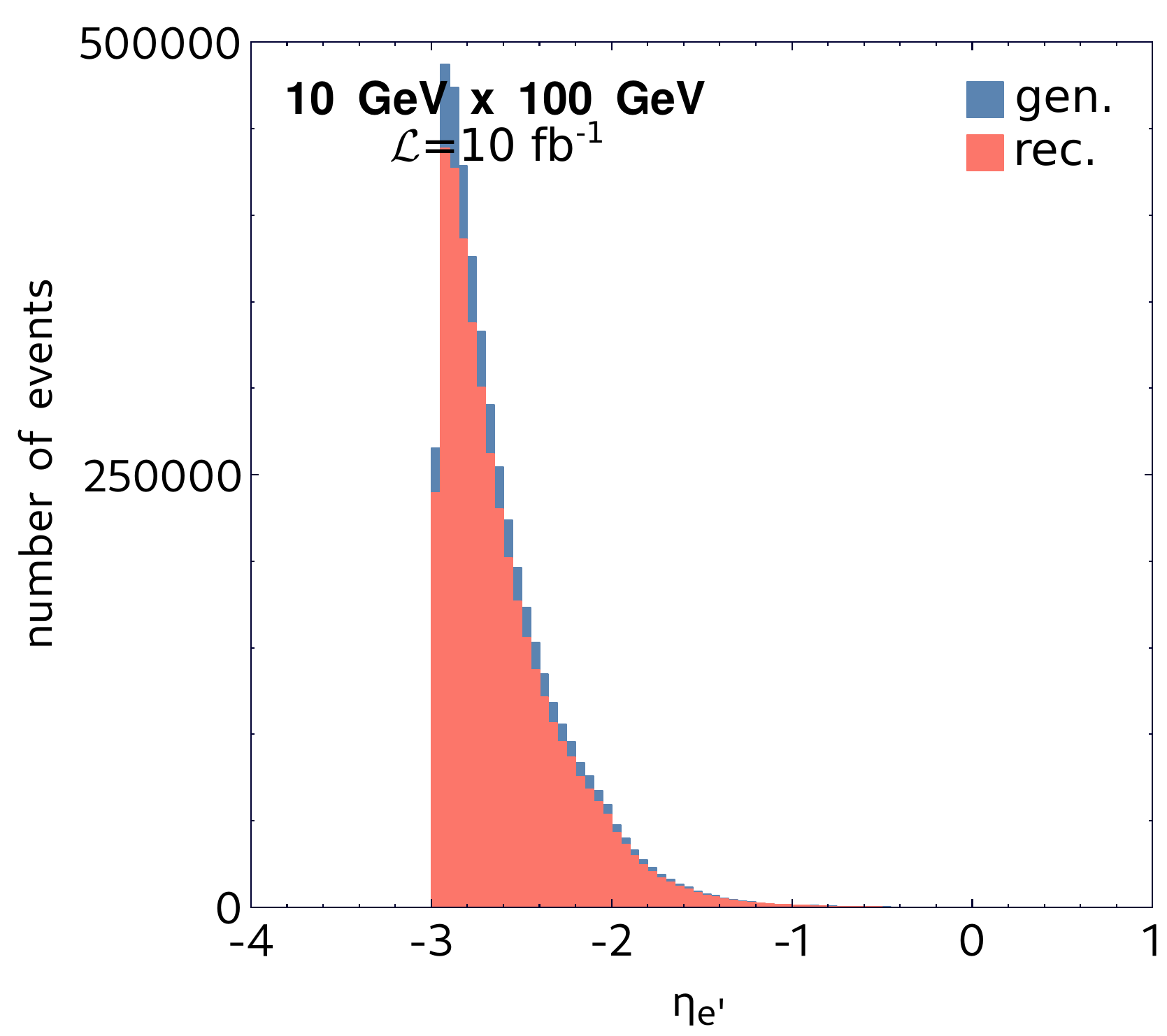}
\includegraphics[width=\plotWidthThree]{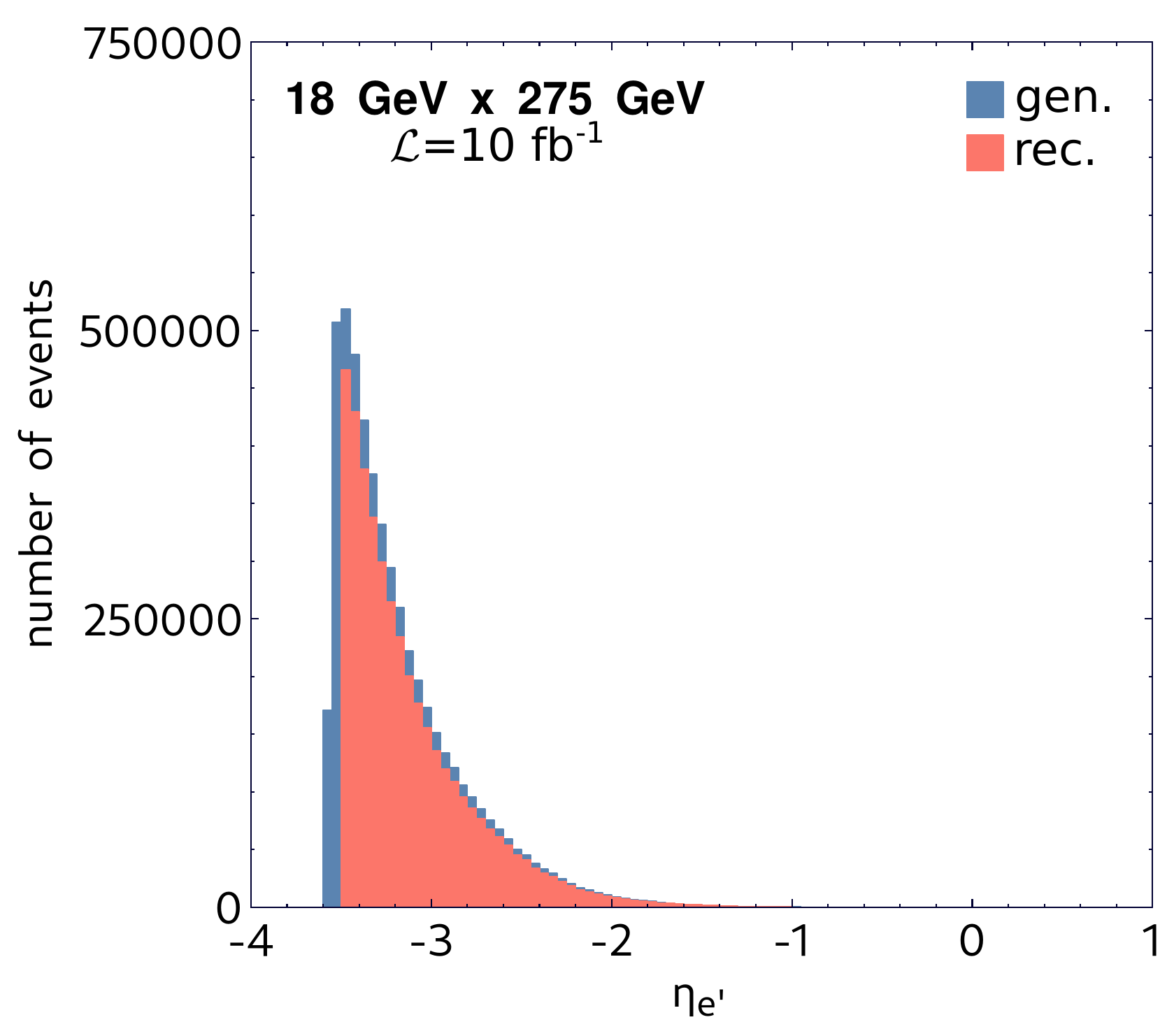} \\ 
\includegraphics[width=\plotWidthThree]{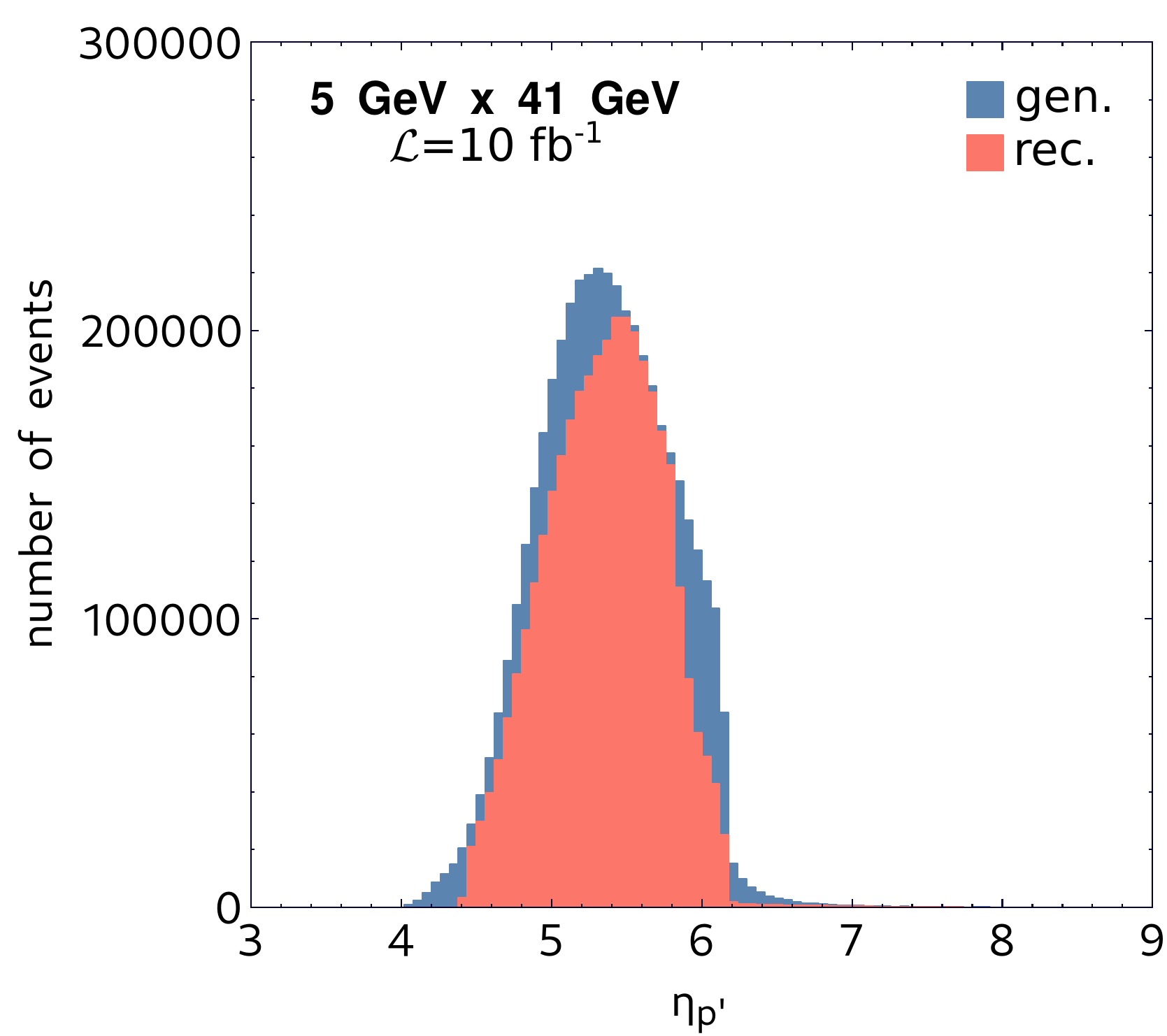}
\includegraphics[width=\plotWidthThree]{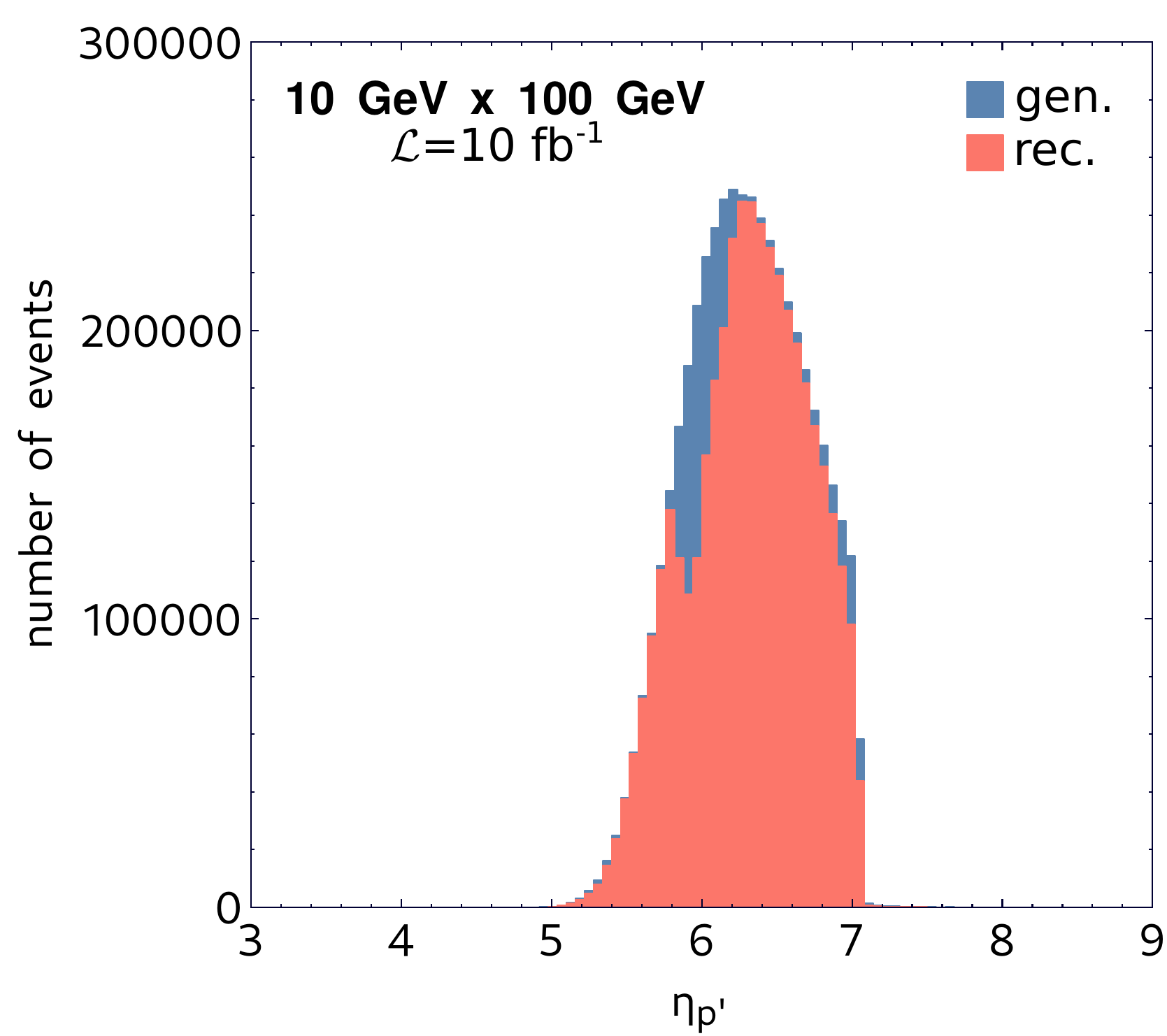}
\includegraphics[width=\plotWidthThree]{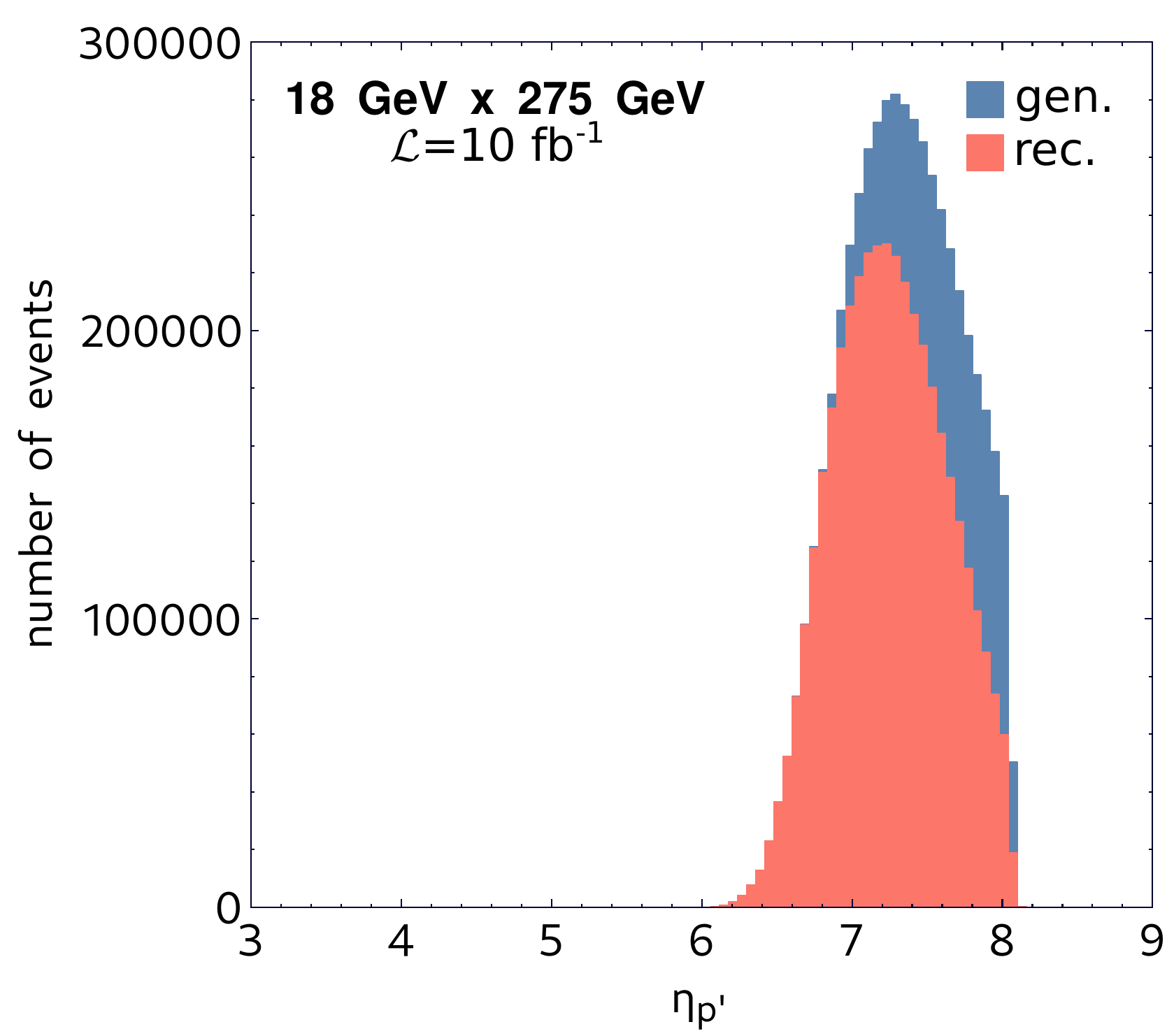} \\ 
\includegraphics[width=\plotWidthThree]{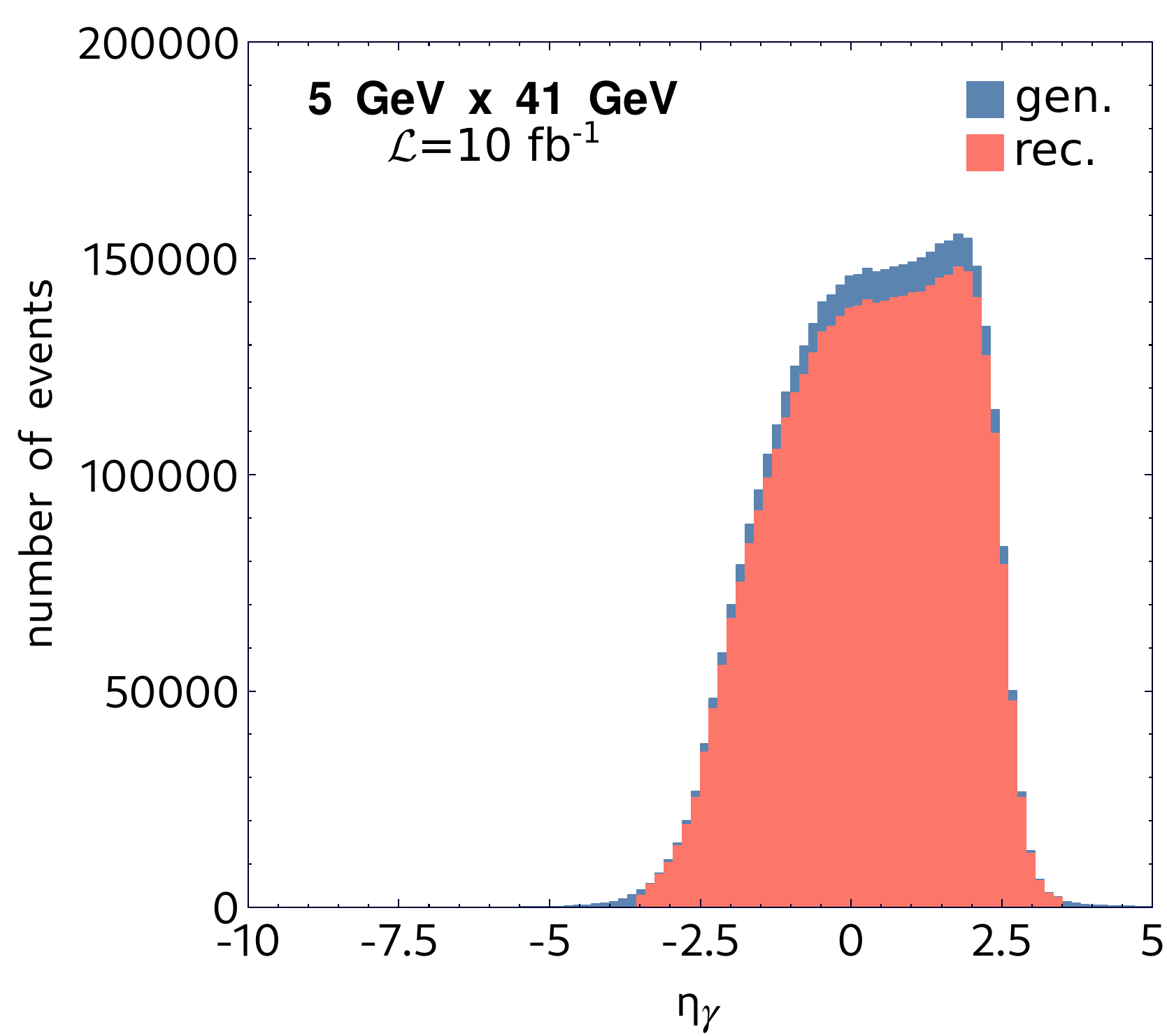}
\includegraphics[width=\plotWidthThree]{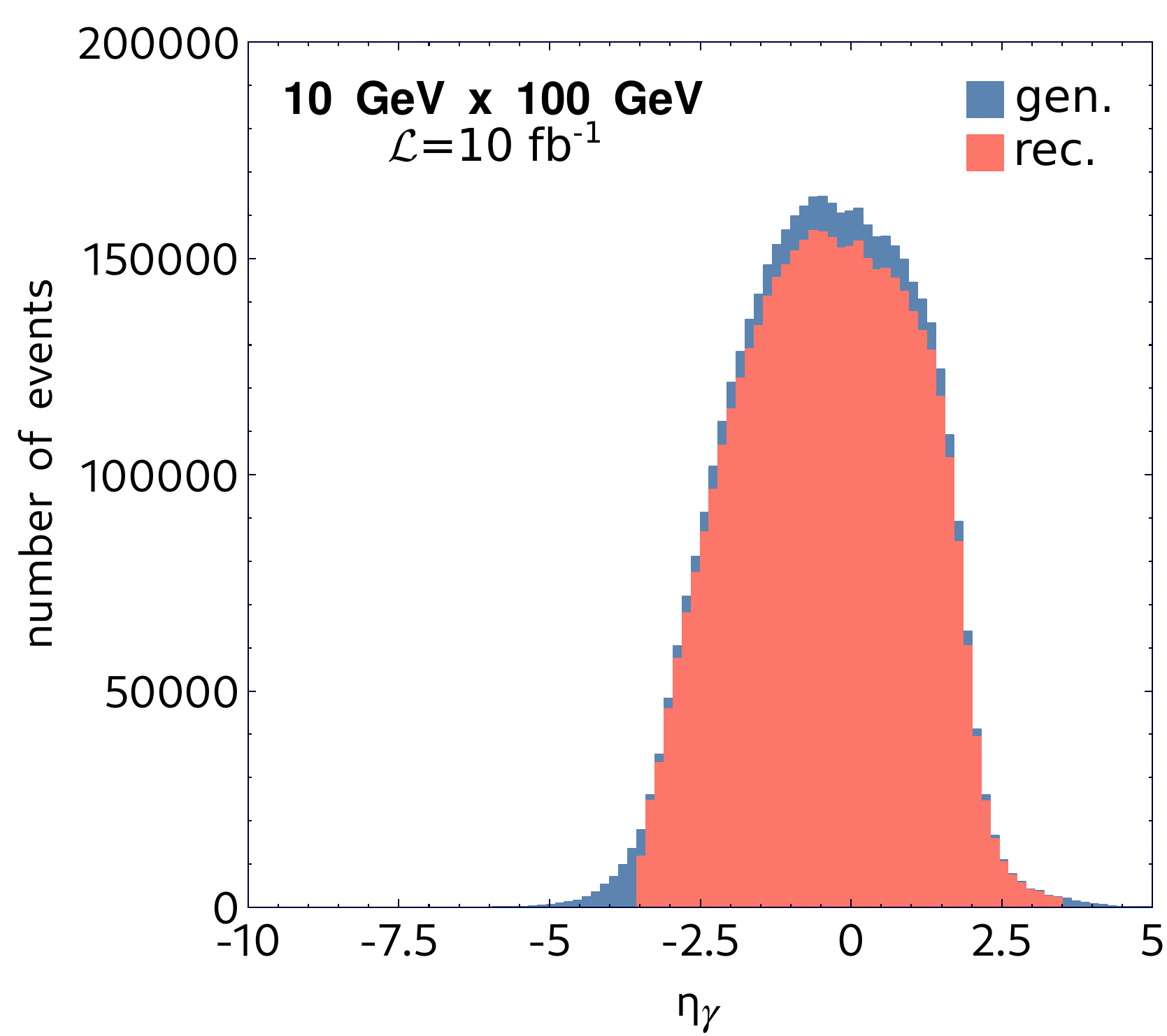}
\includegraphics[width=\plotWidthThree]{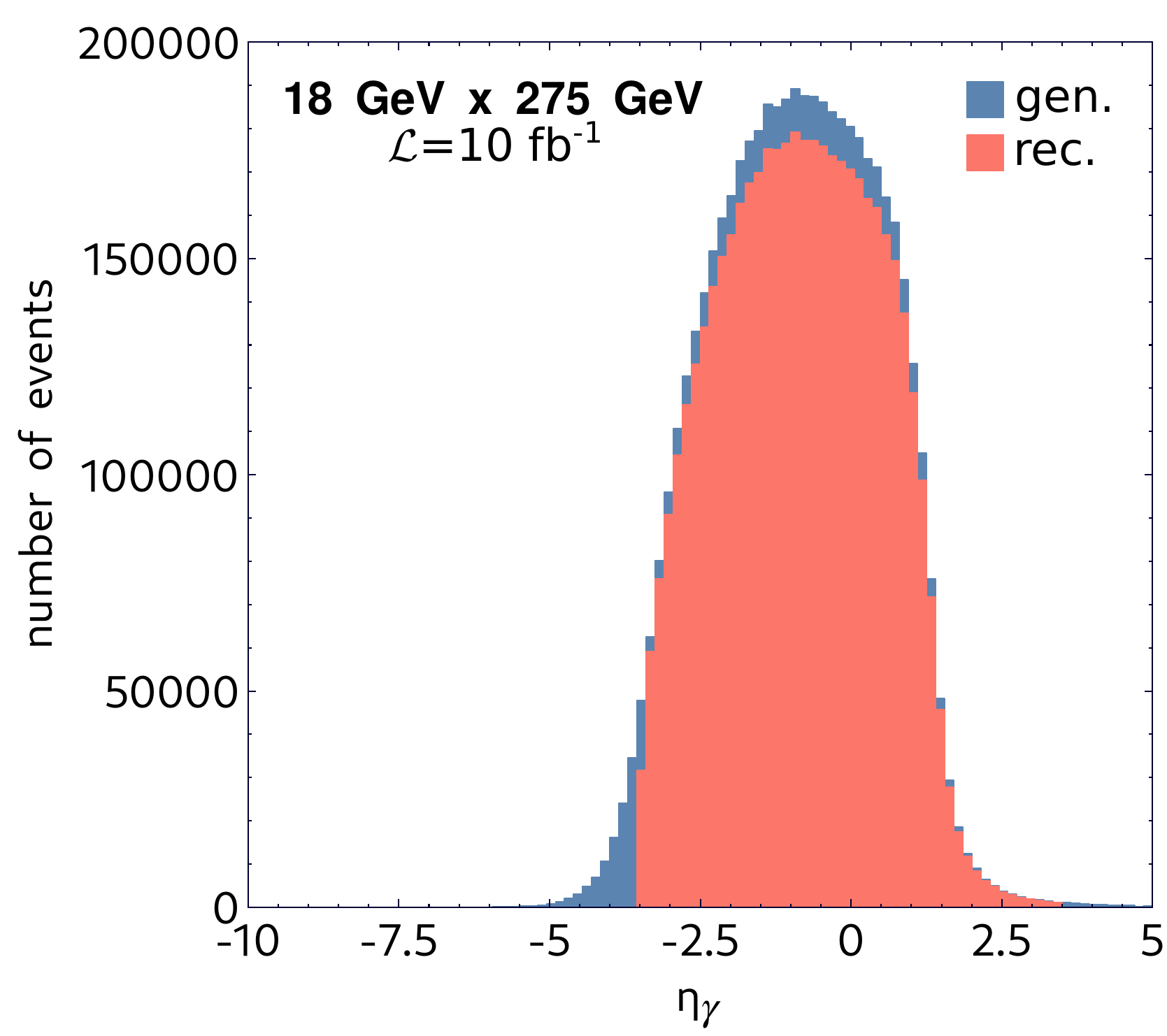} 
\caption{The same as Fig.~\ref{fig:eta}, but only for DVCS events.}
\label{fig:etaDVCS}
\end{figure}

Distributions of pseudo-rapidity for the scattered electron, $\eta_{e'}$, the scattered proton, $\eta_{p'}$, and the produced photon, $\eta_{\gamma}$ are shown in Fig.~\ref{fig:eta} for the mixture of BH and DVCS events, and in Fig.~\ref{fig:etaDVCS} for DVCS events only. To straightforwardly demonstrate the effect of geometric acceptance, which is typically studied with these variables, the distributions shown in Figs.~\ref{fig:eta} and~\ref{fig:etaDVCS} are obtained without including energy or momentum smearing from full reconstruction in the detectors (if a particle is found in the detector, its generator-level kinematics are plotted to isolate the acceptance). One can see that the acceptances change as a function of beam energy, especially between the lowest and highest beam energy configurations. In all cases, the scattered electron has exceptionally high acceptance. This is a fundamental aspect of the ePIC detector design due to the need for scattered electron detection for all deep-inelastic scattering observables. In the case of Fig.~\ref{fig:eta}, the losses in photon acceptance are coming almost entirely from Bethe-Heitler events, which is seen when comparing the same panels in Figs.~\ref{fig:eta} and~\ref{fig:etaDVCS}. This is a consequence of the photon and scattered electron from Bethe-Heitler events being produced with very small opening angles, where they are essentially collinear. The scattered proton is challenging to detect at the lower beam energies due to an acceptance gap between the B0 and Roman Pots detector subsystems, which is a consequence of the required layout for the hadron beam pipe. 

\begin{figure}[!ht]
\centering
\includegraphics[width=\plotWidthThree]{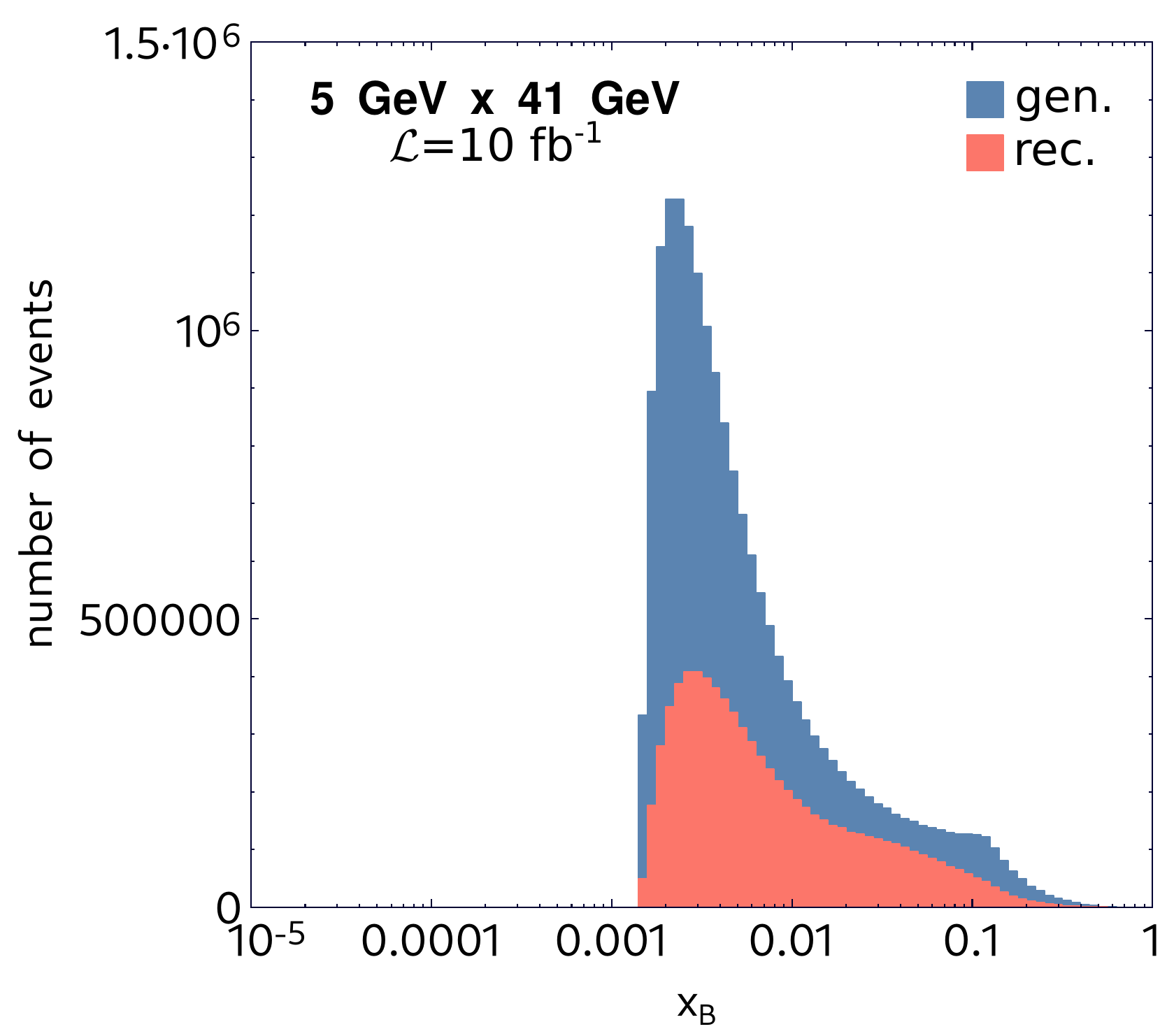}
\includegraphics[width=\plotWidthThree]{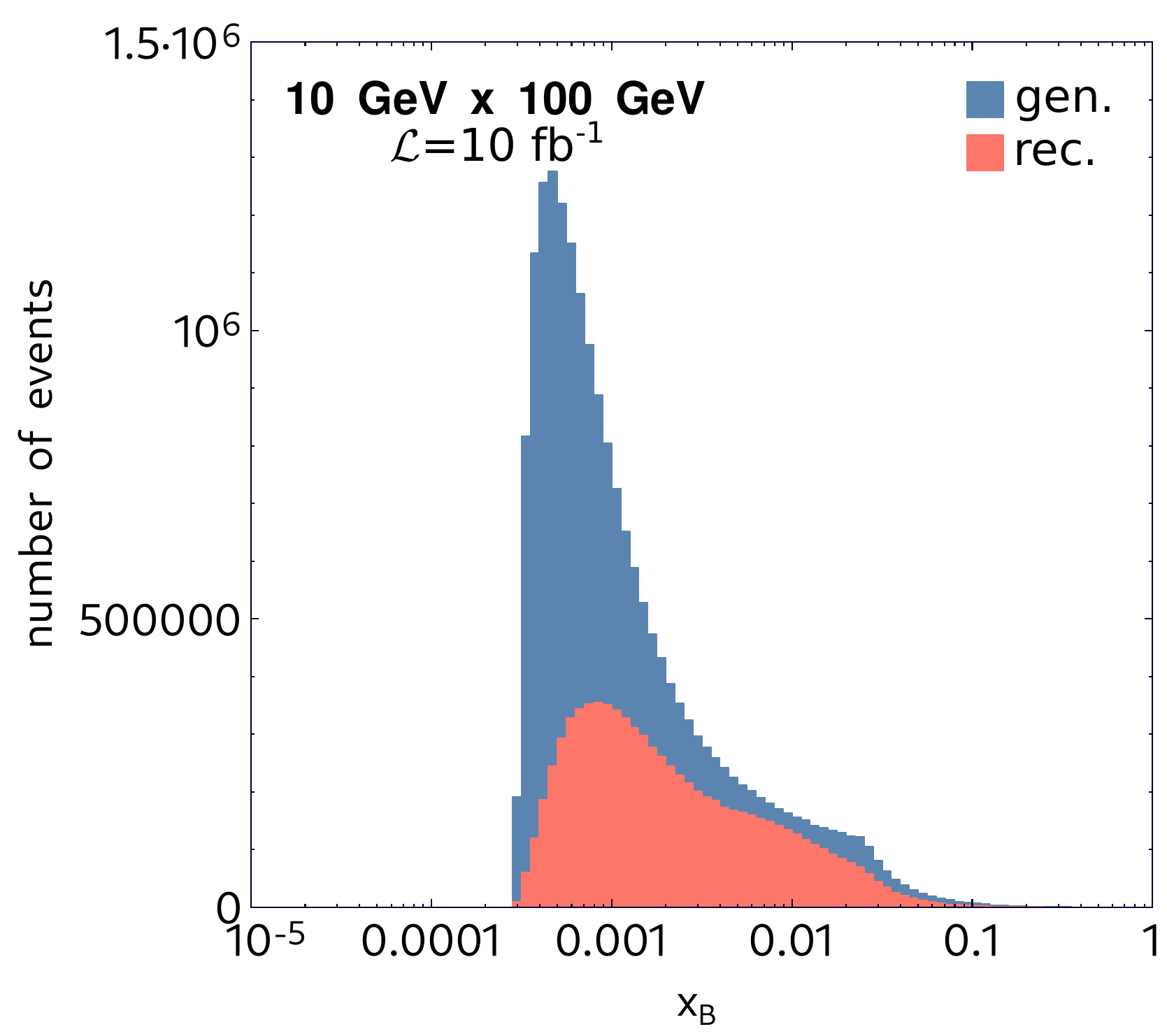}
\includegraphics[width=\plotWidthThree]{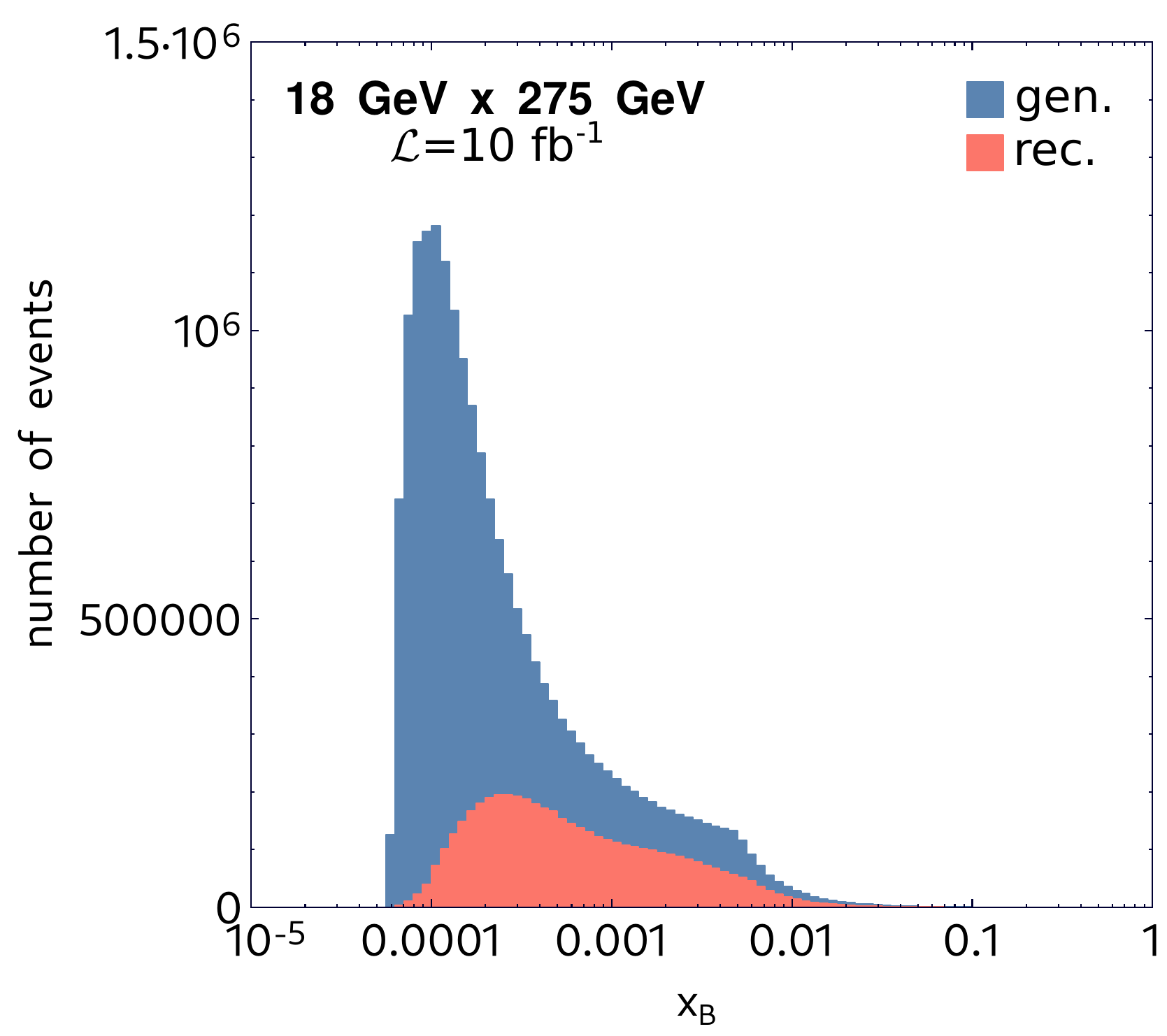}\\
\includegraphics[width=\plotWidthThree]{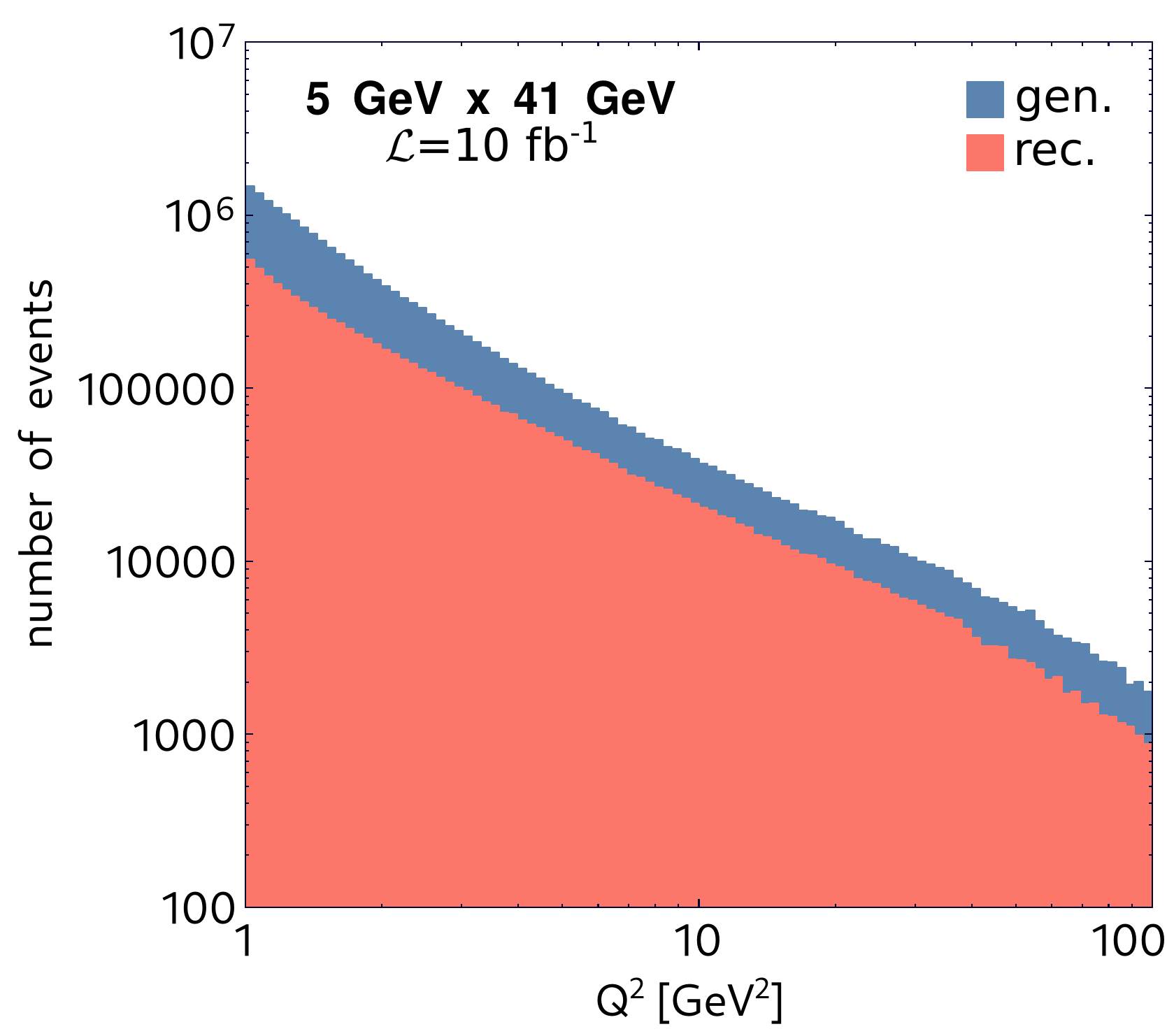}
\includegraphics[width=\plotWidthThree]{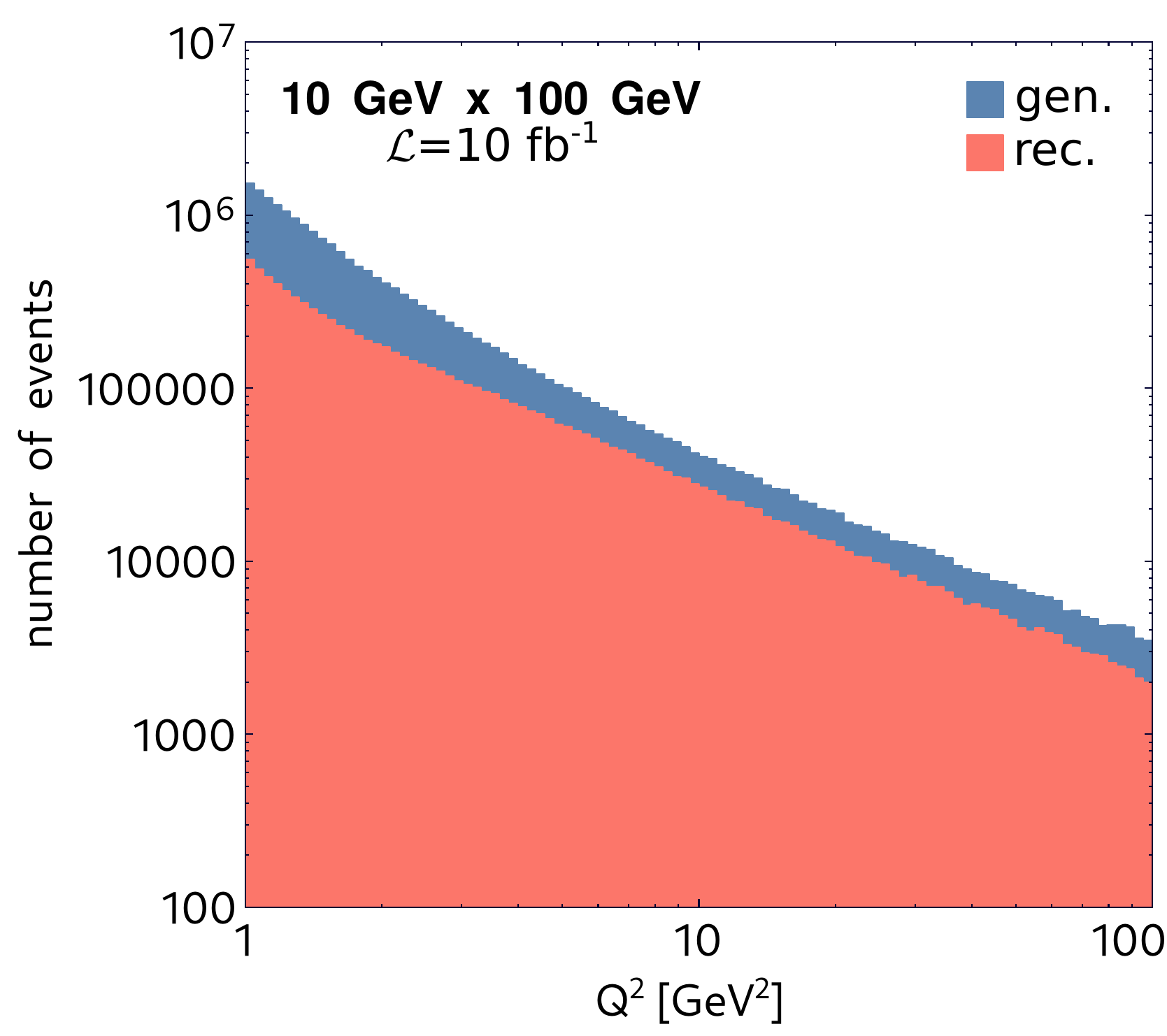}
\includegraphics[width=\plotWidthThree]{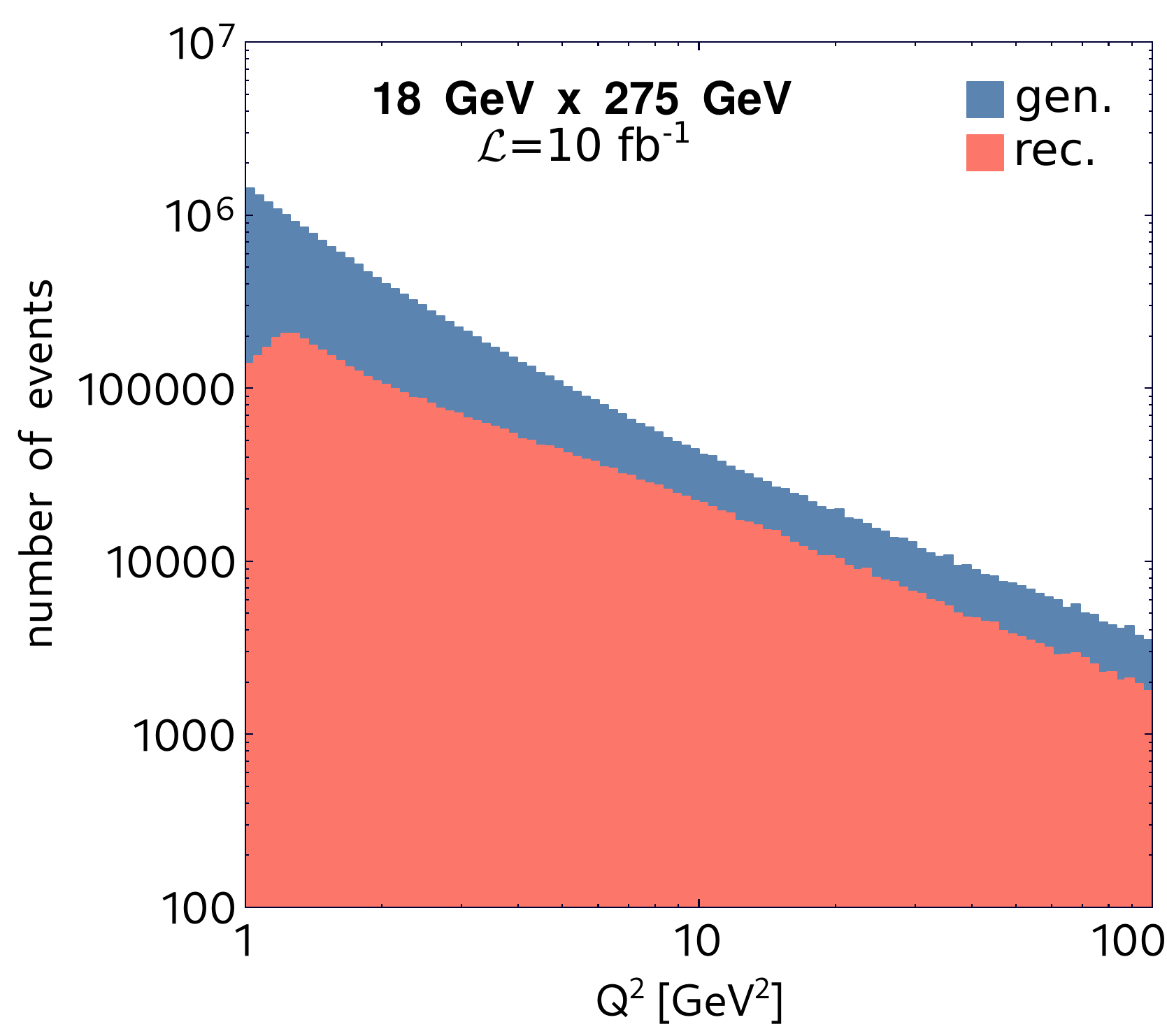}\\
\includegraphics[width=\plotWidthThree]{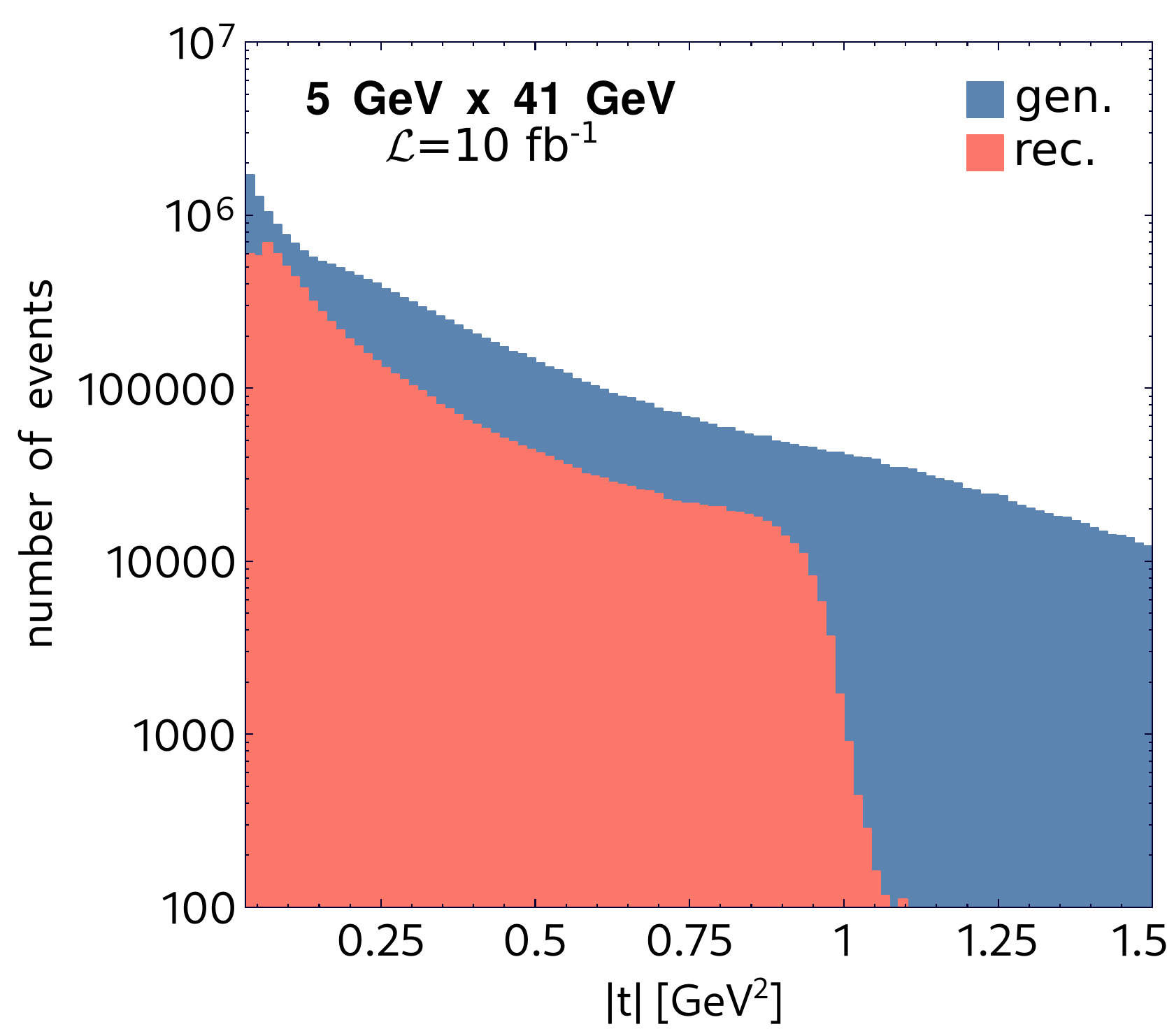}
\includegraphics[width=\plotWidthThree]{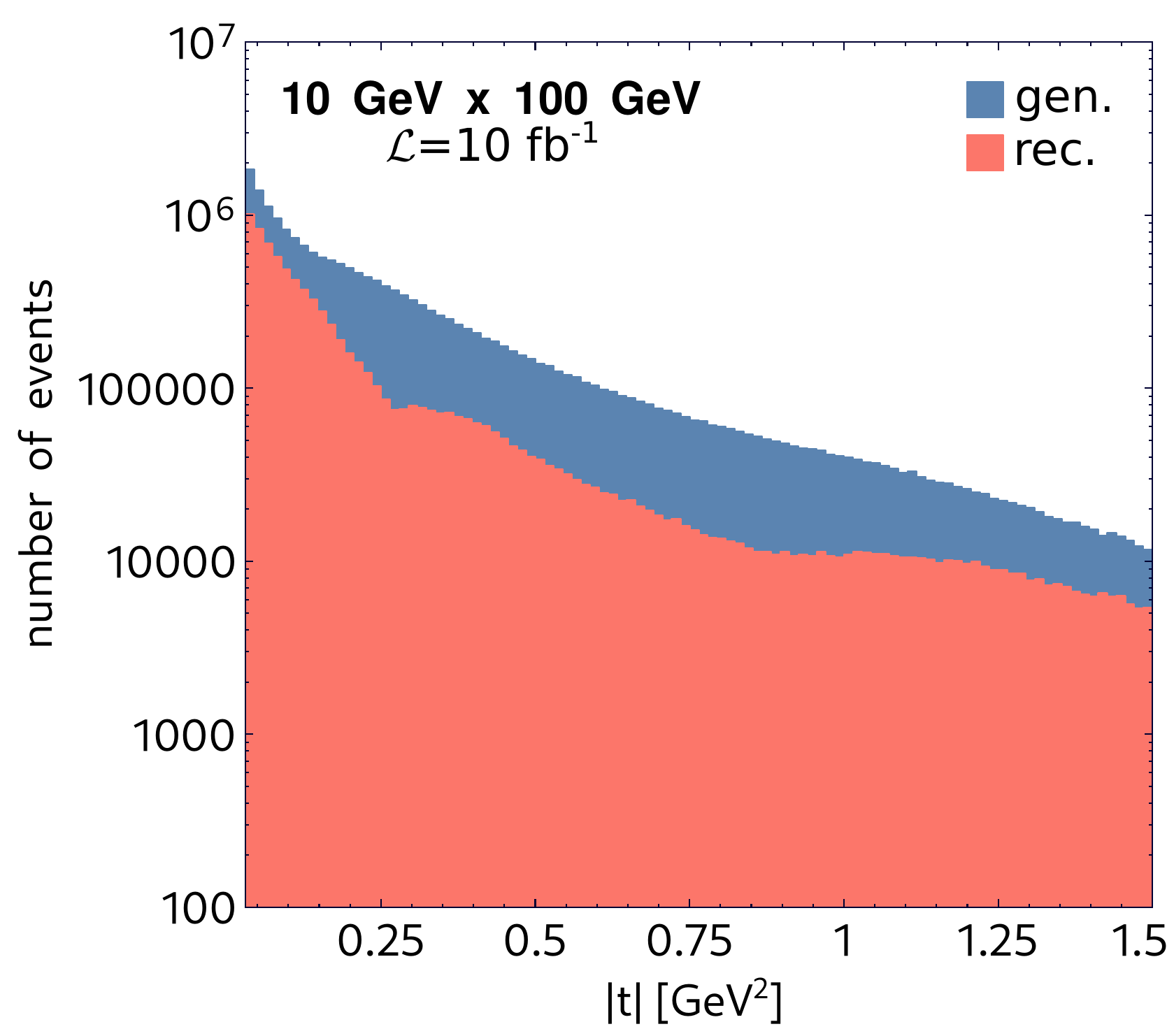}
\includegraphics[width=\plotWidthThree]{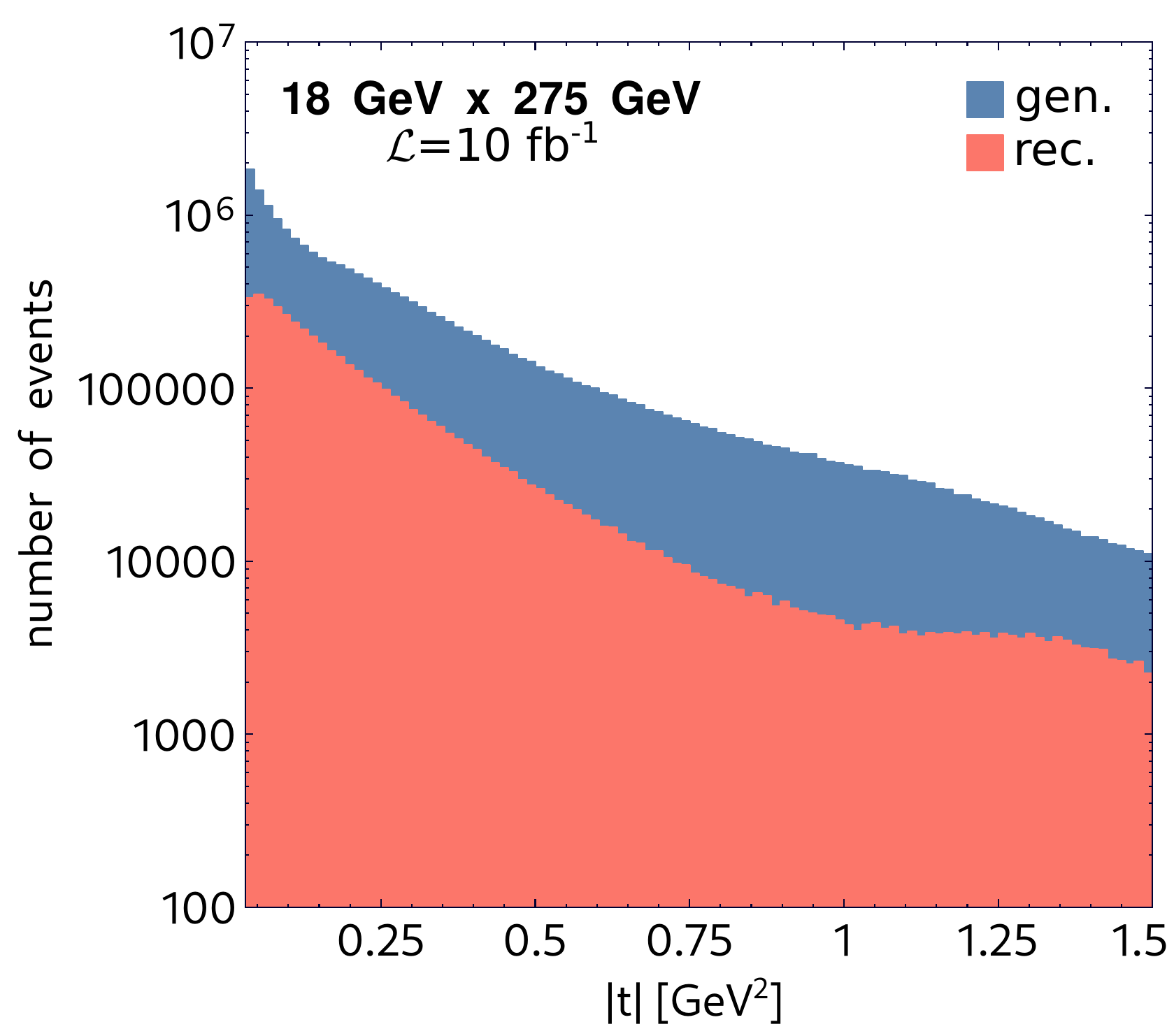}\\
\includegraphics[width=\plotWidthThree]{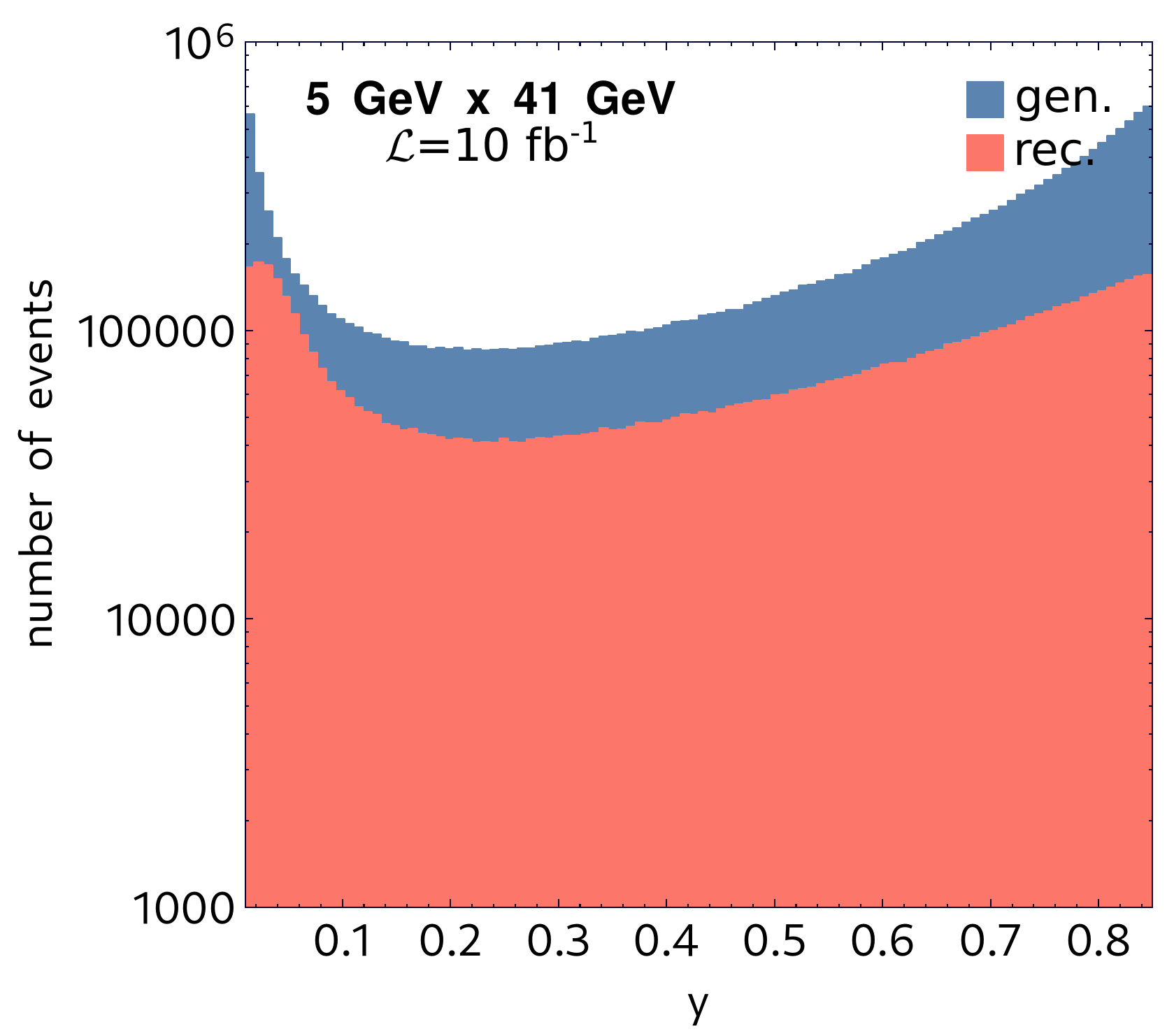}
\includegraphics[width=\plotWidthThree]{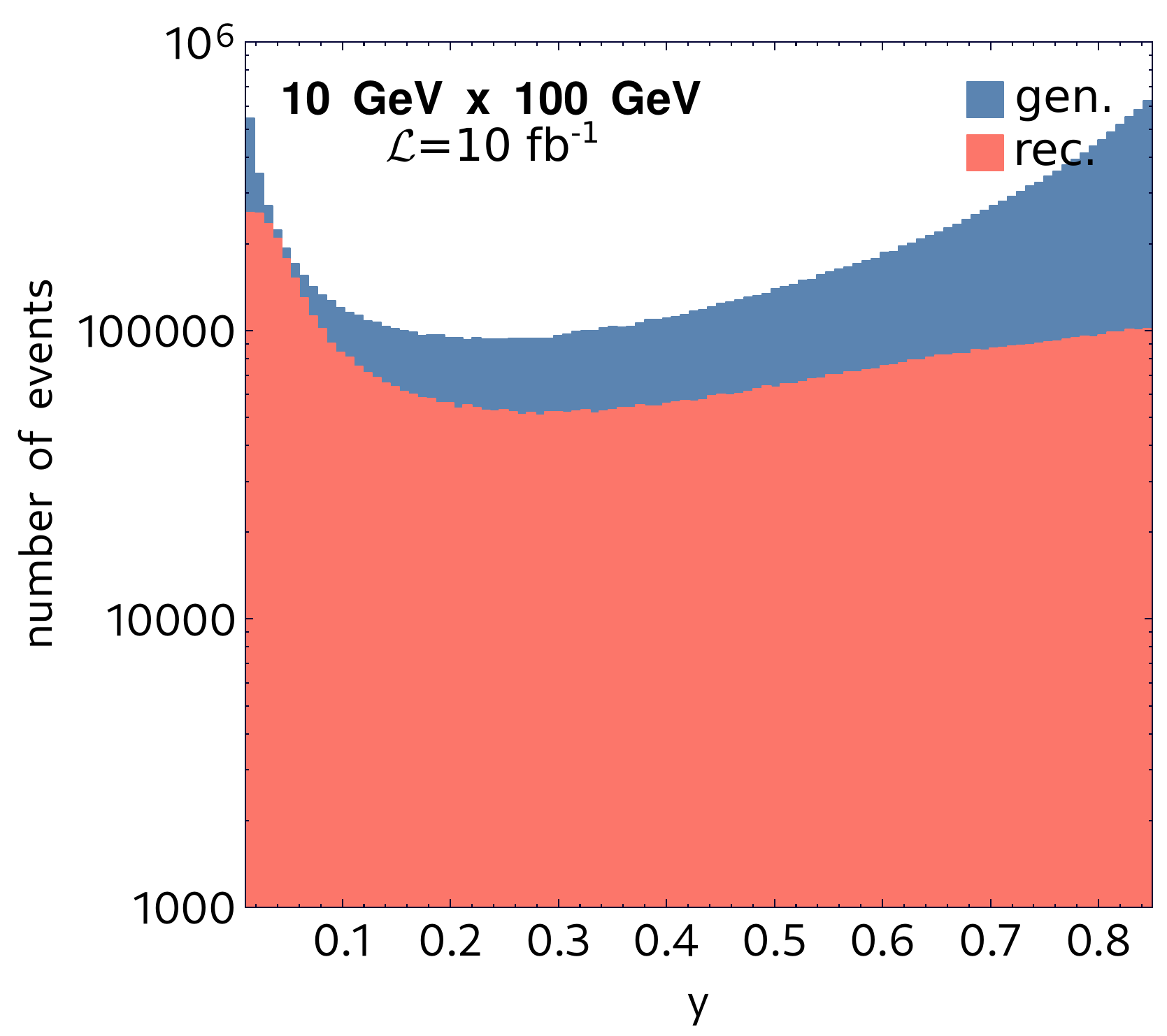}
\includegraphics[width=\plotWidthThree]{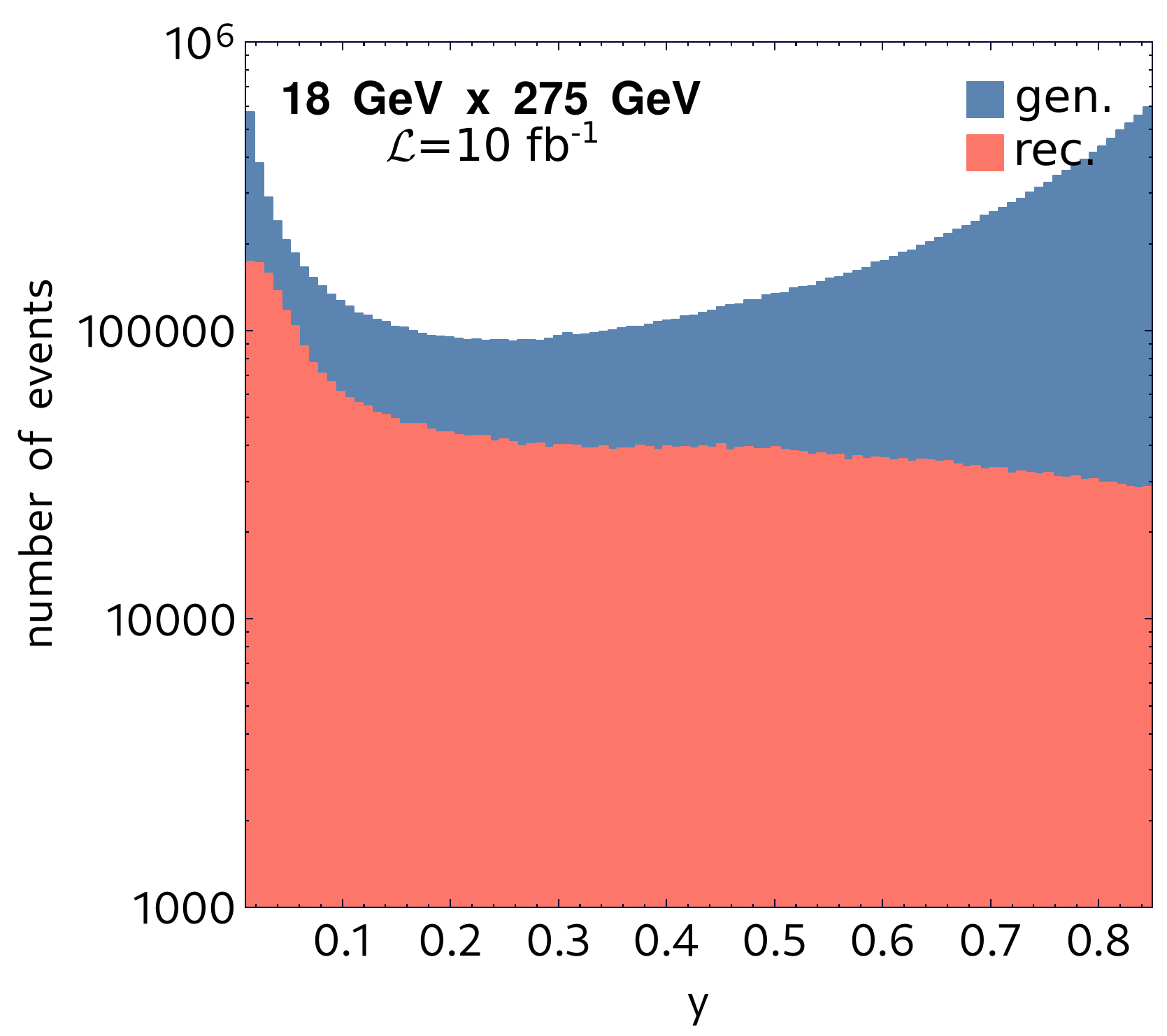}
\caption{Distributions of $\xB$, $Q^2$, $|t|$ and $y$ for generated (blue) and reconstructed (red) MC events containing the DVCS and Bethe-Heitler contributions to the cross section. The various beam energies are indicated in the plots.}
\label{fig:xBQ2ty}
\end{figure}

\begin{figure}[!ht]
\centering
\includegraphics[width=\plotWidthThree]{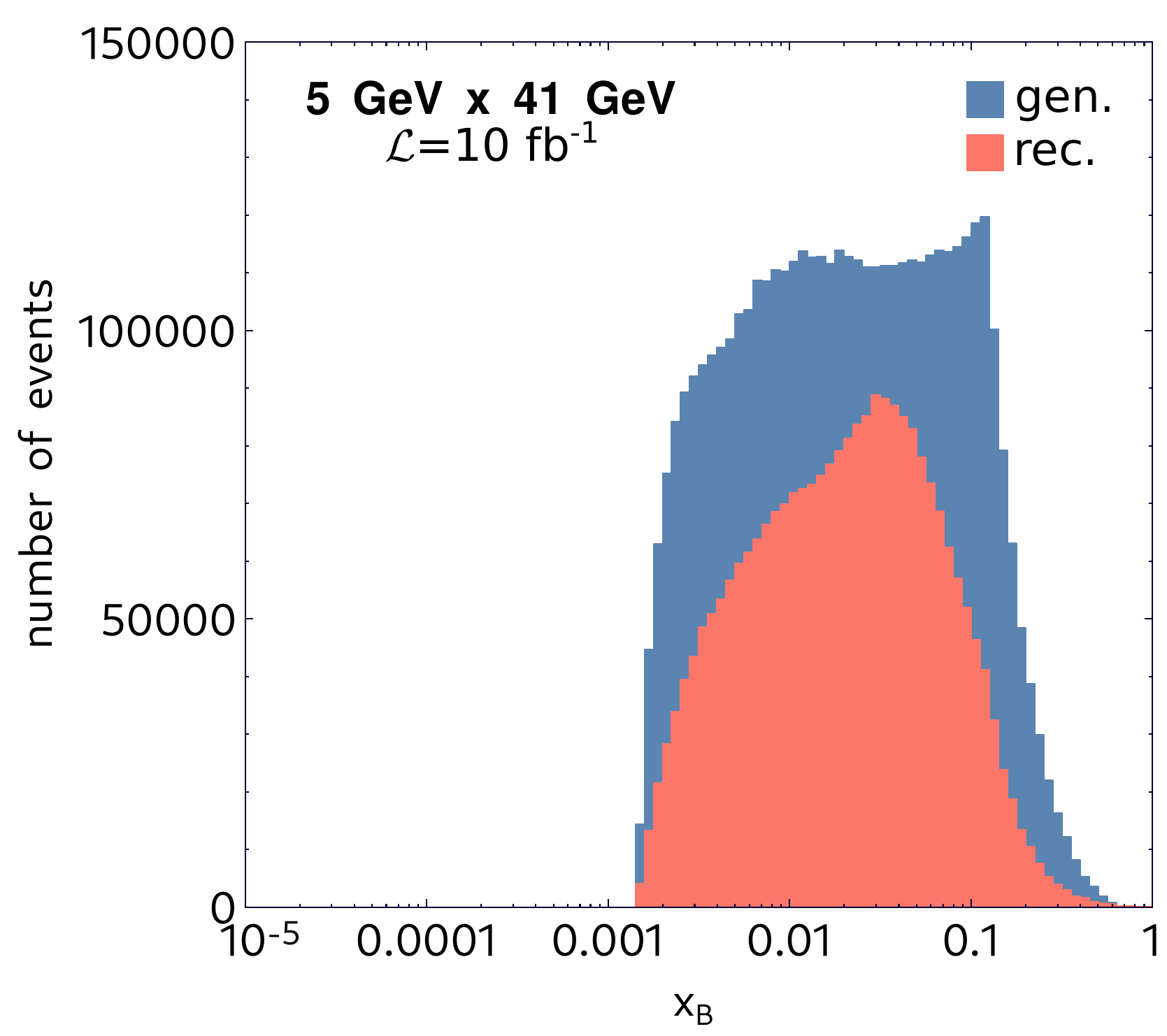}
\includegraphics[width=\plotWidthThree]{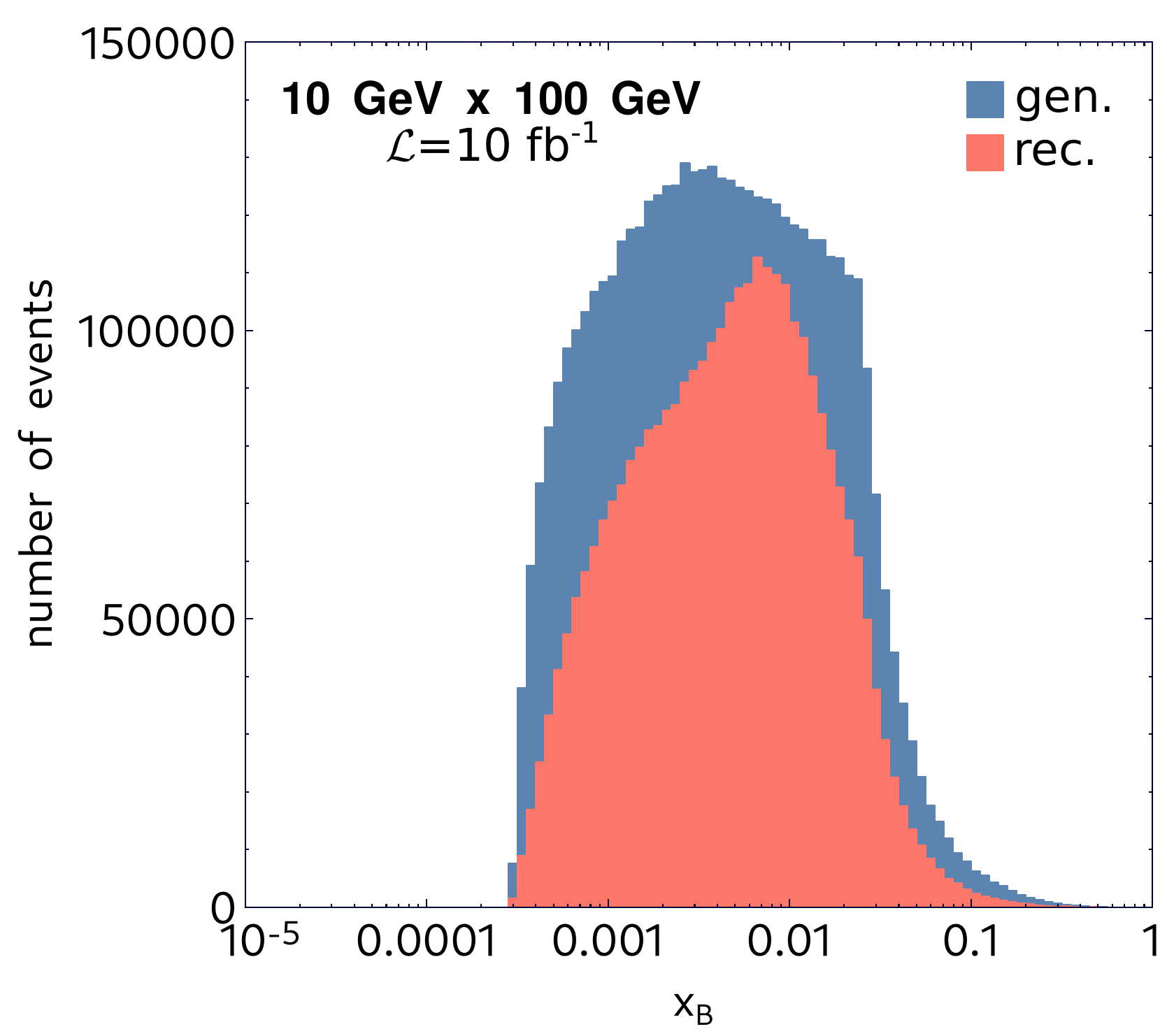}
\includegraphics[width=\plotWidthThree]{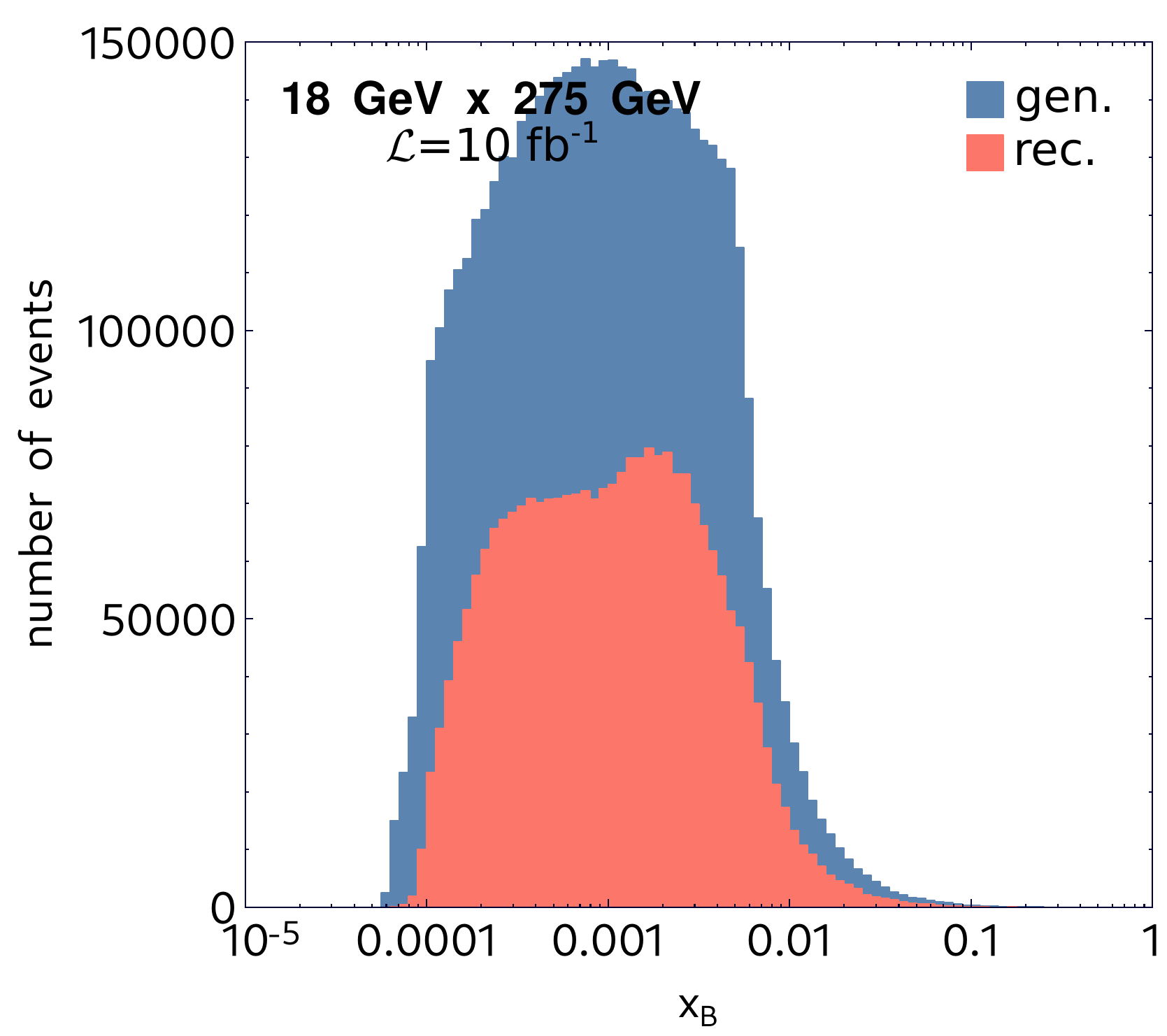}\\
\includegraphics[width=\plotWidthThree]{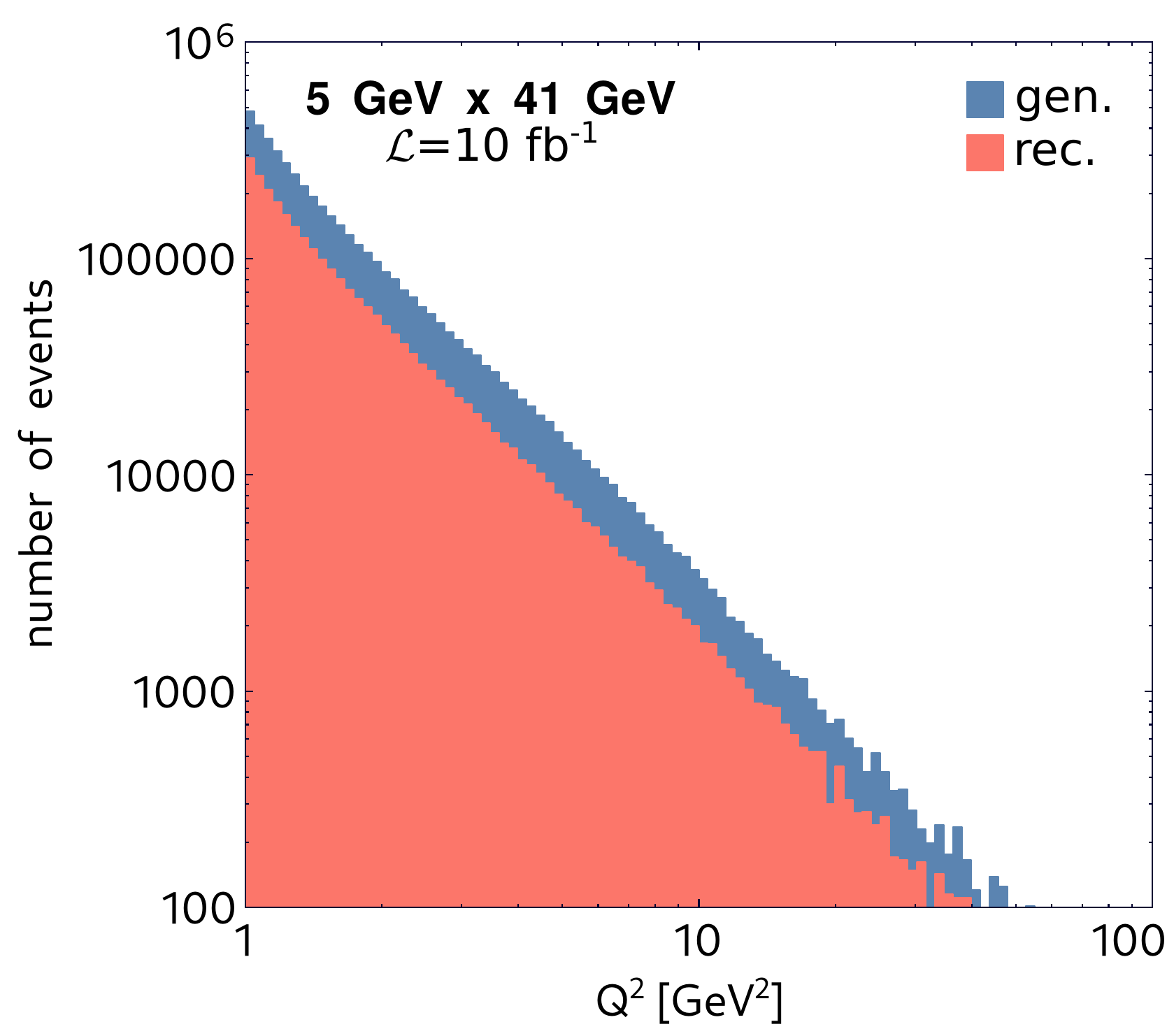}
\includegraphics[width=\plotWidthThree]{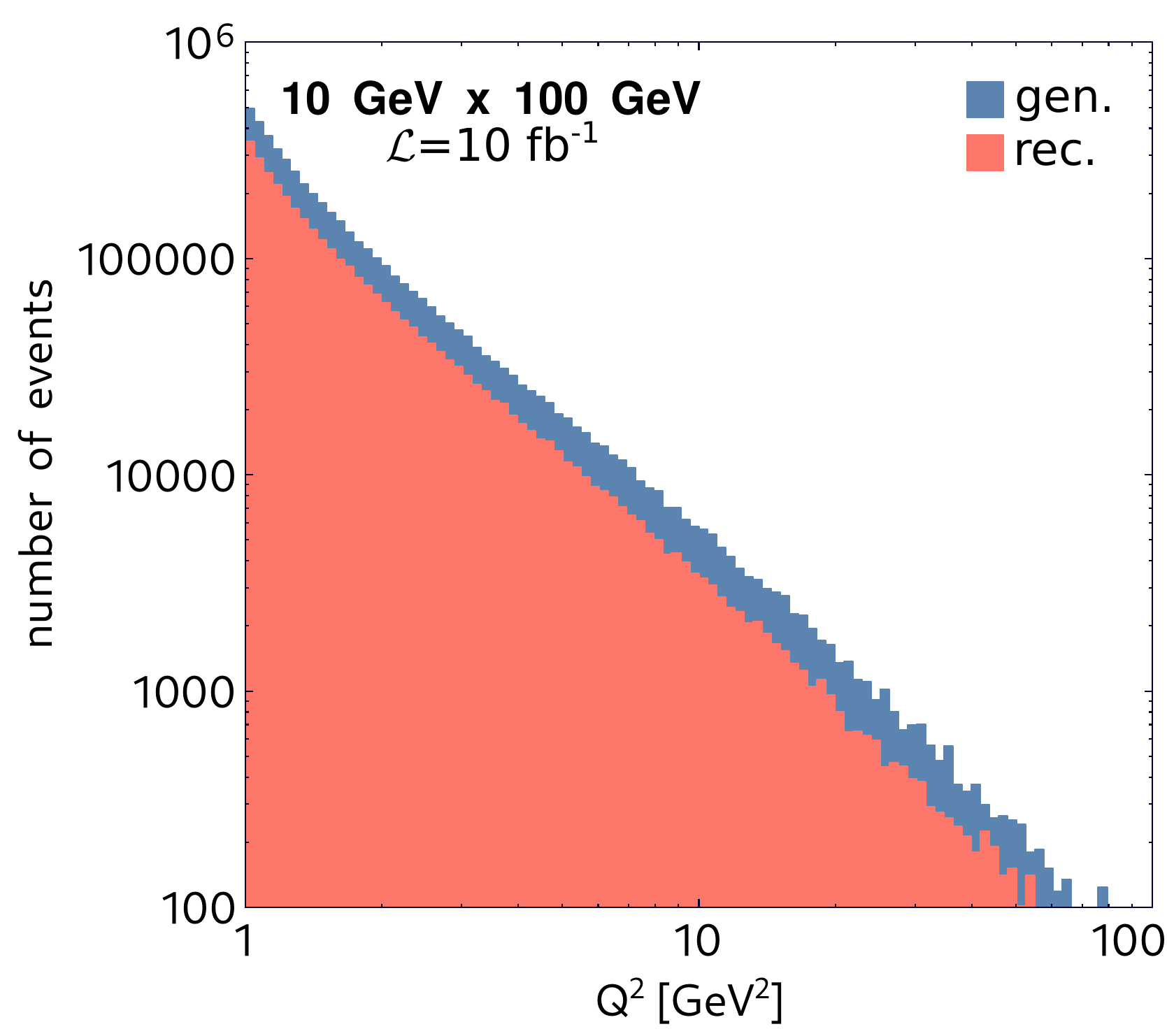}
\includegraphics[width=\plotWidthThree]{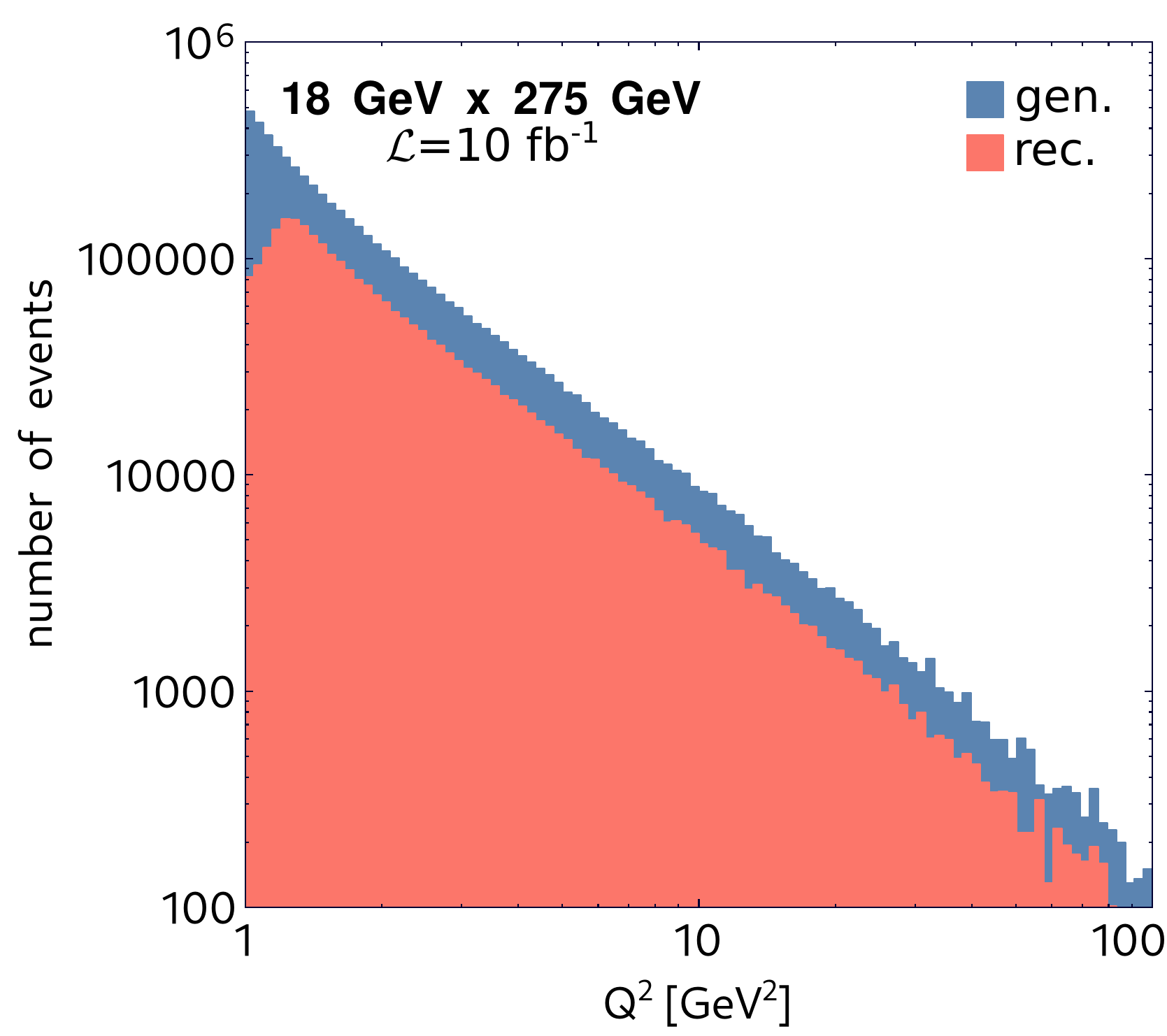}\\
\includegraphics[width=\plotWidthThree]{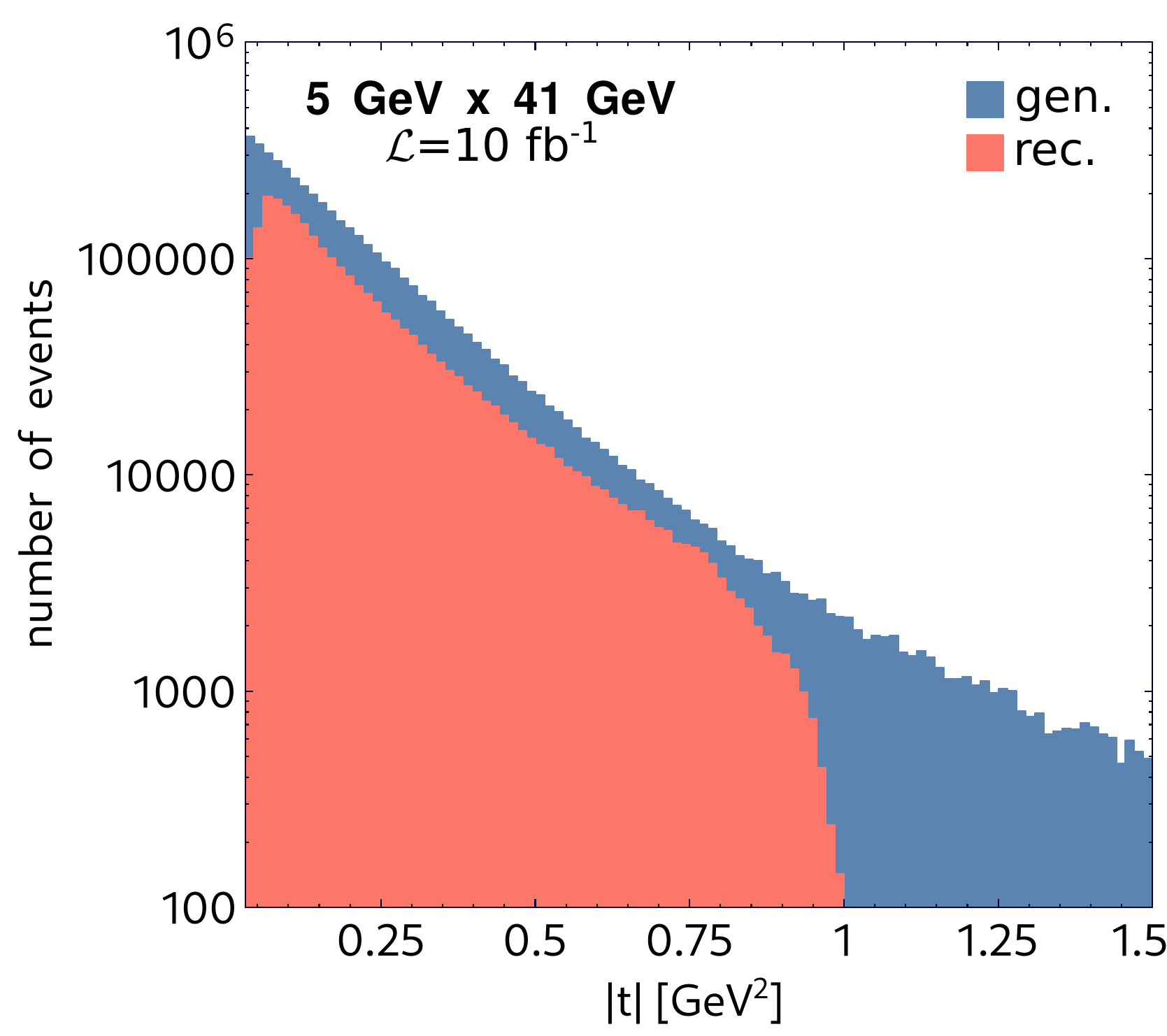}
\includegraphics[width=\plotWidthThree]{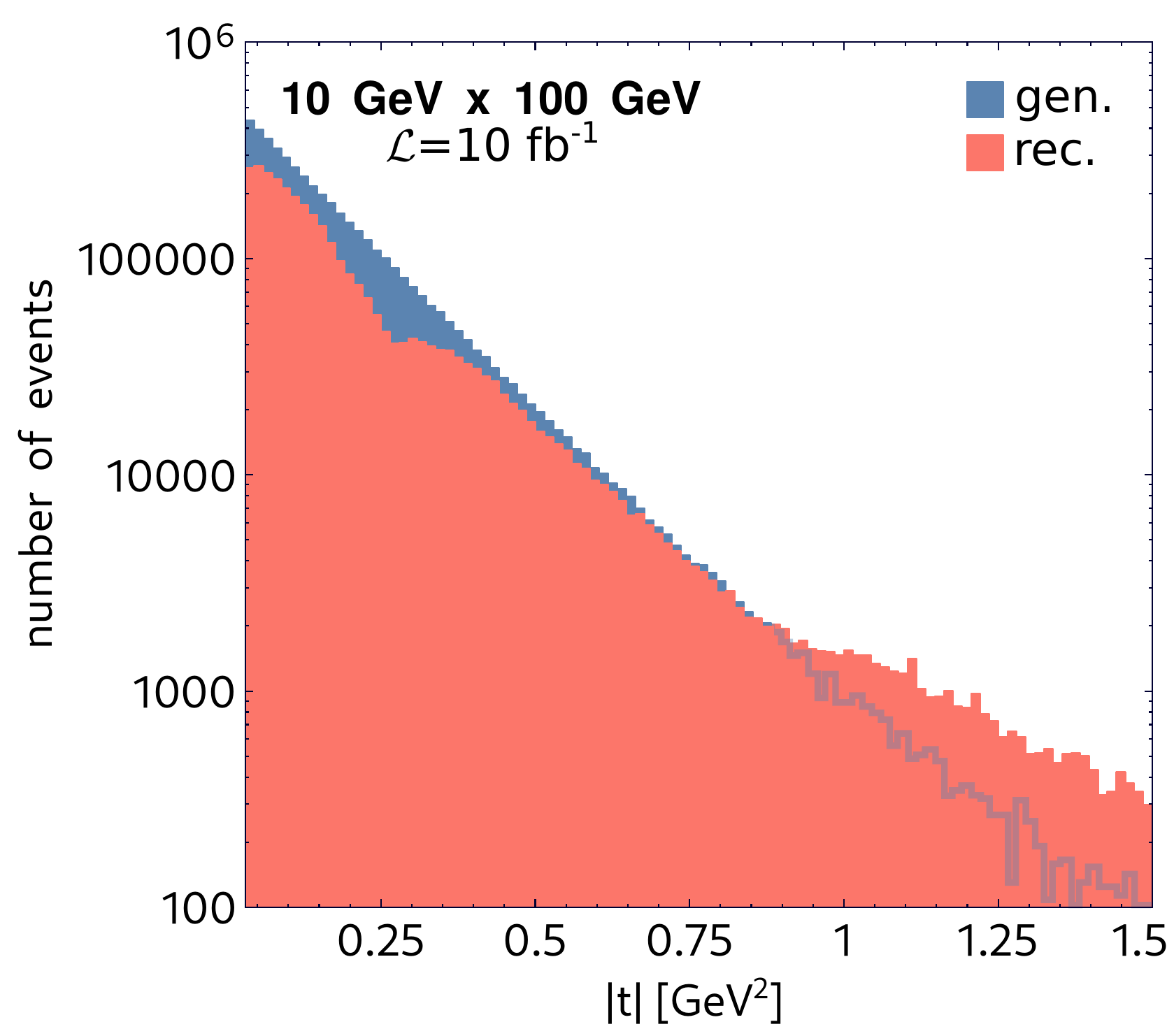}
\includegraphics[width=\plotWidthThree]{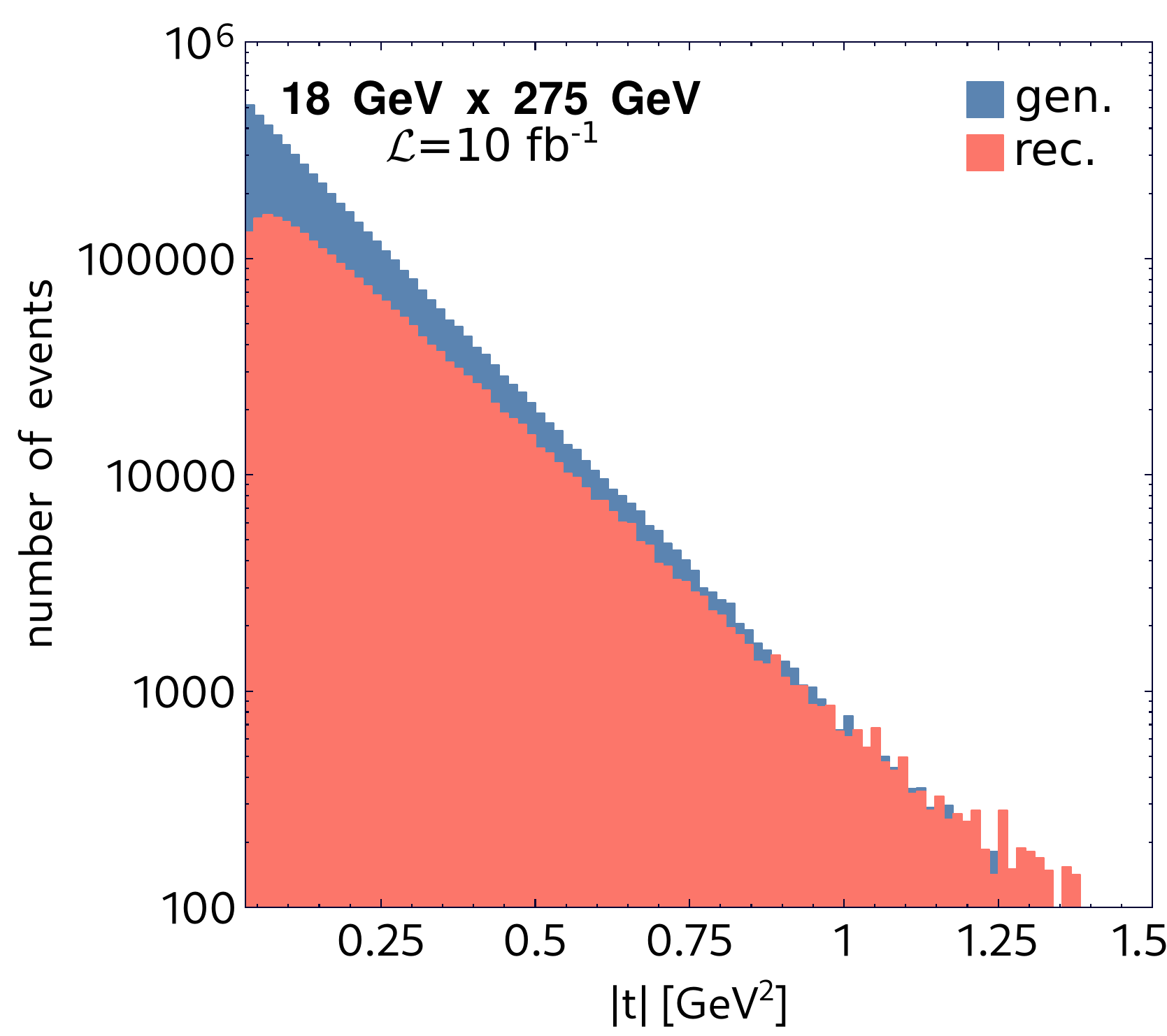}\\
\includegraphics[width=\plotWidthThree]{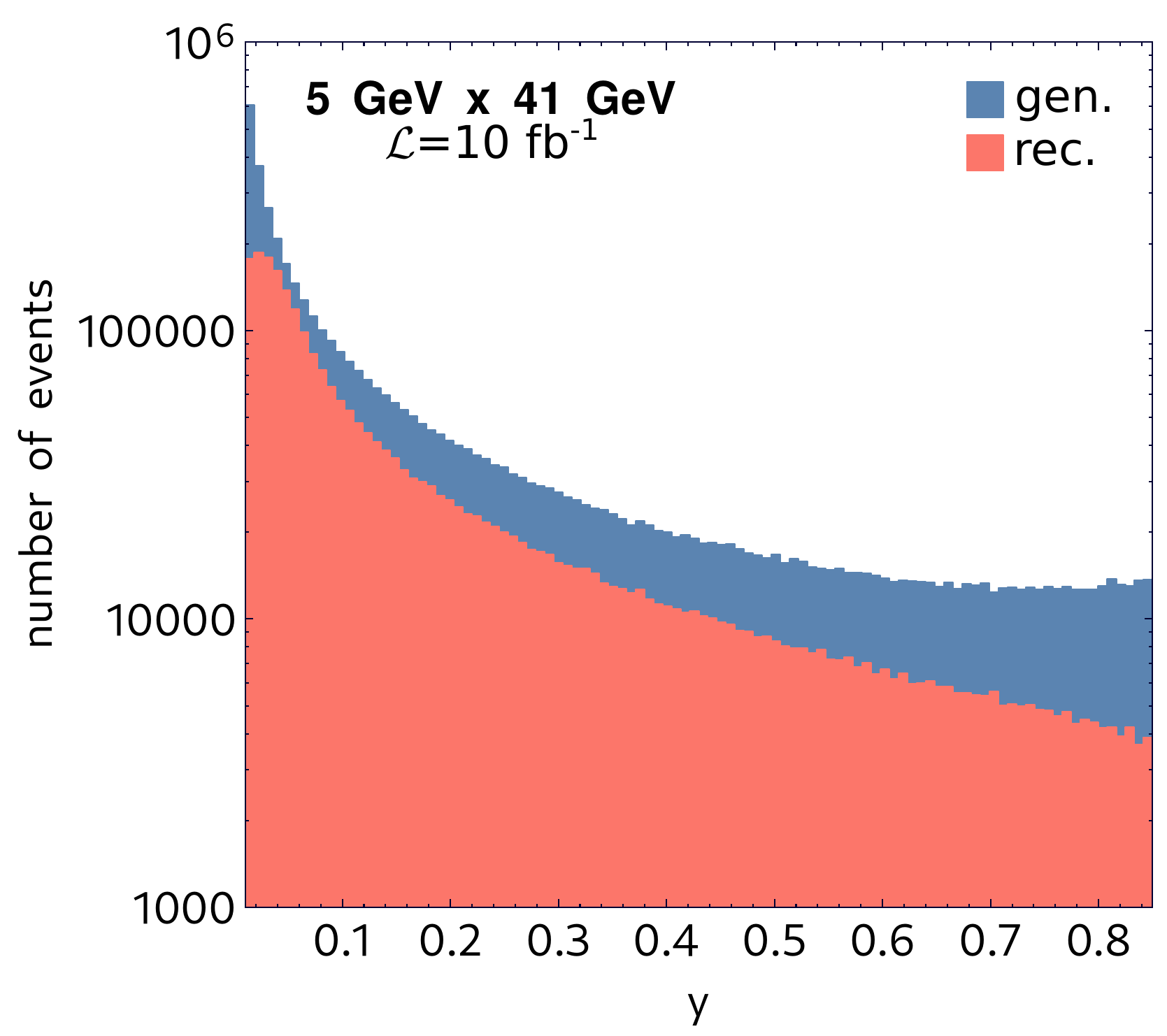}
\includegraphics[width=\plotWidthThree]{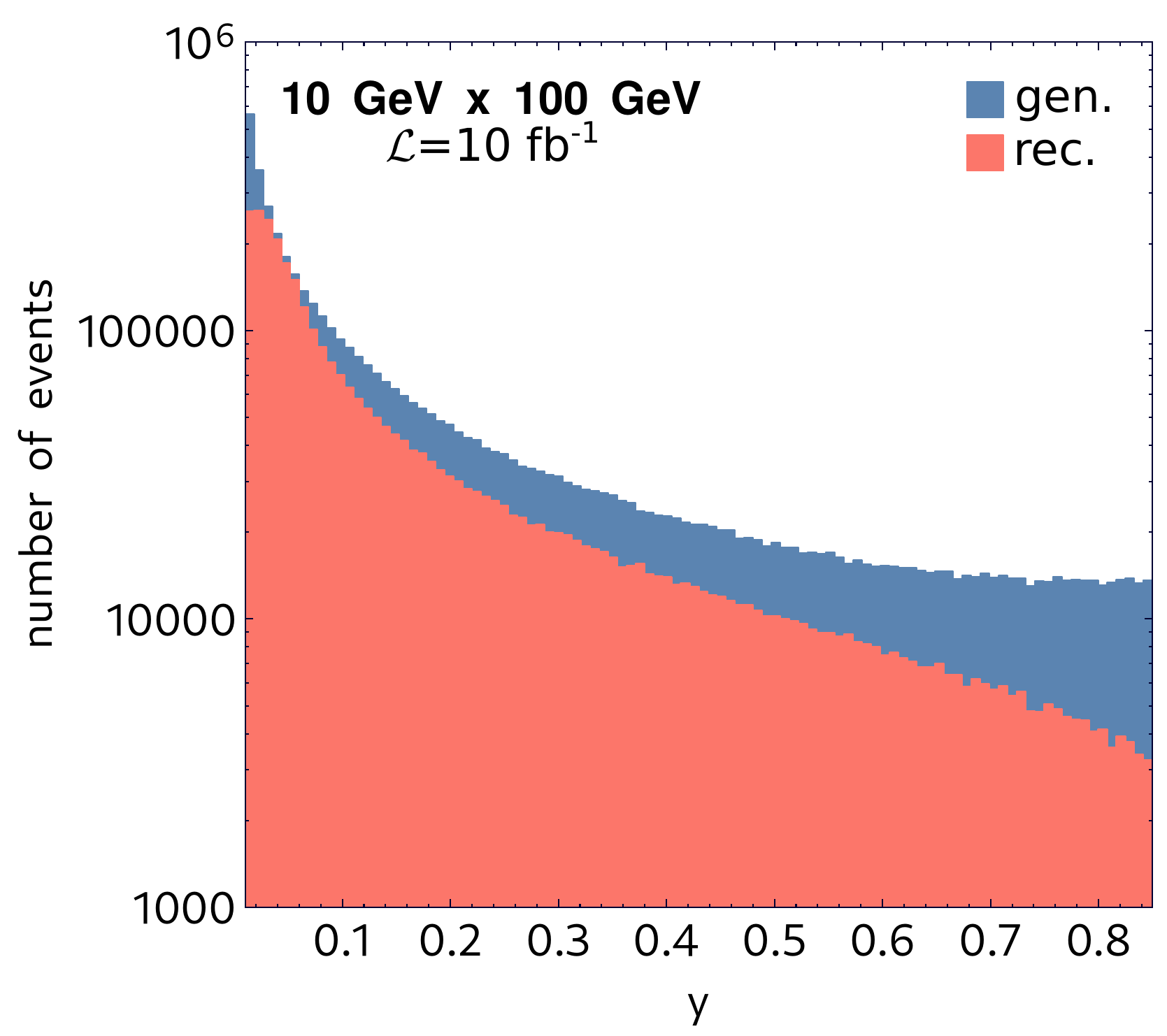}
\includegraphics[width=\plotWidthThree]{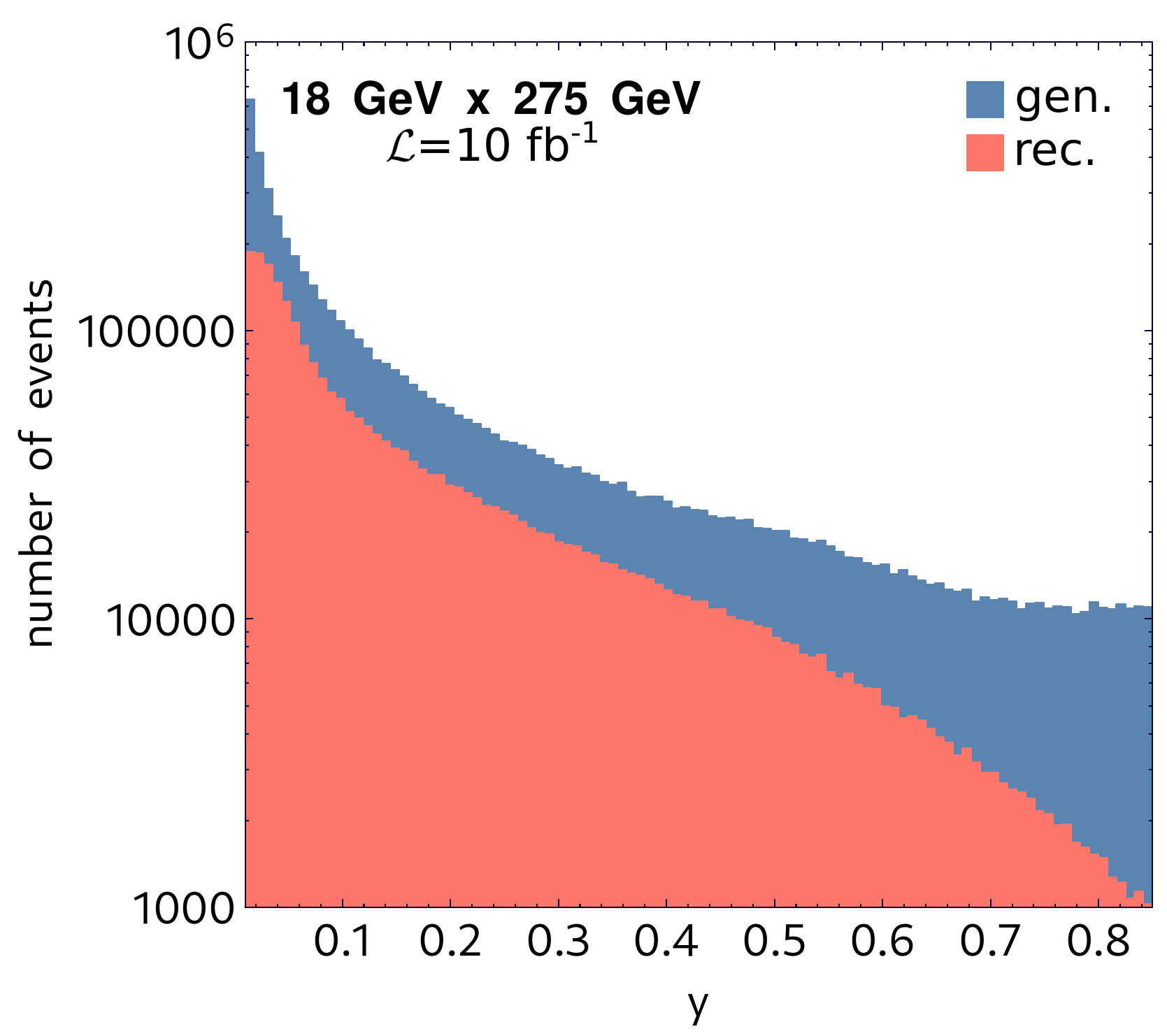}
\caption{The same as Fig.~\ref{fig:xBQ2ty}, but only for DVCS events.}
\label{fig:xBQ2tyDVCS}
\end{figure}

The overall acceptance probabilities are summarized in Table~\ref{tab:acceptance}. A couple of acceptance features are unique to the fiducial coverage of the detectors reconstructing the scattered proton. The 275~GeV beam energy case has essentially full acceptance coverage in the Roman pots, with only the lowest-$|t|$ protons being lost due to their proximity to the center of the proton beam, where the detectors cannot approach at too small a distance. This low-$|t|$ loss is present at all beam energies and is intrinsic to operation of any Roman pots detector in a collider. The 100~GeV and 41~GeV cases become more challenging because the same $|t|$-range corresponds to a larger range of scattering angle, necessitating reconstruction in the B0 tracking spectrometer (higher-$|t|$), along with the Roman pots (lower-$|t|$). The coverage between these detectors is limited by the necessary presence of a beam pipe, resulting in a dip in the acceptance in $|t|$ (see Figs.~\ref{fig:xBQ2ty} and~\ref{fig:xBQ2tyDVCS}, third row, middle plot). The same feature is also present in Figs.~\ref{fig:eta} and ~\ref{fig:etaDVCS} in the center plot.

\begin{table}[!ht]
\centering
\begin{tabular}{P{0.158\textwidth}P{0.158\textwidth}P{0.158\textwidth}P{0.158\textwidth}P{0.158\textwidth}P{0.158\textwidth}}
\toprule
$E_e$ [$\GeV$]   & $E_p$ [$\GeV$] & $p_{e'}$ & $p_{p'}$ & $p_{\gamma}$ & $p_{e'+p'+\gamma}$ \\ \midrule
$5$& $41$& $0.90$& $0.76$& $0.72$& $0.49$\\
$10$& $100$& $0.90$& $0.90$& $0.59$& $0.48$\\
$18$& $275$& $0.87$& $0.81$& $0.46$& $0.29$\\ \bottomrule
\end{tabular}
\caption{Probabilities for reconstructing outgoing particles taking part in the reaction \eqref{eq:process} for given electron, $E_e$, and proton, $E_p$, beam energies and conditions specified in Sect.~\ref{subsec:epic} (including the additional kinematic cuts applied at the level of analysis). The probability of reconstructing all particles in a given event (the exclusivity) is provided in the last column.}
\label{tab:acceptance}
\end{table}

\begin{figure}[!ht]
\centering
\includegraphics[width=\plotWidthThree]{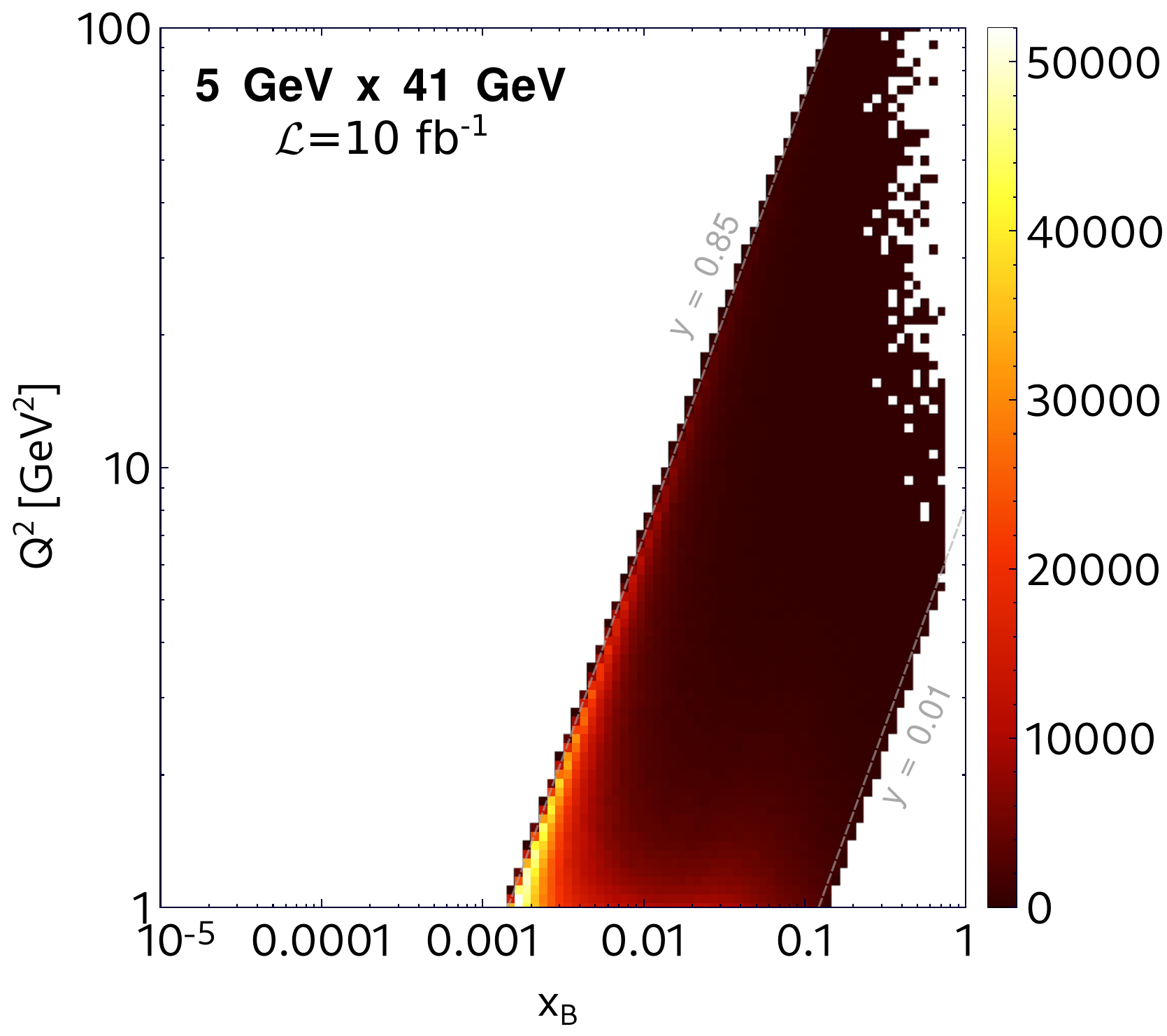}
\includegraphics[width=\plotWidthThree]{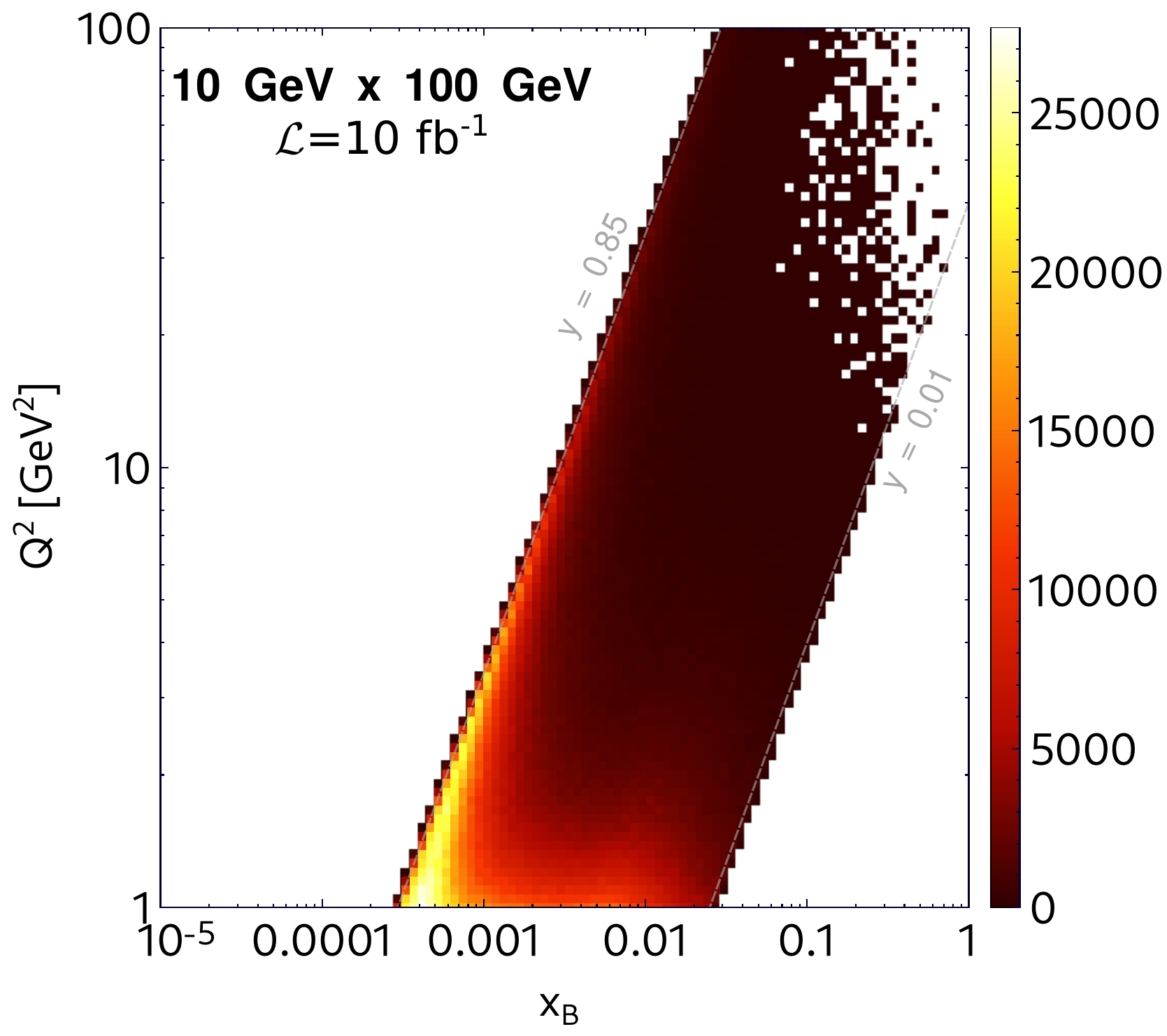}
\includegraphics[width=\plotWidthThree]{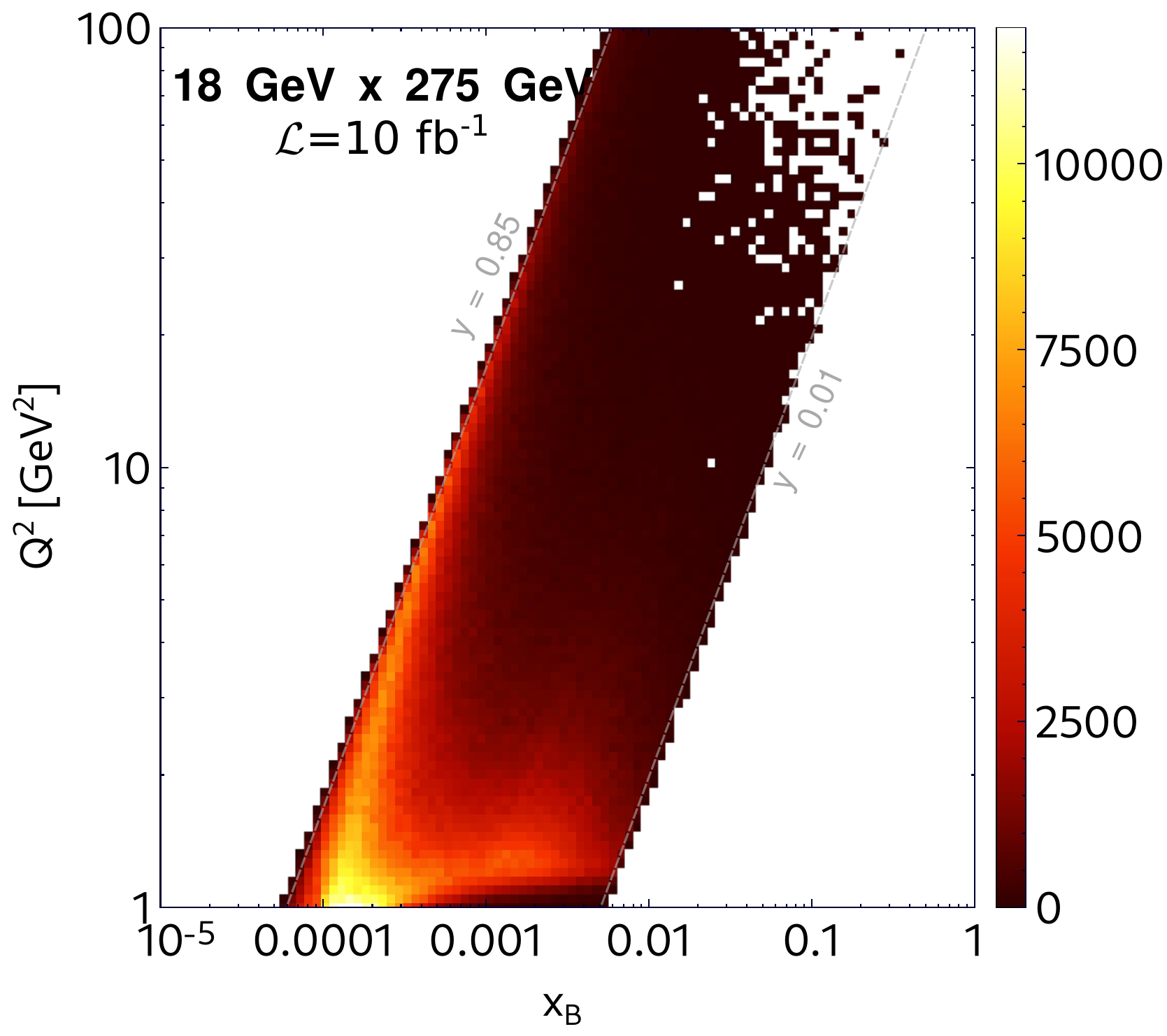}
\caption{Distributions of $\xB$ vs. $Q^2$ for reconstructed MC events and beam energies indicated in the plots.}
\label{fig:xBVsQ2}
\end{figure}

\begin{figure}[!ht]
\centering
\includegraphics[width=\plotWidthThree]{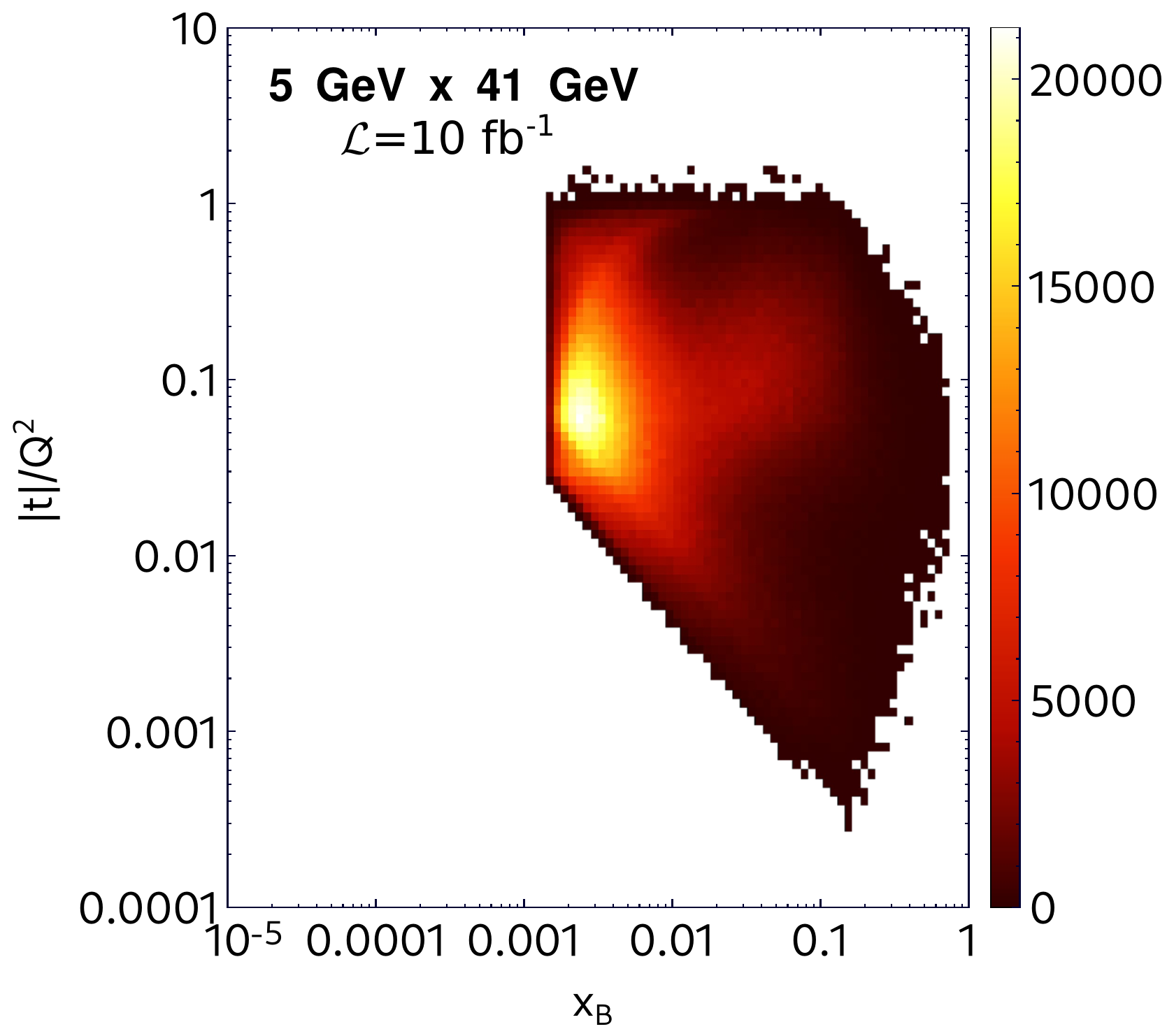}
\includegraphics[width=\plotWidthThree]{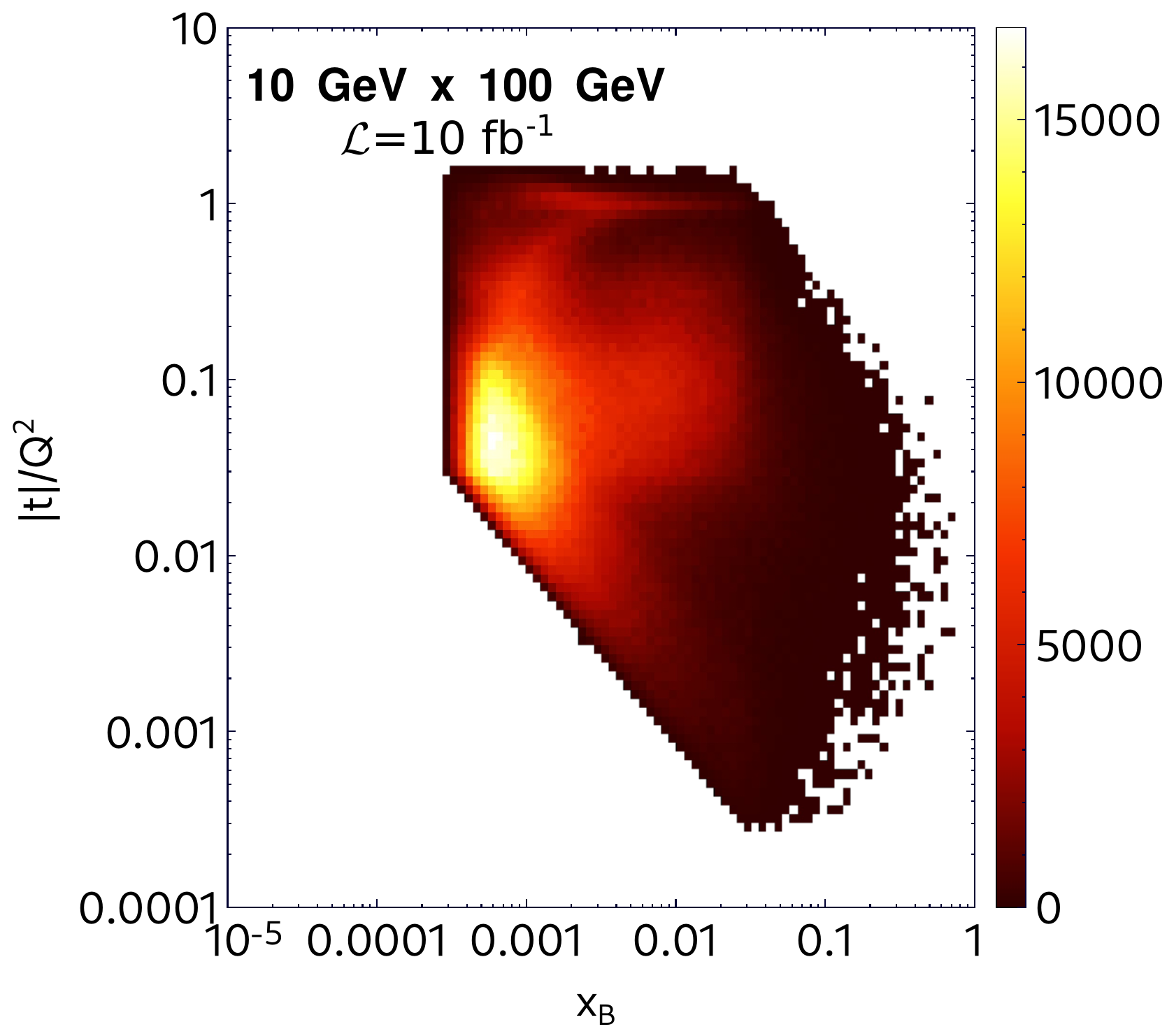}
\includegraphics[width=\plotWidthThree]{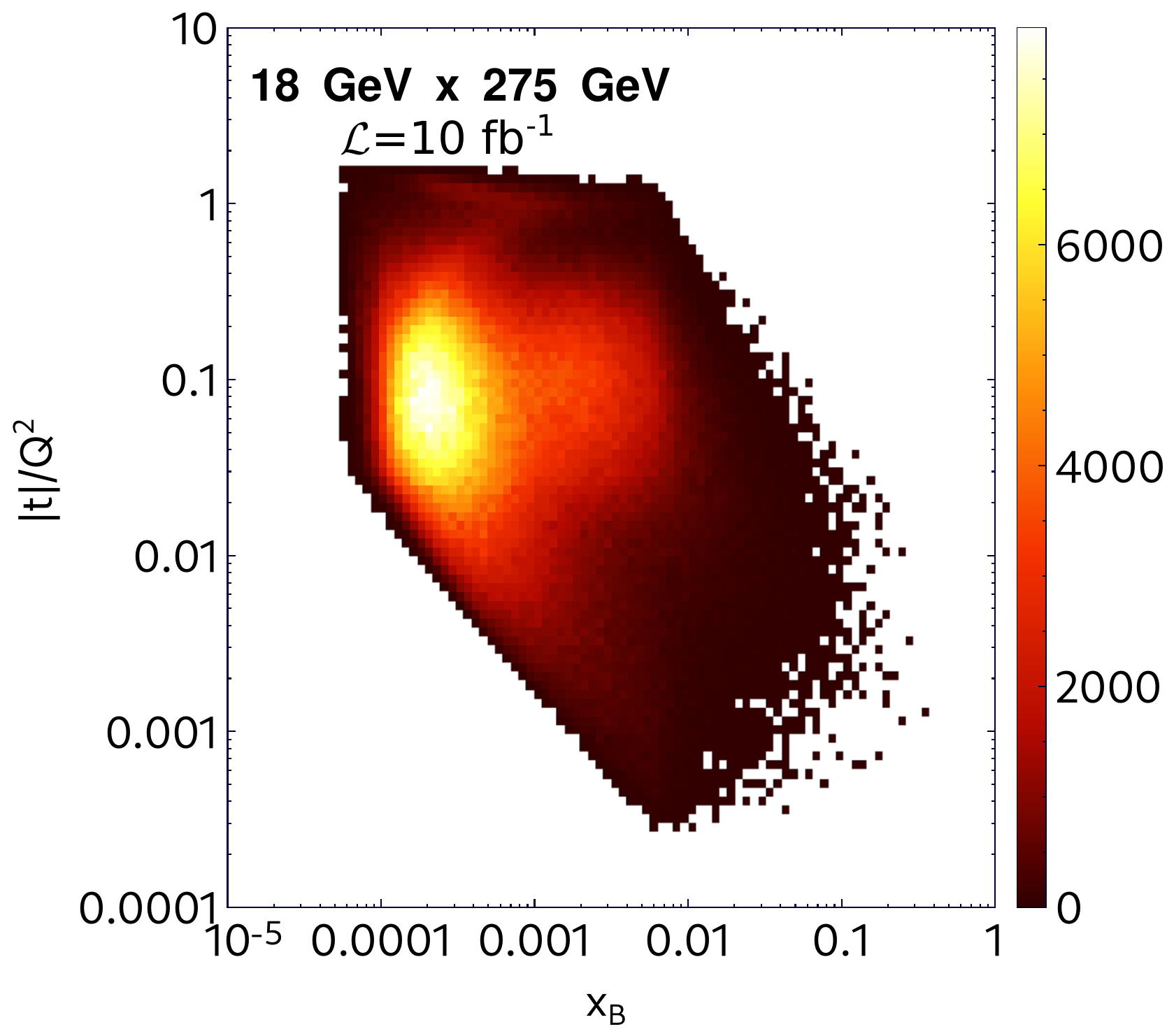}
\caption{Distributions of $\xB$ vs. $|t|/Q^2$ for reconstructed MC events and beam energies indicated in the plots.}
\label{fig:xBVstOverQ2}
\end{figure}

\begin{figure}[!ht]
\centering
\includegraphics[width=\plotWidthTwo]{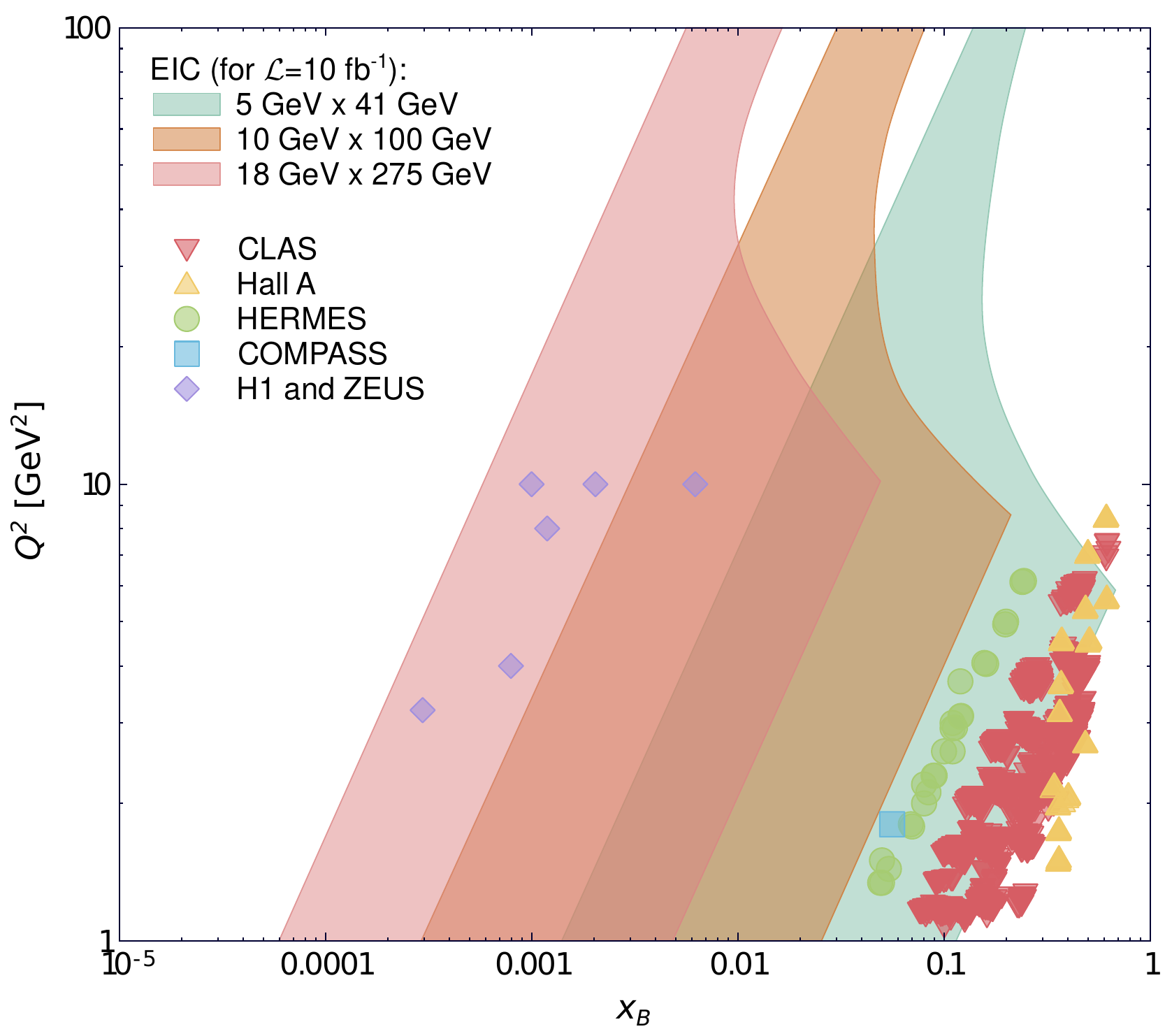}
\includegraphics[width=\plotWidthTwo]{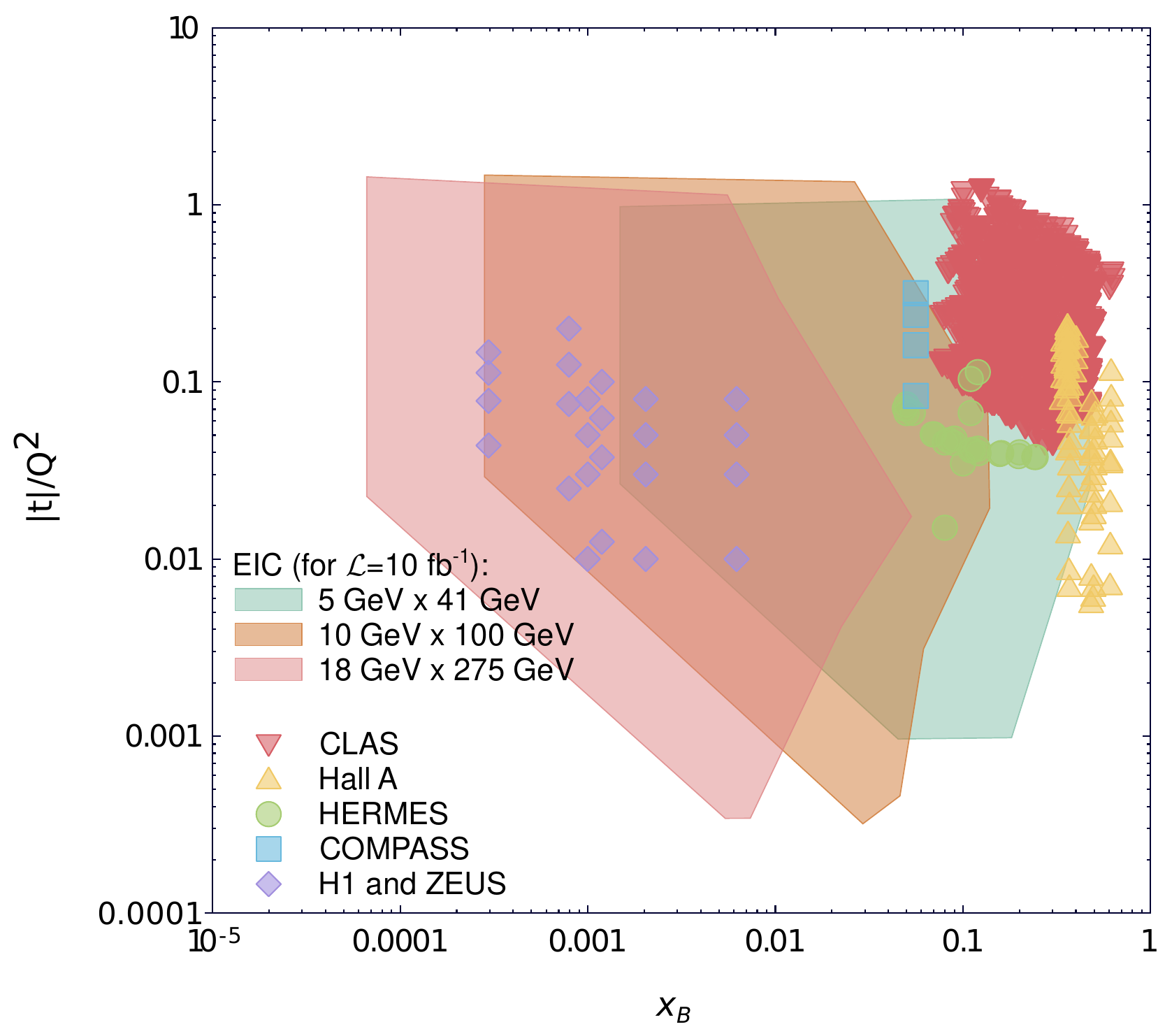}
\caption{Coverage of the $(\xB$,~$Q^2)$  and $(\xB$,~$|t|/Q^2)$ regions by ePIC and existing DVCS data for a proton target.}
\label{fig:xBVsQ2AndxBVstOverQ2AndData}
\end{figure}

\begin{figure}[!ht]
\centering
\includegraphics[width=\plotWidthThree]{plots/kinematics/t_12.pdf}
\qquad
\includegraphics[width=\plotWidthThree]{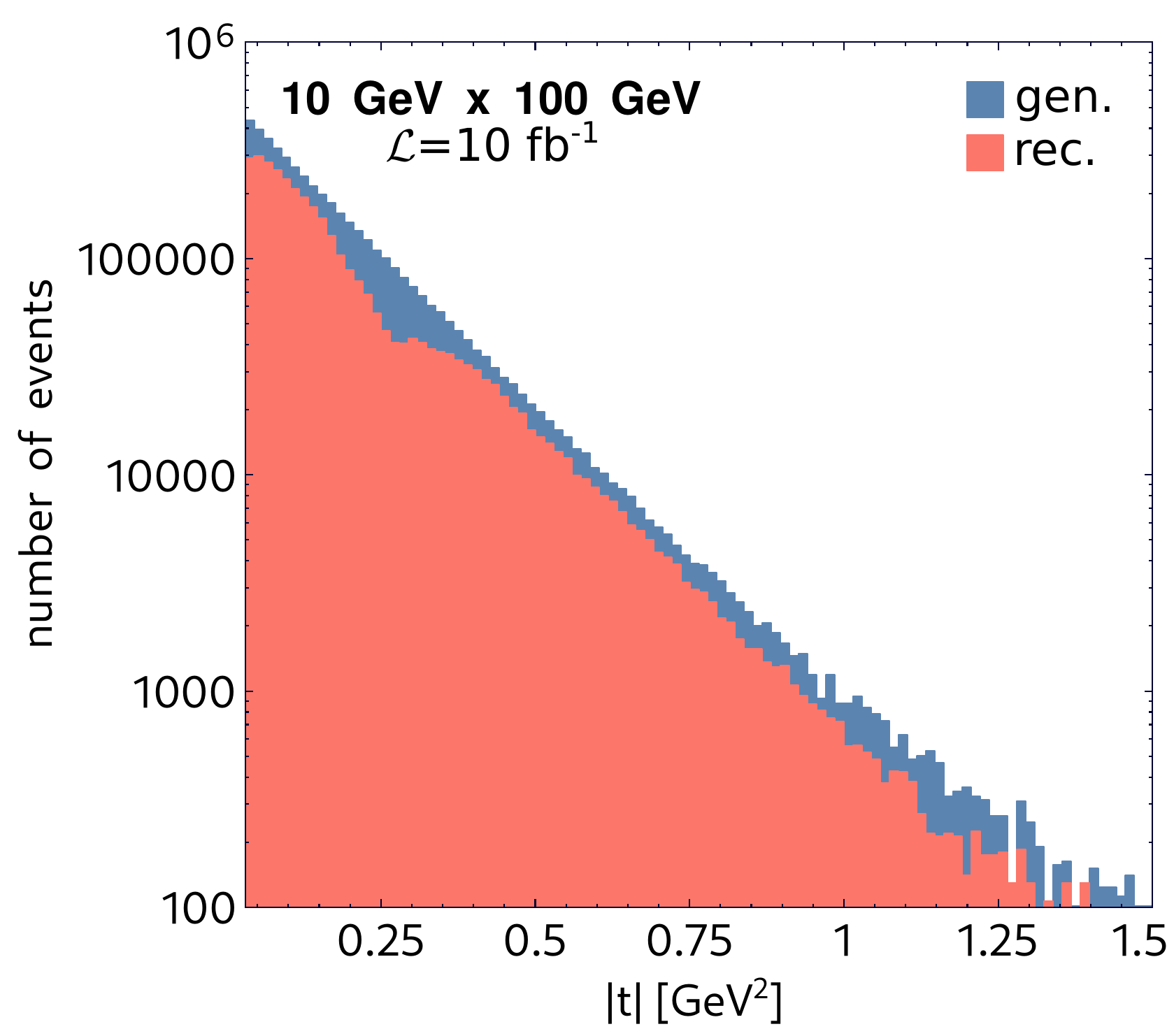}
\caption{Distributions of DVCS events as a function of $|t|$ obtained using two methods for reconstructing kinematic variables: (left) electron method, (right) Jacquet-Blondel method.}
\label{fig:smearing_effect}
\end{figure}

The coverage of $\xB$ vs.\ $Q^2$ and $\xB$ vs.\ $-t/Q^2$ is shown in Figs.~\ref{fig:xBVsQ2} and~\ref{fig:xBVstOverQ2}, respectively. The quantity $-t/Q^2$ is the small parameter in the twist expansion and therefore important for the description of the process~\eqref{eq:process}. We note that the current phenomenological analyses are typically limited to the leading-twist description, see for instance Refs.~\cite{Kumericki:2015lhb,Moutarde:2018kwr}, which restricts the range of usable experimental data to $-t/Q^2 \ll 1$. Although the effect is expected to be less severe compared to fixed-target experiments, the inclusion of higher-twist contributions, calculated for instance in the spirit of Ref.~\cite{Braun:2022qly}, will be crucial for fully utilizing EIC data.

The covered kinematic domains corresponding to Figs.~\ref{fig:xBVsQ2} and~\ref{fig:xBVstOverQ2} are collectively shown in Fig.~\ref{fig:xBVsQ2AndxBVstOverQ2AndData} together with the existing experimental data. To highlight only the relevant domain for ePIC, the marked areas exclude the least populated bins, which sum up to $0.1\%$ of the signal. One-dimensional distributions of the variables $\xB$, $Q^2$, $t$, and $y$ are shown in Fig.~\ref{fig:xBQ2ty} for the combination of BH and DVCS events, and in Fig.~\ref{fig:xBQ2tyDVCS} for DVCS events only. These figures include distributions for both generated (blue histograms) and reconstructed (red histograms) events, illustrating the impact of ePIC's geometrical acceptance and energy and momentum smearing. For instance, a manifestation of the latter effect is an excess for reconstructed events observed at high values of $t$ ($t$ > 1 GeV$^2$) for the sample of DVCS events at the $10\, \textrm{GeV} \times 100\, \textrm{GeV}$ beam energy configuration. This excess of events is a manifestation of the worsening resolution in the inelasticity at low values of $y$ which leads to an enhancement of events at high $\xB$ and low $Q^2$ in the $t$-distribution. The effect becomes so pronounced due to the very flat $t$-dependence (cf. the discussion in Sect.~\ref{subsec:tomography}) at high $\xB$. This effect can be mitigated by applying alternative methods of reconstructing kinematic variables. This is demonstrated in Fig.~\ref{fig:smearing_effect}, where we compare the $t$-distribution reconstructing $x$ and $Q^2$ through the electron method with an alternative method, namely, the so-called Jacquet-Blondel technique (see for instance Ref.~\cite{ZEUS:1993ppj}). This alternative method utilizes information from all other particles in the final state. A detailed study of applying different reconstruction methods is beyond the scope of this analysis. However, we note that regardless of the method used, the effect of smearing must still be corrected when unfolding detector effects in the extraction of observables.

From the comparison of Figs.~\ref{fig:xBQ2ty} and \ref{fig:xBQ2tyDVCS}, we directly see the dominance of BH over DVCS in the high-$y$ domain, corresponding to low $\xB$. This motivates a stricter cut on $y$ in the analysis of nucleon tomography (see Sect.~\ref{subsec:tomography}), which relies on the reconstruction of a pure DVCS signal. The cut is relaxed in the analysis of asymmetries sensitive to the interference between DVCS and BH sub-processes (see Sect.~\ref{subsec:asymmetries}). The distributions of $t$ shown in Figs.~\ref{fig:xBQ2ty} and \ref{fig:xBQ2tyDVCS} are also significantly different. For the mixture of BH and DVCS events, we observe a superposition of distributions describing both sub-processes. In our simulation, we assume a dipole-like distribution for BH and an exponential distribution for DVCS. For DVCS, focusing on the dominant contribution of GPD $H$ for a given quark flavor, the exponential distribution has a slope that depends on $\xB$ and $Q^2$. This can be seen in Fig.~\ref{fig:xBQ2tyDVCS}, as the distributions for generated events do not follow a single exponential shape.


\subsection{Radiative corrections}
\label{subsec:rc}

\begin{figure}[!ht]
\centering
\includegraphics[width=\textwidth]{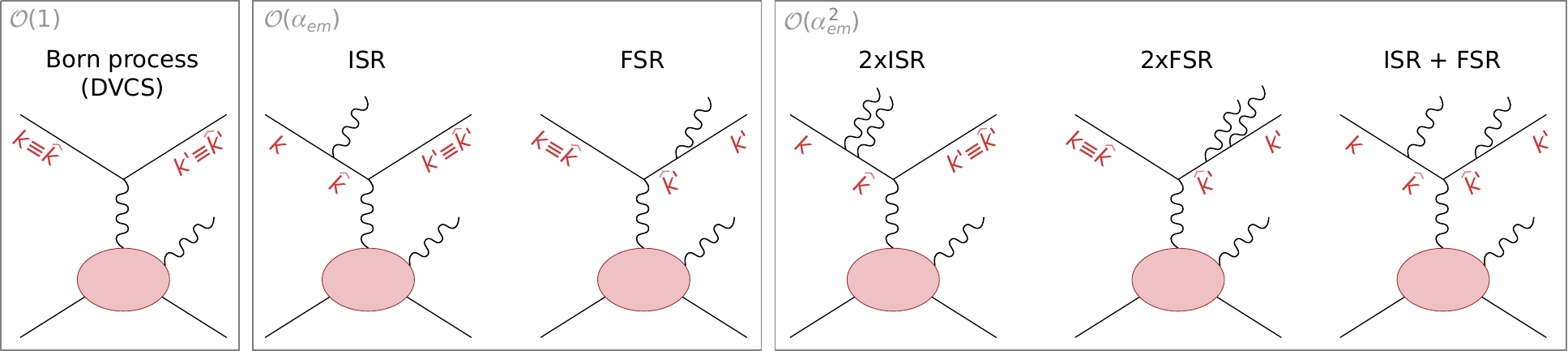}
\caption{Born process (here DVCS), initial (ISR) and final (FSR) state radiation, and combinations of them. Four-momenta of electrons ($k$, $k'$) and those relevant for the Born process ($\hat k$, $\hat k'$) are denoted by red letters. Corresponding plots for the BH sub-process are not shown.}
\label{fig:rc}
\end{figure}

Unfolding radiative corrections is needed for a robust determination of the cross-section for the process of interest from experimental data. If emitted photons are soft or nearly collinear, they can escape the detection system and their contribution has to be counted as a radiative correction. In contrast, wide-angle photons from the BH sub-process contribute to the signal. 
In the following, we show how additional photon radiation from incoming and outgoing electrons affect the kinematics of observables, especially the distribution of the $y$ variable. For this purpose we use a MC sample generated with EpIC, see Sect.~\ref{subsec:epic}, where radiative corrections are simulated based on the collinear approximation~\cite{Kripfganz:1990vm}. In our study we focus on the dominant contributions of initial (ISR) and final (FSR) state radiation from the beam and the scattered electron, respectively, and their interference. We include also second-order corrections with the emission of two photons, see Fig.~\ref{fig:rc}. 


As mentioned above, to assess the effects of radiative corrections, we use the collinear approximation, wherein the transverse component of radiated photons is neglected, and the radiated photons are confined to move collinear to the direction of the electron they are emitted from. In this approximation, two essential parameters that represent the energy of the incoming and outgoing electron are denoted by $z_1$ and $z_3$, respectively, and are given by
\begin{equation}
    z_1 = \frac{E_e - E_\gamma}{E_e}\,, \qquad z_3 = \frac{E_{e^\prime}}{E_{e^\prime} + E_{\gamma^\prime}} \,,
\end{equation}
where $E_e$ and $E_{e^\prime}$ are the energies of incoming and outgoing electrons, respectively, while $E_\gamma$ 
 and $E_{\gamma\prime}$ are the corresponding energies of the radiated photons. ``True'' and ``observed'' kinematic variables are related in the following way   
\begin{equation}
   \hat{x} = \frac{z_1 x y}{z_1 z_3 + y - 1} \,,~ \hat{y} = \frac{z_1 z_3 + y - 1}{z_1 z_3} \,,
\end{equation}
and the kinematic limits enforce that 
\begin{equation}
    z_1^\text{min} = \frac{1-y}{1-x y}, \qquad z_3^\text{min} = 1-y(1-x)\,.
\end{equation}
With the inclusion of radiative corrections, one can express the DVCS cross section as follows
\begin{align}
    \frac{\dd^5\sigma}{\dd x \, \dd Q^2 \, \dd t \, \dd \phi \, \dd \phi_S} = \int_{z_1^\text{min}}^1 \dd z_1 z_1 D(z_1) \int_{z_3^\text{min}}^1 \frac{\dd z_3}{z_3^2} \overline{D}(z_3)\times \frac{y}{\hat{y}}\,  \frac{\dd^5\hat{\sigma}_\text{Born}}{\dd \hat{x} \, \dd \hat{Q}^2 \, \dd t \, \dd \, \phi \, \dd \phi_S}\,. 
\end{align}
Here, one constructs the DVCS cross section $\dd^5\sigma$ from the Born cross section $\dd^5\hat{\sigma}_\text{Born}$ which is calculated using the ``true'' kinematics. This distinction is crucial because at the observed level, the effects of soft photons are not accounted for, and hence, the measured kinematic variables are different from the true kinematic variables that enter at the Born level. 
The effects of the undetected initial and final state photons in the DVCS cross section are encoded in the radiator functions, $D(z_1)$ and $\overline{D}(z_3)$, respectively. They can be expressed order-by-order in terms of the fine structure constant $\alpha$ as $D(z) = \sum\limits_{n=0} D^{(n)}(z)$ for both $D(z_1)$ and $\overline{D}(z_3)$, where at least the first three terms are identical for both. 
Explicit expressions have been extracted from Ref.~\cite{Kripfganz:1990vm} and can be written in the following simple form: 
%
%
\begin{align} 
    D^{(0)}(z) & = \delta(1-z)\,, \label{RCs-expansionparams-0} \\
    D^{(1)}(z) & = \delta(1-z)\Bigg[\frac{\alpha}{2\pi}L\Bigg(2\ln\,\epsilon + \frac{3}{2}\Bigg)\Bigg] + \Theta(1-\epsilon-z)\frac{\alpha}{2\pi}L\frac{1+z^2}{1-z}\,, \label{RCs-expansionparams-1} \\
    D^{(2)}(z) &= \delta(1-z)\bigg(\frac{\alpha}{2\pi}\bigg)^2\,\frac{L^2}{2}\Bigg[\frac{13}{4} + 6\,\text{ln}\epsilon+4\,\text{ln}^2\epsilon+2\,z^2_\text{min}\text{ln}(z_\text{min})-z^2_\text{min}\Bigg] \nonumber \\
    & \qquad + \Theta(1-\epsilon-z)\bigg(\frac{\alpha}{2\pi}\bigg)^2\,\frac{L^2}{2}\Bigg[2\frac{1+z^2}{1-z} \bigg(2\,\text{ln}(1-z) -\text{ln}(z)+\frac{3}{2}\bigg) + (1+z)\text{ln}(z)-2(1-z)\Bigg] \label{RCs-expansionparams-2}
\end{align}
with $L = \ln(Q^2/m_l^2)$ and $\Theta$ denoting the Heaviside function. The parameter $\epsilon$ controls the energy of the generated photons. It sets a lower limit on the energy of generated photons, where $E_\gamma$ and $E_{\gamma^\prime}$ cannot be less than $\epsilon E_e$ and $\epsilon E_{e^\prime}$, respectively. A larger value of $\epsilon$ results in generating higher energy photons and the number of generated events with a radiated photon decreases. These patterns are reflected in the radiator function through two components: the Dirac delta term, which accounts for the virtual corrections and the contribution of soft photons below the threshold determined by the $\epsilon$ parameter, and the $\Theta$-function term, which accounts for events with generated photons with sizable energies. For the sake of this study, we use Eqs.~\eqref{RCs-expansionparams-0}-\eqref{RCs-expansionparams-2} to generate events only up to order $\alpha^2$, as shown in Fig.~\ref{fig:rc}.

\begin{figure}[!ht]
\centering
\includegraphics[width=\plotWidthThree]{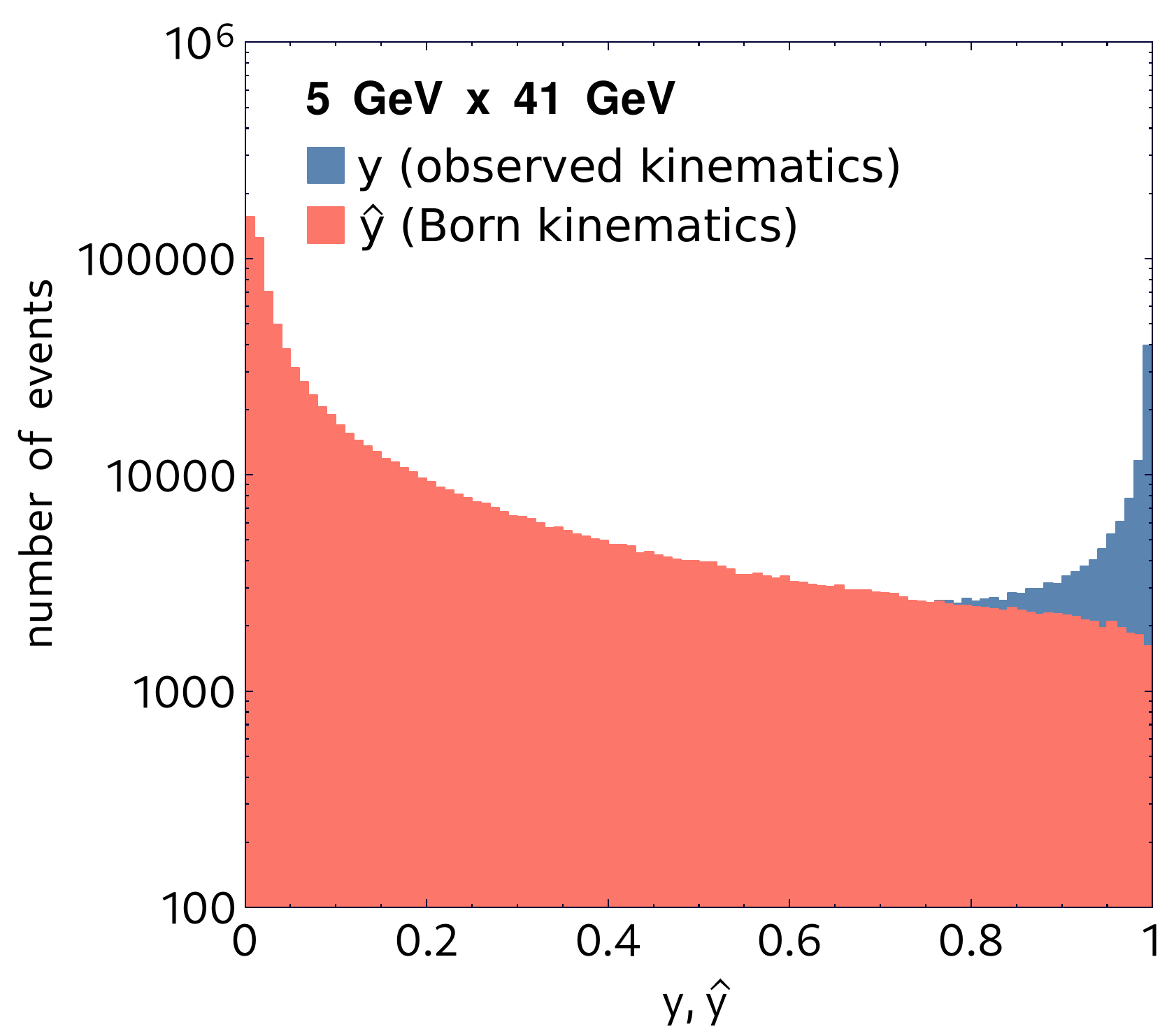}
\includegraphics[width=\plotWidthThree]{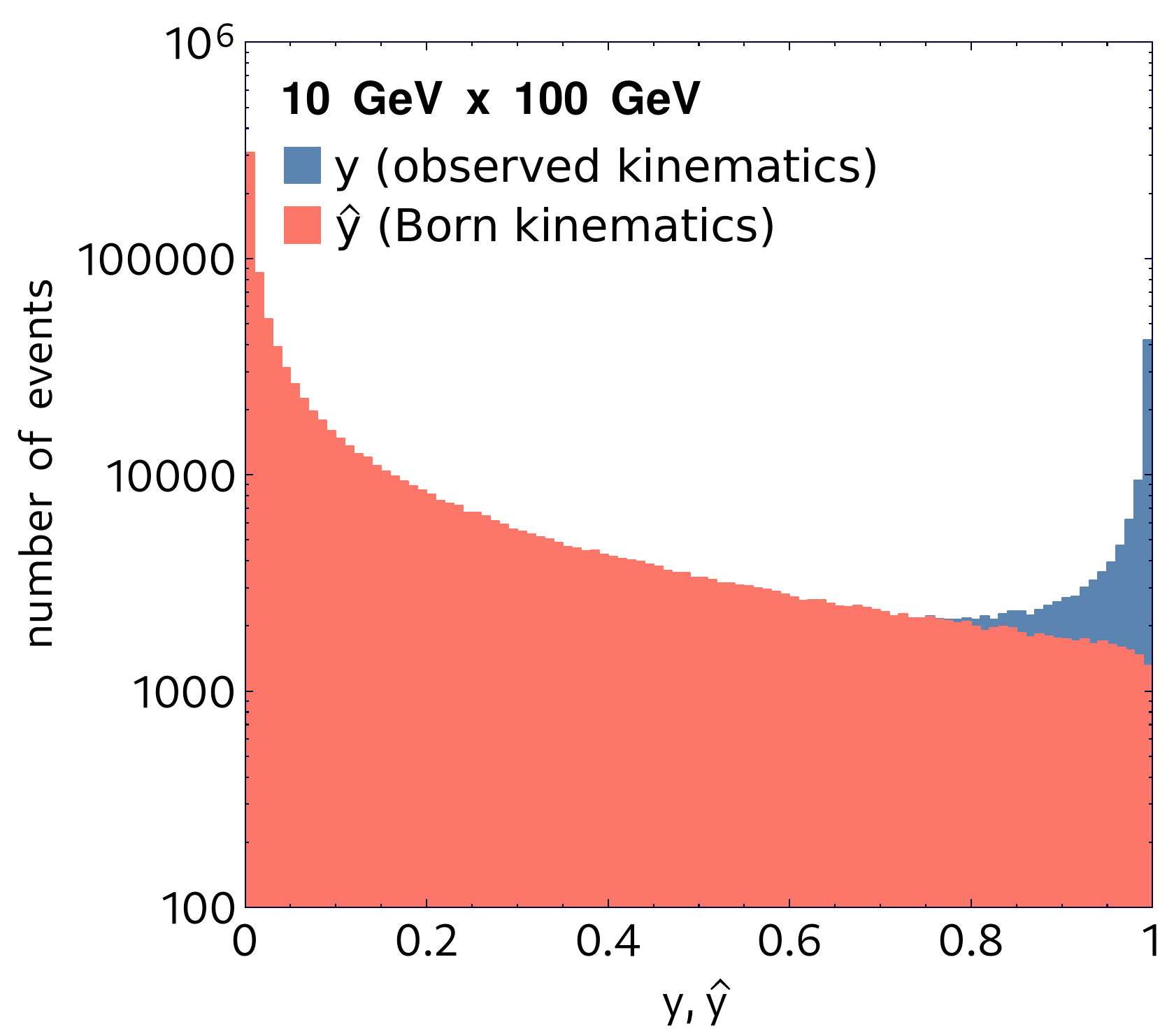}
\includegraphics[width=\plotWidthThree]{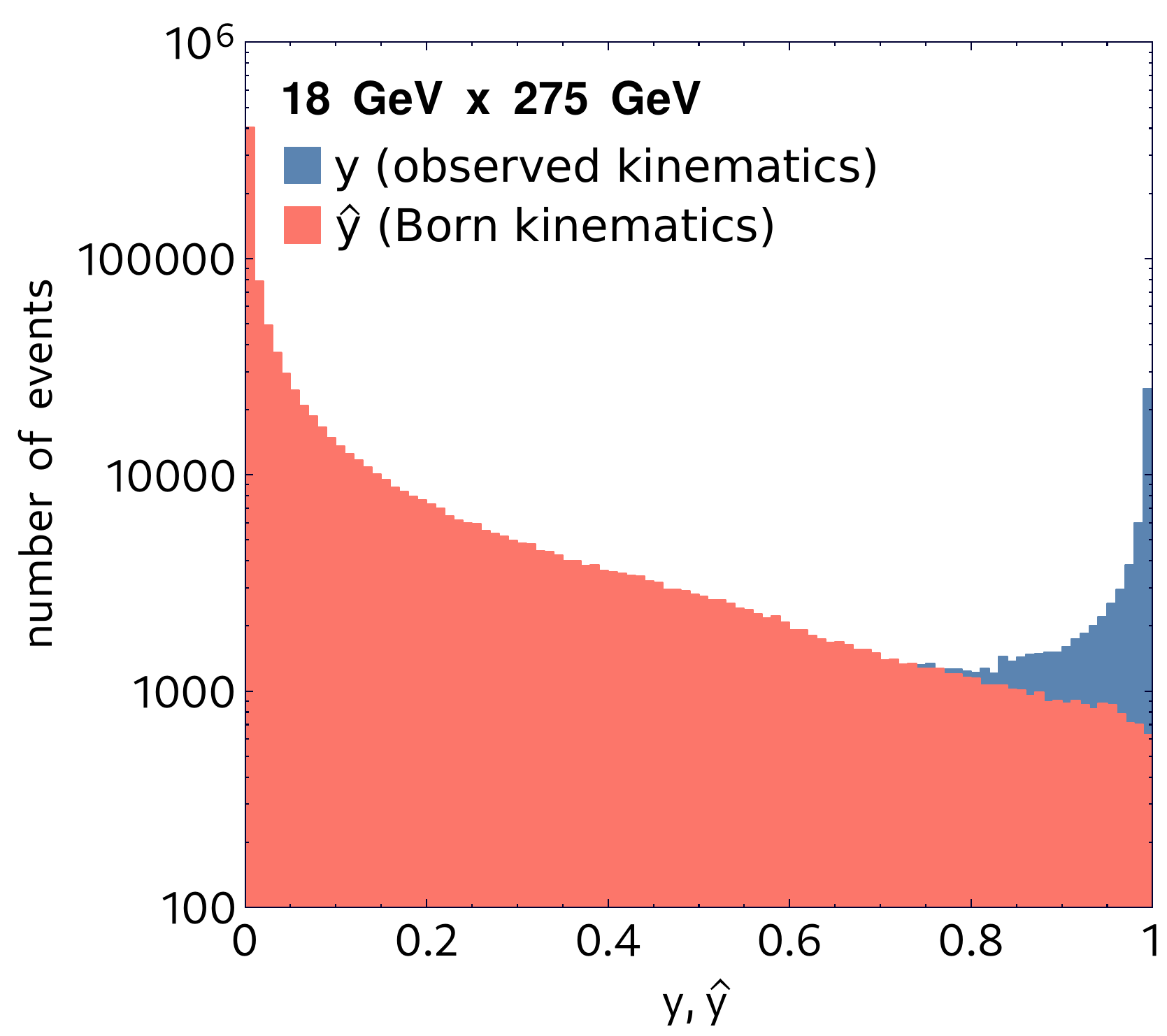}
\caption{Blue: distributions of $y$ evaluated from the observed momenta $k$ and $k'$, see Fig.~\ref{fig:rc}, for $10^6$ MC events generated for the DVCS sub-process only and various beam energy configurations indicated in the labels. Red: corresponding distributions of $\hat y$ evaluated from the momenta $\hat k$ and $\hat k'$ relevant for the Born process.}
\label{fig:rc_y}
\end{figure}

In Fig.~\ref{fig:rc_y}, we show the effect of radiative corrections on the distributions of the inelasticity variable, for all beam energy configurations considered in this study. These distributions are made for the DVCS sub-process only, without including detector effects and without kinematic cuts on $y$ and $\xB$, as specified in Sect.~\ref{subsec:epic}. For a given beam energy configuration, we show two distributions: a) the distribution of $y$ evaluated from the measured four-momenta $k$ and $k'$; b) the distribution of $\hat y$ evaluated from the four-momenta $\hat k$ and $\hat k'$ relevant for the Born process.

From Fig.~\ref{fig:rc_y} one can see that radiative corrections mostly affect the edges of the $y$-spectra, which presents a familiar pattern (see e.g. Fig.~4 of Ref.~\cite{Kripfganz:1990vm}). Since in our analysis of the DVCS signal, the region $y \sim 0$ is suppressed due to large smearing effects, and $y \sim 1$ due to a large contribution of BH, see Sect.~\ref{subsec:epic}, the effect of radiative corrections is already suppressed. In particular, we observe a negligible shift of the mean kinematic variables $\langle \xB \rangle$ and $\langle Q^2 \rangle$ obtained in bins used in the extraction of observables discussed in Sects.~\ref{subsec:tomography} and~\ref{subsec:asymmetries}. The relative magnitude of this shift is at the order of $1\%$.

We note that our study of radiative corrections is based on the collinear approximation. Corrections from non-collinear radiation to the cross section are expected to be in the order of a percent only and will be integrated in the future.


\subsection{Exclusive neutral pion production background}
\label{subsec:pi0}

Exclusive $\pi^0$ production may become a significant source of background to DVCS, as reported, for instance, by the COMPASS~\cite{COMPASS:2018pup} and CLAS~\cite{CLAS:2015uuo} experiments. A possible misidentification can occur when the angle between photons from $\pi^0$ decay is small enough to seemingly create a single electromagnetic cascade, or if one of these photons is lost, e.g.\ due to the limited acceptance of the detector. 

To check this effect in the current analysis, we generate a sample of $\pi^0$ events with the EpIC MC generator. The kinematic conditions and the simulation of the experimental setup are the same as for the DVCS sample, see Sect.~\ref{sec:simulations}. The cross-section used in the simulation has been evaluated according to Ref.~\cite{Goloskokov:2011rd}, and we used GPDs tuned to the COMPASS measurement of exclusive $\pi^0$ production~\cite{COMPASS:2019fea}. We consider the use of this tuning important, as currently COMPASS data provide the only rough guide to the low-$\xB$ kinematic domain. 
It is important to note, that recently the analysis~\cite{Goloskokov:2011rd} has been revisited and improved by including, in contrast to the Wandzura-Wilczek approximation, the complete twist-3 pion contribution~\cite{Duplancic:2023xrt}. The inclusion of these contributions is postponed for future work due to the drastic increase in code complexity and only slight changes in numerical results ($\sim\, \mathrm{few}\, \%$) for the cross-section. 

In \cite{Duplancic:2023xrt} it was also noted that twist-2 next-to-leading order QCD corrections are sizable at low-$\xB$, and therefore are expected to be important for a proper interpretation of future EIC data.

We consider a $\pi^0$ event misidentified as a DVCS one when both the scattered electron and proton are reconstructed and when one of the two following conditions occurs:
\begin{itemize}
    \item if both photons from the $\pi^0$ decay are within the fiducial acceptance of ePIC ECALs , and the separation between them at the calorimeter face is smaller than twice the cell size for the EMCAL (cell size refers to, for instance, the size of an EMCAL crystal or tower); 
    \item if one of the photons is lost, and its energy is smaller than three times the energy resolution associated with the second (detected) photon, see Table~\ref{tab:EMCAL_resolutions}. If the energy of the lost photon was higher than that, we assume such an event would not pass the exclusivity cut. 
\end{itemize}
It should be noted that the latter condition only accounts for $\ll1\%$ of the events. With these conditions implemented, we estimate that the contribution of the $\pi^0$ background to the DVCS sample is of the order of $\lesssim1\%$ for the $5~\GeV \times 41~\GeV$ beam energy configuration, $\lesssim0.5\%$ for the $10~\GeV \times 100~\GeV$ beam energy configuration, and $ < 0.05\%$ for the $18~\GeV \times 275~\GeV$ beam energy configuration. The different levels of contamination with energy are due to the kinematics of the $\pi^{0}$ shifting from the acceptance of the least-precise to the most-precise (in terms of spatial resolution, and energy resolution) electromagnetic calorimetry as one goes from lower to higher beam energy configurations, respectively. It should be noted that these contamination percentages reflect the ``simplest'' approach to reconstructing a $\pi^{0}$ with ePIC calorimeters, and represent an upper-bound for the level of contamination. With modern machine learning based clustering and reconstructing algorithms, the contamination can easily be $\ll 1\%$ for all beam energies, but an updated study of these methods within ePIC needs to be performed. The reasons for such a low contribution are the excellent energy and spatial resolution of the ECALs in ePIC, and the small cross section of exclusive $\pi^0$ production compared to DVCS at small $\xB$, as  the former is not sensitive to the sea contributions~\cite{Goloskokov:2011rd}. We note that the main uncertainty in our study is the knowledge of the exclusive $\pi^0$ cross-section in the considered kinematic region, which, in particular for the highest beam energies, covers a region never mapped by experimental data for this process before.

\subsection{Nucleon tomography}
\label{subsec:tomography}

In this section, we present the procedure for a direct extraction of tomography information from $t$-distributions. ``Direct'' here means that the deconvolution of GPDs from DVCS amplitudes is avoided, with the focus instead on the cross section for this process. The presentation of the extraction procedure is followed by estimates of the precision expected to be achieved with the ePIC experiment.

The procedure of the extraction, which has been utilized in the past by experiments such as H1 \cite{H1:2005gdw, H1:2009wnw}, ZEUS \cite{ZEUS:2008hcd}, and COMPASS~\cite{COMPASS:2018pup}, relies on two assumptions: \emph{i}) the dominance of the imaginary part of the CFF $\mathcal{H}$, allowing to neglect its real part and other CFFs; \emph{ii}) constant skewness, i.e. $H^q(x,x,t)/H^q(x,0,t) = \mathit{const}$ at small $\xB$. With these assumptions, at fixed $\xB$ and $Q^2$, the differential cross section for the $\gamma^* p \to \gamma p$ sub-process can be related to $H^{q(+)}(x, 0, t) = H^{q}(x, 0, t) - H^{q}(-x, 0, t)$ in the following way:
\begin{equation}
\frac{\dd\sigma^{\gamma^* p \to \gamma p}}{\dd t} \propto 
\left(\im\mathcal{H}(\xi, t)\right)^2 \propto
\left(\sum_{q} e_{q}^{2} H^{q(+)}(\xi, \xi, t)\right)^2 \propto 
\left(\sum_{q} e_{q}^{2} H^{q(+)}(\xi, 0, t)\right)^2 \,,
\label{eq:tomography:assumption}
\end{equation}
Here, LO/LT relations between the DVCS cross section and GPDs are used explicitly, enabling for expression of the skewness as $\xi = \xB/(2-\xB)$ and the scale of GPDs (suppressed for brevity) as $\mu^2 = Q^2$ (see Sect.~\ref{sec:theory}). Since the extraction is done in narrow bins of $Q^2$, we do not need an explicit assumption for the evolution of the GPDs with the scale. Keeping in mind Eqs.~\eqref{eq:nucleon_tomography} and~\eqref{eq:tomography:assumption}, the Fourier transform of the square root of ${\dd\sigma^{\gamma^* p \to \gamma p}}/{\dd t}$ cross-section, up to the normalization, gives the spatial distribution of a quark mixture:
\begin{equation}
 \sum_{q} \int \frac{\dd^2 \vec{\Delta}_\perp}{(2\pi)^2} e^{-i\vec{b}_\perp\cdot\vec{\Delta}_\perp} e_{q}^{2} H^{q(+)}(\xi, 0, t=-\vec{\Delta}_\perp^2) =
q^{\mathrm{DVCS}}(\xi,\vec{b}_\perp) = \sum_{q} e_{q}^{2} q^{(+)}(\xi,\vec{b}_\perp)\,,
\label{eq:tomography:assumption_2}
\end{equation}
where $q^{(+)}(\xi,\vec{b}_\perp) = q(\xi,\vec{b}_\perp) + \bar{q}(\xi,\vec{b}_\perp)$ accounts for both quarks and anti-quarks. The normalization is lost as indicated by the proportionality signs in Eq.~\eqref{eq:tomography:assumption}, and, strictly speaking, it cannot be recovered without involving modeling of GPDs. We can, however, fix the normalization at a given $\xi$ with the help of 1D PDFs, since, after integrating $q^{(+)}(\xi,\vec{b}_\perp)$ over the impact-parameter space we know how many partons we should get,
\begin{equation}
\int \dd^2 \vec{b}_\perp\, q^{\mathrm{DVCS}}(\xi,\vec{b}_\perp) = \sum_{q} e_{q}^{2} q^{(+)}(\xi)\,.
\end{equation}
This, of course, requires choosing a specific parametrization of 1D PDFs to represent $q^{(+)}(\xi)$.

We note that the Fourier transform in Eq.~\eqref{eq:tomography:assumption_2} requires the knowledge of $H^{q(+)}(\xi, 0,t)$ in the full domain of $t$, while the DVCS measurement is limited by $t_0$ and $t_1$ values (sometimes denoted as $t_{\min}$ and $t_{\max}$, respectively)~\cite{Zyla:2020zbs}, and usually also by additional kinematic cuts imposed by the experimental apparatus. As a consequence, the extraction of tomography information relies on the parametrization of $\dd\sigma^{\gamma^* p \to \gamma p} / \dd t$ in terms of a specific ansatz, allowing for an extrapolation to unmeasured domains of $t$. The choice of this ansatz introduces a model uncertainty, which, however, can be minimized with the help of non-parametric methods that do not rely on predefined functional forms, but instead adapt to the structure of the data used. Additionally, the extraction of tomography information may be influenced by higher-twist contributions, which scale as $|t|/Q^2$~\cite{Braun:2014sta}, further limiting the usable range of $t$. The impact of higher-twist contributions in the kinematic domain covered by the EIC is expected to be much less severe compared to fixed-target experiments. Still, it requires detailed studies, which are beyond the scope of this analysis, as we primarily focus on detector performance.

\begin{figure}[!ht]
\centering
\includegraphics[width=0.5\plotWidthOne]{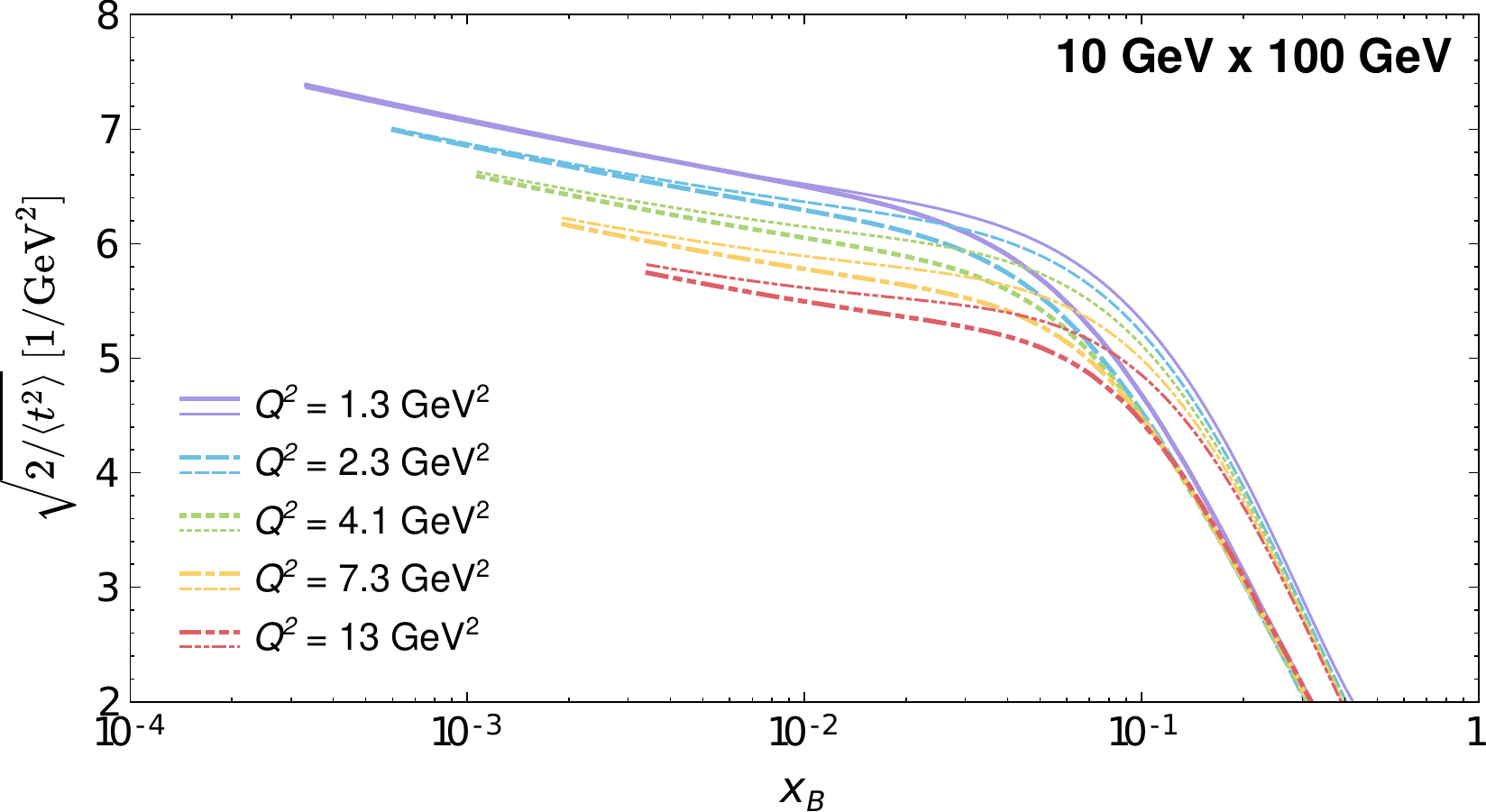}
\caption{The quantity $\sqrt{2/\langle t^2 \rangle}$ evaluated as a function of $\xB$ for $Q^2$ values corresponding to kinematic bins used in this analysis. Here, $\langle t^2 \rangle$ is the variance of the  distribution of the $t$ variable, evaluated either from the $\gamma^* p \to \gamma p$ cross section (thick lines, see Eq.~\eqref{eq:aprox_A}) or from the GPD $H$ (thin lines, see Eq.~\eqref{eq:aprox_B}). Both give the same results only if the assumption \eqref{eq:tomography:assumption} is true.}
\label{fig:nt_approx}
\end{figure}

The dominance of the $\im\mathcal{H}$ and constant skewness assumptions are expected to hold at small $\xB$. Since ePIC will probe both small and moderate $\xB$ with unprecedented precision, it seems appropriate to check in the current study the validity of these assumptions. We present our findings in Fig.~\ref{fig:nt_approx}, which shows the quantity $\sqrt{2/\langle t^2 \rangle}$ as a function of $\xB$ for selected values of $Q^2$, corresponding to the kinematic bins introduced later in this section. Here, $\langle t^2 \rangle$ is the normalized second moment (variance) of the distribution of the variable $t$, evaluated either from the $\gamma^* p \to \gamma p$ cross section,
\begin{equation}
    \langle t^2 \rangle_{\mathrm{DVCS}} = \frac{
        \int_{t_0}^{t_1} \dd t\, t^2\, \dd \sigma^{\gamma^* p \to \gamma p} / \dd t
        }{
        \int_{t_0}^{t_1} \dd t\, \dd \sigma^{\gamma^* p \to \gamma p} / \dd t
        }\,,
        \label{eq:aprox_A}
\end{equation}
or directly from the GPD $H$,
\begin{equation}
    \langle t^2 \rangle_{\mathrm{GPD}} = \frac{
        \int_{t_0}^{t_1} \dd t\, t^2\, \left(\sum_{q} e_{q}^{2} H^{q(+)}(\xi, 0, t)\right)^2
        }{
        \int_{t_0}^{t_1} \dd t\, \left(\sum_{q} e_{q}^{2} H^{q(+)}(\xi, 0, t)\right)^2
        }\,,
        \label{eq:aprox_B}
\end{equation}
where $t_0$ and $t_1$ are minimal and maximum values of $|t|$ allowed by the kinematics. We note that:
\begin{equation}
\frac{
    \int \dd t\, t^2\, \exp (2B t)
    }{
    \int \dd t\,  \exp (2B t)
    } = \frac{1}{2B^2}\,,
\end{equation}
so if, for instance, the underlying GPD is described by a single exponent, $\sum_{q} e_{q}^{2} H^{q(+)}(\xi, 0, t) \propto \exp(Bt)$, the plotted quantity provides a good approximation of its slope, $\sqrt{2/\langle t^2 \rangle_{\mathrm{GPD}}} \approx B$. However, if the underlying GPD is characterized by a different dependence, for instance a dipole, both $\langle t^2 \rangle_{\mathrm{DVCS}}$ and  $\langle t^2 \rangle_{\mathrm{GPD}}$ remain well-defined quantities which can be used to analyze the tomographic properties of the nucleon.

The quantities $\langle t^2 \rangle_{\mathrm{DVCS}}$ and $\langle t^2 \rangle_{\mathrm{GPD}}$ give the same results only if the assumption \eqref{eq:tomography:assumption} is true. From Fig.~\ref{fig:nt_approx} we see that the assumption is well justified in the kinematic domain covered by ePIC running with the $10~\GeV \times 100~\GeV$ beam energy configuration, but is violated for higher values of $\xB$. In this region, a full fit including $\re\mathcal{H}$ and other CFFs, and a de-skewness procedure to recover $H^{q(+)}(\xi, 0, t)$ from $H^{q(+)}(\xi, \xi, t)$ may be required, see for instance Ref.~\cite{Dupre:2017hfs}.

To accurately extract tomography information, one requires the distribution of events, $N(t_{i})$, obtained within a narrow bin of $(\xB, Q^2)$ that faithfully reproduces the shape of the cross section for the $\gamma^* p \to \gamma p$ sub-process.  Here, $t_i$ denotes a bin in $t$. Such distributions can be obtained from the measured one, $N^{\mathrm{exp}}(t_{i})$, by unfolding detector effects and subtracting the BH contribution (we note that the interference between BH and DVCS can be neglected, as the integration over $\phi$ only leaves a constant term predominantly sensitive to the real part of the CFF $\mathcal{H}$~\cite{Belitsky:2001ns}):
\begin{equation}
N(t_{i}) = 
N^{\mathrm{exp}}(t_{i}) / a(t_{i})- N^{\mathrm{gen, BH}}(t_{i}) \,.
\end{equation}
The acceptance factor, $a(t_{i})$, is estimated from a MC sample of events:
\begin{equation}
a(t_{i}) = \frac{N^{\mathrm{rec, ALL}}(t_{i})}{N^{\mathrm{gen, ALL}}(t_{i})} \,,
\end{equation}
where $N^{\mathrm{rec, ALL}}(t_{i})$ and $N^{\mathrm{\mathrm{gen, ALL}}}(t_{i})$ denote samples of reconstructed (``$\mathrm{rec}$'') and generated (``$\mathrm{gen}$'') events, respectively, for the mixture of BH and DVCS sub-processes (``$\mathrm{ALL}$''). The sample of MC events used for BH subtraction, $N^{\mathrm{gen, BH}}(t_{i})$, must be normalised according to the integrated luminosity corresponding to the $N^{\mathrm{exp}}(t_{i})$-distribution. Formally, $N(t_{i})$ should be corrected for the photon flux~\cite{Hand:1963bb}. However, since it does not modify the shape of the $t$-distribution (assuming that kinematic bins are sufficiently narrow), the flux can be considered as a multiplicative factor, which is not relevant for the direct extraction of tomography information. 

One can expect a twofold effect from radiative corrections. On the one hand, radiative corrections may change the shape of $t$-distributions, directly affecting the extraction of tomography information. Since $t$ is evaluated from the four-momenta of hadrons, the effect of radiative corrections is expected to be small. A precise estimation would require an improved description of radiative corrections than that used in this study (see Sect.~\ref{subsec:rc}). On the other hand, radiative corrections affect the estimation of the mean values of kinematic variables attributed to the extraction of tomography information, namely the values of $\langle \xB \rangle$ and $\langle Q^2 \rangle$. Within the description of radiative corrections used in this study, we found this effect to be negligible. 

Our impact study is based on a MC sample obtained for $\mathcal{L}=10~\fb^{-1}$. The analysis is done in 2-dimensional bins of $(\xB, Q^2)$  defined by the following limits: 
$\xB: \{ 
      0.0001, 0.00016,\allowbreak
      0.00025, 0.00040,\allowbreak
      0.00063, 0.0010,\allowbreak
      0.0016, 0.0025,\allowbreak
      0.0040, 0.0063,\allowbreak
      0.010, 0.016,\allowbreak
      0.025, 0.040,\allowbreak
      0.063, 0.10,\allowbreak
      0.16, 0.25,\allowbreak
      0.40, 0.7 \}$, 
$Q^2/\GeV^2: \{
      1.0, 1.78,\allowbreak
      3.16, 5.62,\allowbreak
      10, 18,\allowbreak
      32, 56,\allowbreak
      100\}$. 
Bins with fewer than $2000$ reconstructed events are discarded in the analysis. 

In order to extract tomographic pictures, i.e.\ a visual representation of $q^{\mathrm{DVCS}}(\xi,\vec{b}_\perp)$, we parameterize the corrected distribution of events, $N(t_{i})$, with the following ansatz:
\begin{equation}
f(t) = \left(\sum_{i=0}^{N_{\mathrm{max}}}
A_{i}\exp(B_{i}t)\right)^2\,,
\label{eq:exp_expansion}
\end{equation}
where $A_i > 0$ and $B_{i} > 0$, and where the square of the sum corresponds to the square of GPDs in Eq.~\eqref{eq:tomography:assumption}. The flexibility of the model is controlled by $N_{\mathrm{max}}$, which in this analysis is set to $5$. The benefit of using a sum of exponents is threefold: i) its Fourier transform is straightforward, ii) the outcome of this transform is always positive, and iii) the ansatz can be used to approximate other popular functions used in the context of nucleon tomography, such as a dipole. The latter can be demonstrated using the Laplace transform: 
\begin{equation}
\left(1-\frac{t}{m}\right)^{-n} = \int_0^{\infty}dt'\,
\frac{\exp(-m t') (m t')^n}{t' \Gamma(n)}\exp(t'\,t)\,
\end{equation}
which, despite the infinite integration limit, converges quickly and can be efficiently approximated by the sum in Eq.~\eqref{eq:exp_expansion}.

To propagate statistical uncertainties, we employ the replication method. The fitting of \eqref{eq:exp_expansion} to $N(t_{i})$ is repeated multiple times ($100$ in this analysis), with each repetition involving random alterations to the fitted distribution. To generate a single altered distribution, we randomly generate a new number of events in each bin of $N^{\mathrm{exp}}(t_{i})$ from the Poisson distribution. The result of fitting to a single altered distribution is referred to as a replica. We utilize a collection of replicas to estimate the uncertainty at a given point, based on their spread. This approach is also employed for quantities derived from \eqref{eq:exp_expansion}, such as the tomographic pictures.  

\begin{figure}[!ht]
\centering
\includegraphics[width=0.7\textwidth]{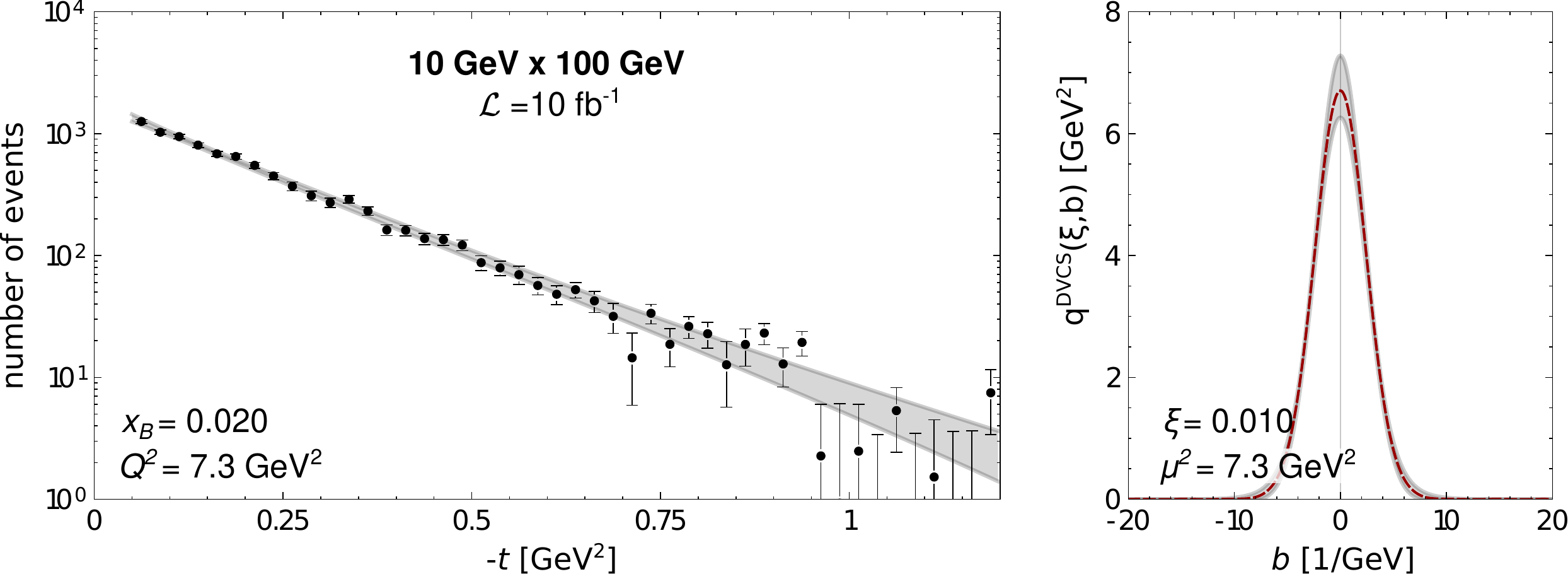}
\caption{Left: Distribution of events corrected for acceptance and after subtraction of the BH contribution as a function of $t$ for $\xB = 0.020$ and $Q^{2} = 7.3~\GeV^2$. The gray band represents the result of the fit described in the text, corresponding to the $95\%$ confidence level. Right: Resulting tomographic picture, with the red dashed curve representing the reference values given by the GK model.}
\label{fig:nt_ab}
\end{figure}

An exemplary result is shown in Fig.~\ref{fig:nt_ab} for a single kinematic bin. The figure presents the distribution of events fitted with ansatz \eqref{eq:exp_expansion}, utilizing the replication prescription for estimating uncertainties. The resulting tomographic picture, i.e., the Fourier transform of the fitted $t$-profile normalized to the PDF (the same as used in the GK model), is also shown. One can notice an increase in uncertainty at $b=0$, where $b=\|\vec{b}_\perp\|$, caused by the poor knowledge of the domain with $|t| \gg 1\GeV^2$. This effect is expected and would not be visible with simple ans{\"a}tze, like a single exponential fit, due to the model bias. This demonstrates the usefulness of flexible ans{\"a}tze like that given by Eq.~\eqref{eq:exp_expansion}, which due to a large number of free parameters can be considered as a good approximation of a non-parametric method. The result agrees with the GK model, which was used in the generation of the Monte Carlo sample, proving the validity of the extraction procedure.

\begin{figure}[!ht]
\centering
\includegraphics[width=0.6\plotWidthOne]{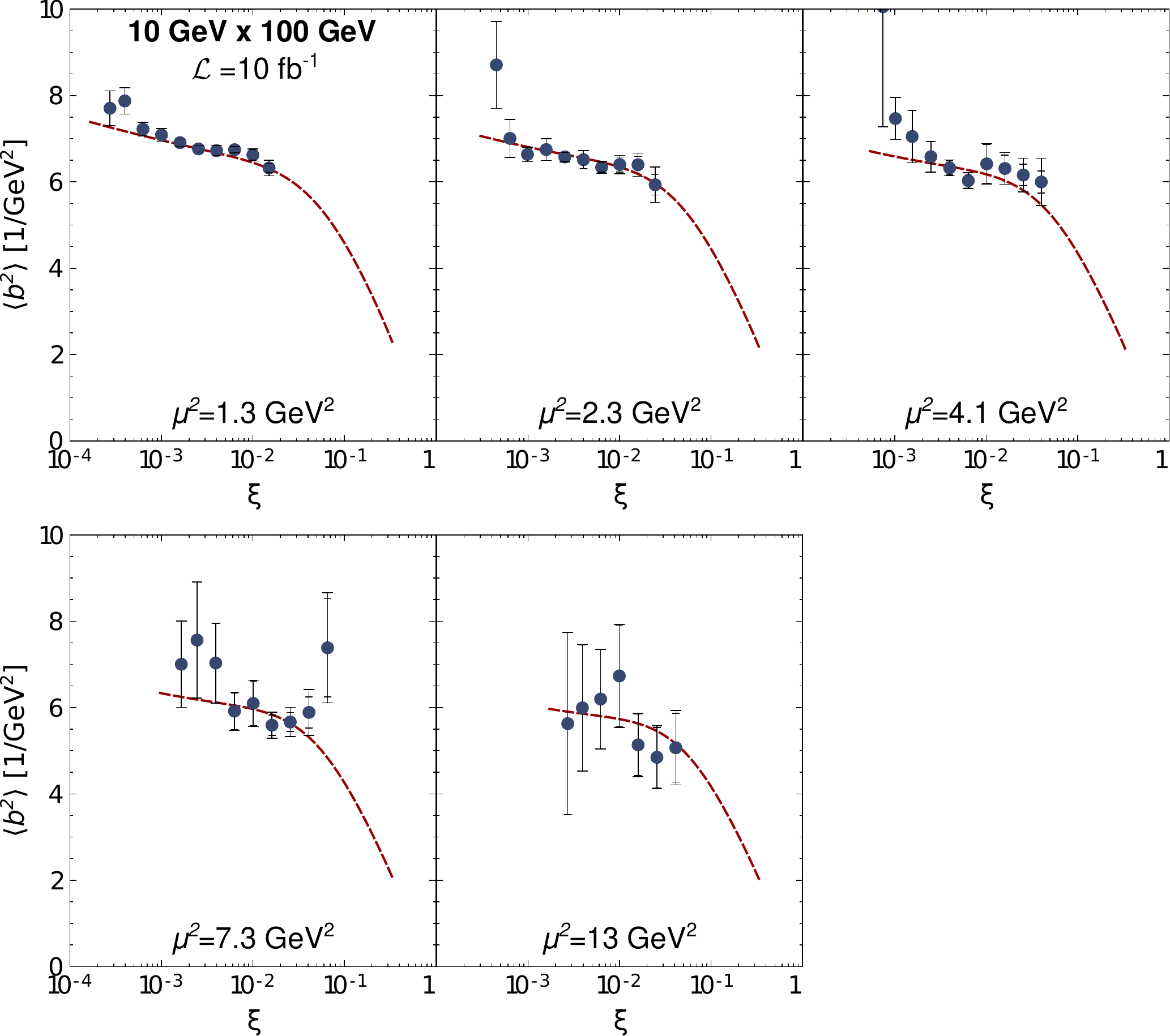}
\caption{Second moments of $q^{\mathrm{DVCS}}(\xi,b)$, see Eq.~\eqref{eq:averageb2}, estimated from distributions of events as a function of $\xi$ for the $Q^2 \equiv \mu^2$ bins used in this analysis. The reference values obtained from the GK model are denoted by the red dashed curves. The inner error bars represent statistical uncertainties, while the outer error bars also account for uncertainties related to the application of the direct extraction method (see text for details).}
\label{fig:nt_sum}
\end{figure}

\begin{sidewaysfigure}[p]
\centering
\includegraphics[width=\textwidth]{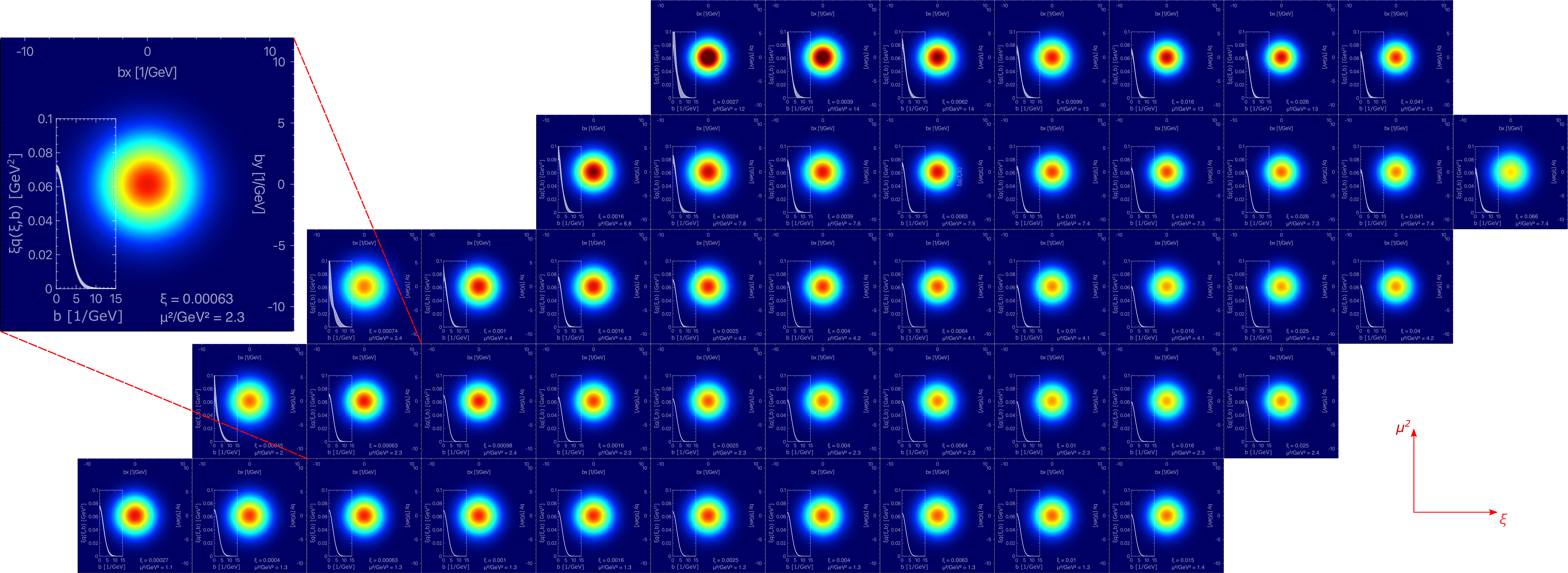}
\caption{2-dimensional tomographic images obtained from EIC pseudo-data for DVCS, corresponding to $\mathcal{L}=10,\fb^{-1}$. Each image represents a single kinematic bin used in this analysis and includes information about the average kinematics and uncertainties of the estimated charge-weighted quark flavor spatial profile. One image is zoomed in for better~readability.}
\label{fig:nt_2d}
\end{sidewaysfigure}

The extracted average transverse sizes of the $q^{\mathrm{DVCS}}(\xi,b)$ distributions,
\begin{equation}
    \langle b^2 \rangle = \frac{\int\dd b\,  b^2\,q^{\mathrm{DVCS}}(\xi,b)}{\int\dd b\, q^{\mathrm{DVCS}}(\xi,b)}\,,
    \label{eq:averageb2}
\end{equation}
are collectively shown in Fig.~\ref{fig:nt_sum} for all kinematic bins used in this study. The results are grouped according to $Q^2$ values and compared with the GK model to verify the correctness of the analysis and extraction procedure once again. On top of systematic uncertainties, this figure also marks uncertainties related to the applicability of the direct extraction method. Specifically, for each bin, we estimate, within the GK model, the difference between $\langle b^2 \rangle$ evaluated from the cross-section and that evaluated directly from the GPD $H$, cf.\ Fig.~\ref{fig:nt_approx}. The estimated uncertainties are tiny for low-$\xi$ bins, while for the highest-$\xi$ bins they do not exceed the statistical uncertainties, demonstrating that the direct extraction method can still be used in that kinematic domain. Additionally, Fig.~\ref{fig:nt_2d} presents 2D profiles, including the estimated uncertainties. Once again, one can see that the distribution of partons becomes narrower as $\xi$ and/or $\mu^2$ grows. While such behavior in $\xi$ is generally expected, that in $\mu^2$ is not constrained by theory and is a consequence of the modeling assumptions in the GK model. The figure stresses the high potential of the EIC for nucleon tomography, with its fine multi-dimensional kinematic binning and the high statistical precision of the expected measurements. This feature of EIC will be used to reveal the detailed dependence of 2D profiles on $\xi$ and $\mu^2$.

\subsection{Extraction of Compton form factors}
\label{subsec:asymmetries}

\begin{figure}[!ht]
\centering
\includegraphics[width=0.7\plotWidthOne]{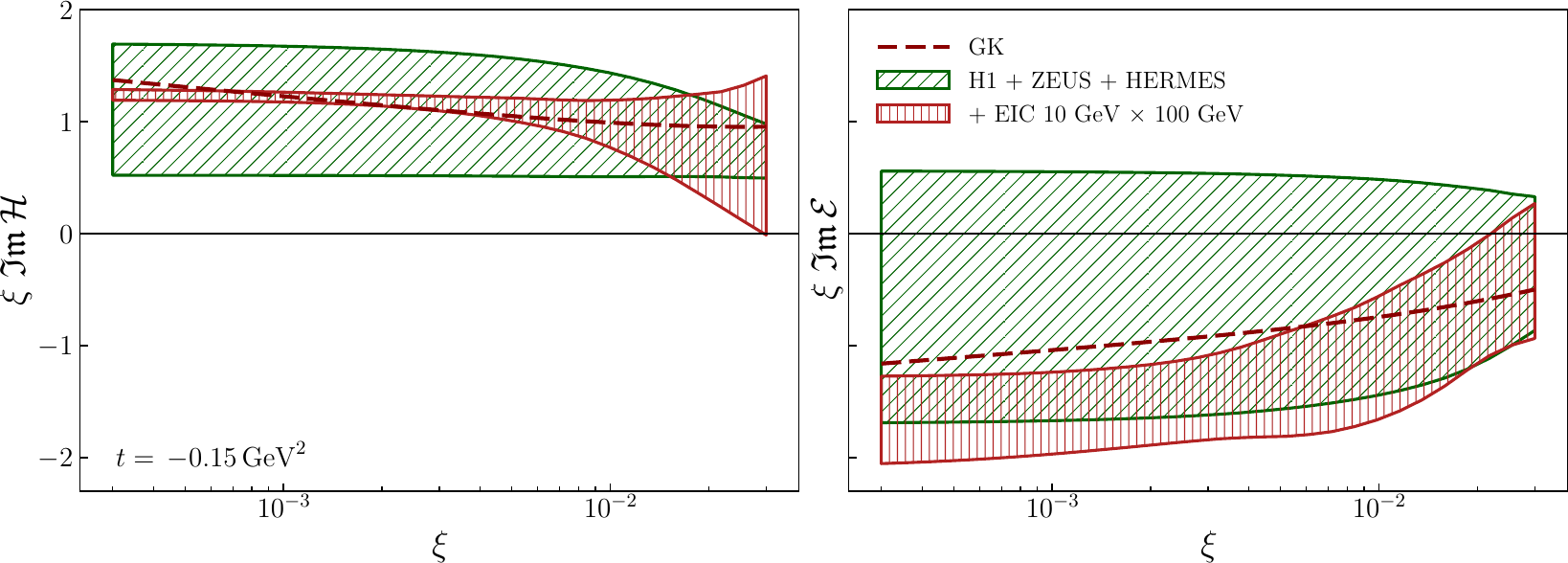}
\includegraphics[width=0.7\plotWidthOne]{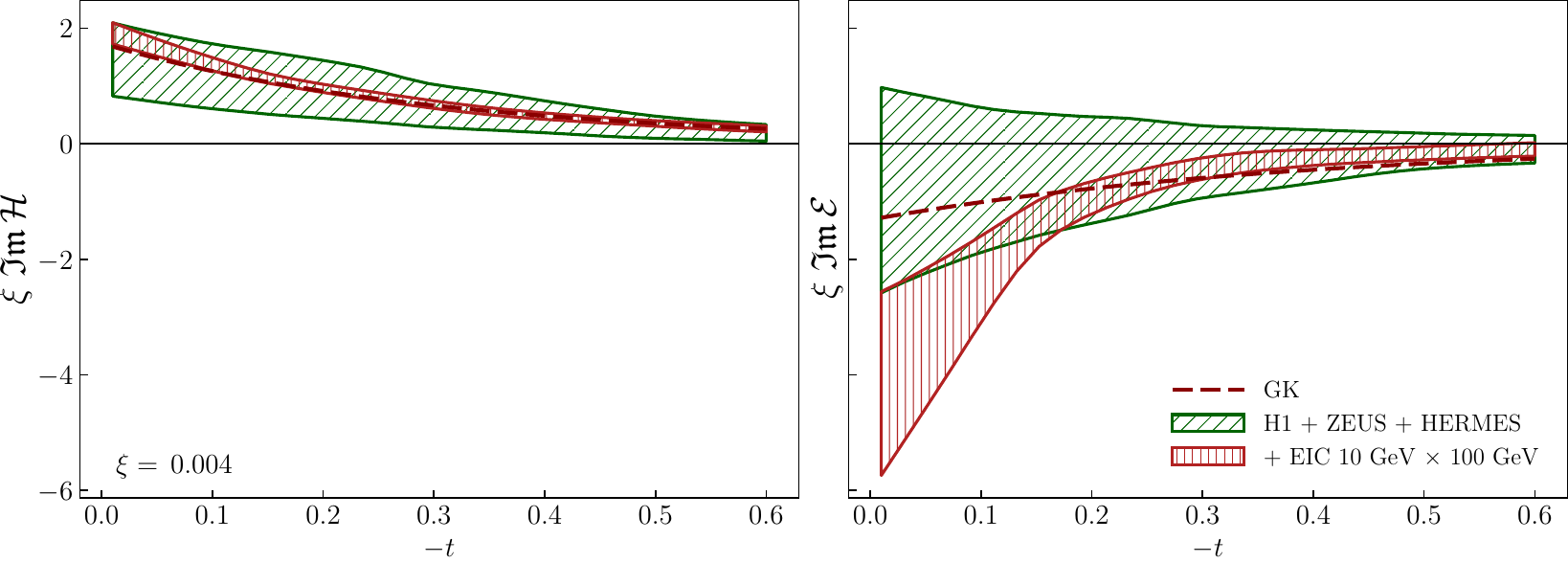}
\caption{CFFs $\im\mathcal{H}$ (left) and $\im\mathcal{E}$ (right) in dependence on $\xi$ (first row)
    and $t$ (second row), as extracted by training an ensemble of neural nets to only old HERA data
(green slanted dashes) and additionally to simulated EIC data (red vertical dashes) at $Q^2 = 4\,\GeV^2$.
GK model values are plotted for comparison (red dashed line).}
\label{fig:nncff}
\end{figure}

In this section, we present the extraction of CFFs from the polarization asymmetry $A_{\mathrm{LU}}$,
\begin{equation}
   A_{\mathrm{LU}}(\phi) = \frac{\dd^4 \sigma^{+}(\phi)-\dd^4 \sigma^{-}(\phi)}{\dd^4 \sigma^{+}(\phi)+\dd^4 \sigma^{-}(\phi)}\,,
   \label{eq:ALU}
\end{equation}
where $\dd^4 \sigma^{+}(\phi)$ and $\dd^4 \sigma^{-}(\phi)$ denote the differential cross-section~\eqref{eq:process_cs} for positively and negatively polarized electron beams, respectively. In Eq.~\eqref{eq:ALU}, only the $\phi$-dependence is shown explicitly. The asymmetry is sensitive to the interference between the BH and DVCS sub-processes, and at EIC kinematics it mostly probes the imaginary parts of the CFFs $\mathcal{H}$ and $\mathcal{E}$~\cite{Belitsky:2001ns} as:
\begin{equation}
\dd^4 \sigma^{+}(\phi)-\dd^4 \sigma^{-}(\phi) \propto \sin\phi \times
\im\left(
F_{1}\mathcal{H} + \frac{\xB}{2-\xB}(F_{1} + F_{2})\widetilde{\mathcal{H}} - \frac{t}{4m^2}F_{2}\mathcal{E}
\right)\,,
\label{eq:imCFFs}
\end{equation}
where $F_{1}$ and $F_{2}$ are the Dirac and Pauli form factors, respectively. The sensitivity on the CFF $\mathcal{E}$ is due to the long lever arm in $t$ available at EIC because of the carefully designed interaction region. 

The analysis is done in 3-dimensional bins of $(\xB, t, Q^2)$  defined by the following limits: 
$\xB: \{ 
      0.0001, 0.00016,\allowbreak
      0.00025, 0.00040,\allowbreak
      0.00063, 0.0010,\allowbreak
      0.0016, 0.0025,\allowbreak
      0.0040, 0.0063,\allowbreak
      0.010, 0.016,\allowbreak
      0.025, 0.040,\allowbreak
      0.063, 0.10,\allowbreak
      0.16, 0.25,\allowbreak
      0.40, 0.7 \}$, 
$|t|/\GeV^2: \{
      0,\allowbreak
      0.05, 0.1,\allowbreak
      0.2, 0.3,\allowbreak
      0.4, 0.5,\allowbreak
      0.6, 0.7,\allowbreak
      0.8, 0.9,\allowbreak
      1.0, 1.1,\allowbreak
      1.2\}$,
$Q^2/\GeV^2: \{
      1.0, 1.78,\allowbreak
      3.16, 5.62,\allowbreak
      10, 18,\allowbreak
      32, 56,\allowbreak
      100\}$.
Bins with fewer than $500$ reconstructed events are discarded from the analysis. In a real experiment, the asymmetry is extracted in a given bin of $(\xB, t, Q^2)$ from two distributions of events, $N^{\mathrm{exp},+}(\phi_i)$ and $N^{\mathrm{exp},-}(\phi_i)$, corresponding to the two polarization states:
\begin{equation}
    A(\phi_i) = \frac{1}{|P_{e}|}\frac{N^{\mathrm{exp},+}(\phi_i)-N^{\mathrm{exp},-}(\phi_i)}{N^{\mathrm{exp},+}(\phi_i)+N^{\mathrm{exp},-}(\phi_i)}\,,
\end{equation}
where $\phi_i$ denotes a bin in $\phi$, while $P_{e}$ is the beam polarisation. 

In our analysis, $N^{\mathrm{exp},+}(\phi_i)$ and $N^{\mathrm{exp},-}(\phi_i)$ are represented by two MC samples $N^{\mathrm{rec,ALL},+}(\phi_i)$ and $N^{\mathrm{rec,ALL},-}(\phi_i)$, respectively. Each of these samples corresponds to $\mathcal{L}=5~\fb^{-1}$ and has been obtained by processing MC events generated with EpIC through the simulation of the detector setup, see Sect.~\ref{sec:simulations}. The assumed degree of beam polarization that affects the uncertainties of the extracted asymmetries is $|P_{e}|=0.8$.

Similarly to the extraction of nucleon tomography information, one may expect two effects from radiative corrections: a modification of the asymmetry distributions, $A(\phi_i)$, and a shift of the values of mean kinematic variables attributed to the measurement in bins (here: $\langle \xB \rangle$, $\langle Q^2 \rangle$ and $\langle t \rangle$). The first effect is expected to be small, and since the radiatiator function is independent of the beam polarization. The second effect, as with the extraction of tomography information, has been found to be negligible.


To extract the CFFs we trained an ensemble of 200 neural networks to model the CFFs as:
\begin{equation}
\im \mathcal{F}(\xi, t, Q^2) = \xi^\alpha \cdot \mathrm{ANN}(\xi, t, Q^2) \,, \qquad 
     \mathcal{F} = (\mathcal{H}, \mathcal{E}, \ldots) \,,
\end{equation}
where the inputs to the neural network are the kinematic values $\xi$, $t$ and $Q^2$, which
are then transformed by three hidden layers with 12, 37 and 17 neurons each, resulting
in the net output $\mathrm{ANN}(\xi, t, Q^2)$. For details of our procedure we refer to
previous publications dedicated to the neural network method for CFF extraction from data
\cite{Kumericki:2011rz,Moutarde:2019tqa}. The meta-parameter (i.e., one whose value is set before the learning process begins) $\alpha$ improves
the convergence of the training procedure and the precision of the fit.
After repeating the procedure for a sequence of
values, we determined that the best results are obtained using $\alpha=-1$, and this
value is then used when training the final ensemble.
This value is expected since the GK model is dominated by
$\im\mathcal{H}$ with a Pomeron-like behavior $\sim \xi^{-1}$ for small $\xi$.

To establish the baseline of the impact of the EIC, we first extracted the CFFs from the data on the DVCS cross-section as measured
by the H1 and ZEUS collaborations \cite{H1:2005gdw,H1:2009wnw,ZEUS:2008hcd}, and asymmetry data from longitudinally polarized electron beams
$A_{\mathrm{LU}}$ and from longitudinally and transversely polarized targets ($A_{\mathrm{UL}}$ and $A_{\mathrm{UT}}$) as measured by HERMES \cite{HERMES:2012gbh,HERMES:2010dsx,HERMES:2008abz}.
These are the observables providing presently
the strongest experimental constraints on the imaginary parts of the CFFs 
$\mathcal{H}$, $\mathcal{E}$ and $\widetilde{\mathcal{H}}$ at the EIC kinematics.

Using the same procedure as above, we performed the CFF extraction by adding simulated $A_{\mathrm{LU}}$ EIC data for the bins
listed above. We started by using only the dominant $\im\mathcal{H}$ CFF (with other CFFs set to zero) and then added first $\im\mathcal{E}$ and finally $\im\widetilde{\mathcal{H}}$. As expected from the $\xB$ suppression in (\ref{eq:imCFFs}), there is no statistically significant effect of $A_{\mathrm{LU}}$ EIC data on the extraction of $\im\widetilde{\mathcal{H}}$.
The two CFFs that \emph{are} constrained by EIC $A_{\mathrm{LU}}$ data,
$\im\mathcal{H}$ and $\im\mathcal{E}$,
are shown in Fig.~\ref{fig:nncff}.  It is clearly seen that the improvement in the uncertainty
of $\im\mathcal{H}$ is excellent, and the constraints on $\im\mathcal{E}$
are significantly increased. The shift in the uncertainty bands after including the simulated
EIC data, which is particularly evident in $\im\mathcal{E}$, is partly due to statistical
fluctuations inherent in this method and partly due
to the HERA datasets not being in perfect statistical agreement with the EIC datasets.
In particular, the GK model used to generate the simulated EIC data undershoots
the HERMES $A_\mathrm{LU}$ by about two standard deviations, creating tension
that ultimately leads to an artificial increase in the uncertainty of $\im\mathcal{H}$
for $\xi > 0.01$.

\section{Conclusions} 
\label{sec:conclusions}

In this study, we realistically assessed the impact of future EIC measurements on the analysis of nucleon tomography and CFFs by taking into account a wide range of factors that could potentially influence the extraction procedure. To achieve this goal, we have utilized state-of-the-art simulation software and the most recent design of the ePIC detector and the EIC interaction region to evaluate the impact of the apparatus on the anticipated measurements.


Despite focusing only on certain observables, our study clearly demonstrates that thanks to the high luminosity, polarized beams and the careful design of the detectors and the interaction region, the EIC will provide unprecedented access to the spatial and spin structure of the proton. The EIC will cover a wide kinematic domain, ranging from low $\xB$, previously probed only by HERA, to intermediate $\xB$, covered by COMPASS, HERMES and future JLab experiments. The long lever arm in $Q^2$ will allow for a detailed study of evolution effects, particularly those that impact nucleon tomography, which remain largely unknown to this day. The measurement of CFFs will provide much-needed constraints on various types of GPDs. Overall, the anticipated, precise measurements in poorly covered kinematic domains will contribute significantly to the understanding of hadronic structure, marking the EIC as the future QCD laboratory.

Our study presents the coverage of the kinematic region by the ePIC detector and the basic kinematic distributions one expects to see after accumulating $10~\fb^{-1}$ of integrated luminosity. A factor that can influence future DVCS measurements is the QED radiation of photons from incoming and outgoing electrons. We show that this radiation is restricted primarily to the lower and higher ranges of the inelasticity variable, $y$, which are anyway discarded in the analysis. Another potential factor we considered in our study is the background coming from misidentified exclusive $\pi^0$ events. The misidentification occurs in situations when the opening angle between the $\pi^0$ decay photons is too small to separate the two electromagnetic showers in the calorimeters, or when one of these photons is lost in the measurement. Our MC analysis indicates that at the lowest considered beam energy configuration, $5~\GeV \times 41~\GeV$, the contribution of the $\pi^0$ background to the DVCS sample does not exceed $\sim1\%$. For higher beam energy configurations, we observe a significant drop in the background by up to an order of magnitude. Our analysis suggests that the small background effect of $\pi^0$ in DVCS can be attributed to the exceptional energy and spatial resolution of the ECALs in ePIC, and the small cross section of exclusive $\pi^0$ production compared to DVCS at small $\xB$.

The analysis for directly extracting nucleon tomography information from data clearly demonstrates the high precision of future measurements performed at ePIC. We benchmark the extraction procedure, particularly highlighting the assumptions one needs to apply, including assessing their validity in the kinematic domain covered by the EIC. The use of flexible ans{\"a}tze close to non-parametric methods allows for reduction of model dependency of future extractions, and enables reduction of biases in the propagation of the uncertainties.  

The presented example of extracting CFFs from pseudo $A_{\mathrm{LU}}$ asymmetry data provides another instance of how precise future EIC measurements will be. This extraction utilizes an artificial neural network technique, which reduces model bias and improves uncertainty estimation. We report a significant expected impact of EIC $A_{\mathrm{LU}}$ asymmetry measurements on our understanding of CFFs $\mathcal{H}$ and $\mathcal{E}$, both of which are particularly important for estimating the total angular momentum of partons via the Ji sum rule. Although we present only a glimpse of what will be possible at the EIC, we clearly demonstrate its potential to become the future QCD laboratory, contributing crucially to our understanding of GPDs and more.
\begin{acknowledgments}
The work of E.~Aschenauer and A.~Jentsch is supported by U.S. Department of Energy under Contract No.
de-sc0012704. H.~Spiesberger has received funding from the European Research Council (ERC) under the European Union’s Horizon Europe research and innovation programme (Grant Agreement No. 101142600) and by the Deutsche Forschungsgemeinschaft (DFG, German Research Foundation) -- Project No. 514321794 (CRC1660: Hadrons and Nuclei as Discovery Tools). This work was supported by 
the Croatian Science Foundation project IP-2019-04-9709,
and by the EU Horizon 2020 research and innovation program, STRONG-2020
project, under grant agreement No 824093. The work of P.~Sznajder
was supported by the Grant No.~2024/53/B/ST2/00968
of the National Science Centre, Poland.
\end{acknowledgments}
\bibliography{bibliography}

\end{document}